\documentclass[
  journal=pasa,
  manuscript=research-paper,
  year=2024,
  volume=41,
]{cup-journal}

\usepackage[nopatch]{microtype}
\usepackage{booktabs}
\usepackage{orcidlink}
\DeclareUnicodeCharacter{2212}{-}
\usepackage{amsmath,amssymb} 
\usepackage{natbib,twoopt}
\usepackage[english]{babel}
\usepackage{caption}
\usepackage{subcaption}
\usepackage{enumitem}
\hypersetup{colorlinks, anchorcolor=blue, linkcolor=blue, urlcolor=blue, citecolor = blue}

\title{The ORT and the uGMRT Pulsar Monitoring Program : \\Pulsar Timing Irregularities \& the Gaussian Process Realization}

\author{Himanshu Grover\,\orcidlink{0009-0004-1150-6151}}
\affiliation{Department of Physics, Indian Institute of Technology Roorkee, Roorkee 247667, India}
\email[H. Grover]{himanshu\_g@ph.iitr.ac.in}

\author{Bhal Chandra Joshi\,\orcidlink{0000-0002-0863-7781}}
\affiliation{National Centre for Radio Astrophysics, TIFR, P Bag 3, Ganeshkhind, Pune - 411007 India}
\alsoaffiliation{Department of Physics, Indian Institute of Technology Roorkee, Roorkee 247667, India}

\author{Jaikhomba Singha\,\orcidlink{0000-0002-1636-9414}}
\affiliation{High Energy Physics, Cosmology \& Astrophysics Theory (HEPCAT) Group, Department of Mathematics and Applied Mathematics, University of Cape Town, Cape Town 7700, South Africa}
\alsoaffiliation{Department of Physics, Indian Institute of Technology Roorkee, Roorkee 247667, India}
        
\author{Erbil Gügercinoğlu}
\affiliation{National Astronomical Observatories, Chinese Academy of Sciences, 20A Datun Road, Chaoyang District, Beijing, 100101, China} 
 
\author{Paramasivan Arumugam\,\orcidlink{0000-0001-9624-8024}}
\affiliation{Department of Physics, Indian Institute of Technology Roorkee, Roorkee 247667, India}	
        
\author{Debades  Bandyopadhyay\,\orcidlink{0000-0003-0616-4367}}
\affiliation{Department of Physics, Aliah University, New Town - 700160, India}

\author{James O. Chibueze\,\orcidlink{0000-0002-9875-7436}}
\affiliation{Department of Mathematical Sciences, University of South Africa, Cnr Christian de Wet Rd and Pioneer Avenue, Florida Park, 1709, Roodepoort, South Africa}
\alsoaffiliation{Department of Physics and Astronomy, Faculty of Physical Sciences, University of Nigeria, Carver Building, 1 University Road, Nsukka 410001, Nigeria}
        
\author{Shantanu Desai\,\orcidlink{0000-0002-0466-3288}}
\affiliation{Department of Physics, Indian Institute of Technology Hyderabad, Kandi, Telangana 502284, India}   
        
\author{Innocent O. Eya\,\orcidlink{0000-0002-9693-7804}}
\affiliation{Physics/Electronics Technique – Department of Science Laboratory Technology, University of Nigeria, Nsukka, Nigeria}   
\alsoaffiliation{Department of Physics and Astronomy, Faculty of Physical Sciences, University of Nigeria, Carver Building, 1 University Road, Nsukka 410001, Nigeria}   
        
\author{Anu Kundu\,\orcidlink{0000-0003-2128-1414}}
\affiliation{Centre for Space Research, North-West University, Potchefstroom 2520, South Africa}   

\author{Johnson O. Urama}
\affiliation{Department of Physics and Astronomy, Faculty of Physical Sciences, University of Nigeria, Carver Building, 1 University Road, Nsukka 410001, Nigeria}


\begin{document}

\begin{abstract}
The spin-down law of pulsars is generally perturbed by two types of timing irregularities: glitches and timing noise. Glitches are sudden changes in the rotational frequency of pulsars, while timing noise is a discernible stochastic wandering in the phase, period, or spin-down rate of a pulsar. We present the timing results of a sample of glitching pulsars observed using the Ooty Radio Telescope (ORT) and the upgraded Giant Metrewave Radio Telescope (uGMRT). Our findings include timing noise analysis for 17 pulsars, with seven being reported for the first time. We detected five glitches in four pulsars and a glitch-like event in PSR J1825--0935. The frequency evolution of glitches in pulsars, J0742--2822 and J1740--3015, is presented for the first time. Additionally, we report timing noise results for three glitching pulsars. The timing noise was analyzed separately in the pre-glitch region and post-glitch regions. We observed an increase in the red noise parameters in the post-glitch regions, where exponential recovery was considered in the noise analysis. Timing noise can introduce ambiguities in the correct evaluation of glitch observations. Hence, it is important to consider timing noise in glitch analysis. We propose an innovative glitch verification approach designed to discern between a glitch and strong timing noise. The novel glitch analysis technique is also demonstrated using the observed data.
\end{abstract}

\section{Introduction}
Pulsars are neutron stars with exceptionally stable rotation and good timekeepers due to their huge moment of inertia, magnetic fields, and compact sizes. In particular, millisecond pulsars rival the best atomic clocks. After accounting for the steady slowdown of the underlying neutron star due to the magneto-dipole radiation or particle outflows in the form of wind, the accuracy of their spin period is even better than one part in $10^{11}$. This allows for predicting the lengthening of the period of a given pulsar once its timing solution is obtained for a reference epoch by the pulsar timing technique. Regular spin evolution of a pulsar is occasionally interrupted by variations and rare discontinuities that cause phase deviations, the so-called timing irregularities. The long-term timing observations have revealed that the rotations of many pulsars are affected by two types of irregularities \citep{Alessandro_1996}: glitches and timing noise. 

A glitch is a sporadic event where a pulsar shows an abrupt change in its rotation rate. Glitches in the rotation rate is usually accompanied by even larger step increases in the spin-down rate. Both of these changes tend to relax back toward the original pre-glitch values on timescales ranging from several days to a few years. In contrast, timing noise refers to a random variation in the rotational parameters of the neutron star. It is usually quasi-periodic in nature, and its frequency spectrum can be characterized by a power-law.

The first detected rotational spin-up in the Vela pulsar occurred on Modified Julian Date (MJD) 40280, February 28, 1969 \citep{Radhakrishnan_Manchester_1969, Reichley_Downs_1969}, followed by a similar event detected in the Crab pulsar in the same year on MJD 40491.80, November 1969 \citep{Downs_1981}. These events were called  “pulsar glitches"
and motivated a deeper exploration of neutron stars' internal structure and dynamics and shedding light on dense matter under extreme conditions. Soon after the detection of the glitches, several models were developed to explain the phenomenon; refer to reviews \citet{Haskell_2015, zhou_erbil_glitch_review, Antonopoulou_2022} and \citet{Antonelli_2022} for more details. The most successful model explaining glitches and their recoveries is based on the superfluid vortex pinning and creep mechanisms \citep{AndersonItoh1975, AlparvortexcreepI1984, Gugercinoglu_2020}, which states that internal superfluid state leads to the development of a differential rotation between the observed crust and stellar interior by immobilization of vortex lines via pinning to lattice nuclei and flux tubes in the outer core. Hence, a large amount of angular momentum gets stored. The crust spins up whenever this angular momentum is abruptly released to produce a glitch. The much slower post-glitch recovery takes place as a result of the weak interaction between the superfluid components and normal matter crust. Pulsar glitches thus probe the superfluid dynamics inside the star.

It has been more than 55 years since the first detection, and only about 700 glitches in around 230 pulsars (combined data from the Jodrell Bank Observatory\footnote{\label{note1}\url{http://www.jb.man.ac.uk/pulsar/glitches/gTable.html}} \citep{basu2022} and the ATNF pulsar catalogue\footnote{\label{note2}\url{https://www.atnf.csiro.au/research/pulsar/psrcat/}} \citep{Manchester_ATNF}) have been recorded, making them rare events. Several types of spin variations have been observed, resulting in the classification of glitches into four categories, namely, conventional glitches, slow glitches, glitches with delayed spin-ups, and anti-glitches, e.g., see the review \citep{zhou_erbil_glitch_review}. 
A conventional glitch is a glitch that shows an abrupt spin-up and exhibits a typical exponential recovery followed by a linear decay of the increase in the spin-down rate. In some events, gradual increments in the frequency have been observed. These gradual change events are referred to as slow glitches \citep{Shabanova1998_slowglitch}. The typical timescale for the formation of a slow glitch can vary from several months to years \citep{Cordes1980}. Only 31 slow glitches in seven pulsars have been reported so far \citep{Zhou2019_slowglitch}. Some of the glitches seen in the Crab pulsar, including its largest 2019 glitch, have displayed a unique behaviour; glitch magnitude attained its maximum value with a delay of about a few days from the beginning of the spin-up-event, forming a category called delayed-spin up glitches \citep{Lyne_2015, Gugercinoglu_2019, Basu2020}. A sudden negative step jump in the spin frequency is called an anti-glitch \citep{Archibald2013_anti-glitch}. Such kind of events have been identified several times, predominantly occurring in magnetars \citep[e.g.,][]{Icdem2012_anti-glitch, Archibald2013_anti-glitch, Sinem2014_anti-glitch, Pintore2016_anti-glitch, Ray2019_anti-glitch, Younes2020_anti-glitch}.

The migration and repinning of the vortices define the post-glitch behavior. It generally exhibits an exponential recovery followed by a linear decay of the increase in the spin-down rate \citep{Yu2013}. Post-glitch recovery behavior can be utilized to study and evaluate various statistical parameters that probe the interior structure and superfluid/superconducting dynamics inside the neutron stars. The post-glitch behavior provides insights into the following scientific aspects: (i) the fractional Moment of Inertia of various layers that participated in a glitch, (ii) the re-coupling timescale of the crustal superfluid, (iii) the theoretical prediction of the time to the next glitch, (iv) constraining internal magnetic field configuration and equation of state of neutron stars, (v) finding a correlation between glitch observables, (vi) the range of proton effective mass in the core, and (vii) a robust internal temperature estimate for neutron stars and many more \citep{Gugercinoglu_2020,erbil2022}.

Various statistical studies have utilized the dataset of pulsar glitches to understand glitches and their correlations with various pulsar parameters. It has been observed that glitch size is at least bimodal \citep{Eya_2019,Celora2020,Arumugam_Desai_2023}, categorizing glitches into two groups: small and large glitches. It is not well known whether the triggering mechanisms for large and small glitches are the same. Many argue that a similar mechanism is responsible for the occurrence of small glitches and timing noise. A proportional behaviour has been observed between glitch activity and spin-down rate \citep{Urama1999, Lyne_2000, Fuentes_2017}. Glitches are observed to be more dominant in younger pulsars. This is partly due to the fact that in young pulsars for which the spin-down rate is comparatively higher, the critical lag for collective vortex unpinning avalanches is easier to achieve \citep{erbil2016}.

Unlike pulsar glitches, timing noise is less understood, partly as characterizing timing noise not only requires long timing baselines but can often be corrupted by noise introduced by the ionized interstellar medium (IISM), particularly in young distant pulsars, which are often found in complex environments. Nevertheless, there have been several studies to characterize the timing noise in pulsars. One such study showed that a power law adequately describes the timing noise in the millisecond and canonical pulsars~\citep{Shannon_2010}. However, for magnetars, the same power law does not hold due to the potential influence of magnetic field decay and frequently occurring crustquakes on the generation of timing noise, giving a hint that the strength of timing noise can be expressed as a combination of the rotational frequency of the pulsar and its derivatives. It was observed that the strength of timing noise differs significantly among canonical pulsars, millisecond pulsars, and magnetars, with variations spanning over eight orders of magnitude \citep{parthasarathy2019_TimingI}. It is conjectured that there is a correlation between the strength of timing noise and the rotational frequency of the pulsar and its derivatives \citep{Shannon_2010, parthasarathy2020_TimingII, lower2020_utmost2}. Thus, detailed studies on these aspects are needed to understand the relationship between timing noise with the various properties of the pulsar, like rotation, magnetic field, age, etc.

It has been a mystery whether all types of reported glitches are real glitches or a manifestation of timing noise. The traditional glitch analysis does not consider the possible irregularities that can occur because of timing noise, which can masquerade as a small glitch. The advancements in timing noise modelling in recent years, primarily driven by Gravitational wave analysis, motivated us to develop an innovative glitch analysis methodology that is capable of differentiating a glitch from a manifestation of the timing noise. Previously, a semi-automated approach for glitch detection with a hidden Markov model (HMM) has been developed \citep{Melatos2020_HMM} as an alternative to the traditional glitch analysis. This method has been utilized for glitch detection and defining the upper limit on the glitch size \citep{Dunn2022_HMM_upper_limits, Dunn2023_HMM_Vela}. The HMM incorporates both glitches and timing noise; however, HMM models timing noise as white noise in the torque derivative (an approximation). A majority of pulsars with glitches show timing noise with power-law spectra, which motivates including a stationery timing noise model derived from long-term timing baselines to evaluate if a small glitch is real. This is particularly important when there are large data gaps in data, making phase tracking a challenging task, or there is no apparent recovery, distinguishing a glitch from timing noise. 

In this paper, we present the results of our monitoring program using the Upgraded Giant Metrewave Radio Telescope (uGMRT) \citep{ugmrt2017} and the Ooty Radio Telescope (ORT) \citep{swarup1971ort}. We detect several glitches in our data, obtain timing noise models for pulsars in our monitoring program, and evaluate these glitches against the timing noise model using a different approach developed by us. The outline of the paper is as follows: We describe the pulsar sample and observations using the uGMRT and the ORT in Section~\ref{PulsarObservations}. In Section~\ref{Analysis}, we explain our analysis methods. Section~\ref{GP_realization} introduces the novel glitch verification methodology. The glitch detection report, results of timing noise analysis, and novel glitch analysis demonstration are discussed in Section~\ref{ResultsandDiscussions}. The conclusions and prospects of future work are given in Section~\ref{conclusion}. 

\section{Pulsar Sample and Observations}\label{PulsarObservations}

We regularly monitor a sample of 24 pulsars using the ORT and the uGMRT to investigate timing irregularities in pulsars. Our sample of pulsars is listed in Table~\ref{sample}, consisting of pulsars of different ages (listed in column 4 of Table~\ref{sample}) that have displayed several glitches of various amplitudes in the past. We started with a smaller sample of frequent glitching pulsars, following the selection criteria described in \citep{Basu2020}. Later, we increased our sample to include pulsars that exhibit glitches of varying amplitudes. This selection criterion enables us to test the occurrence of small glitches and to distinguish timing noise from real glitches. 

\begin{table*}
\centering
\caption{List of pulsars observed using ORT and uGMRT. The columns from left to right represent the pulsar Jname, pulsar period ($P$), the time derivative of the period ($\dot{P}$), characteristic or spin-down age ($\tau$), the surface magnetic field of the pulsar as inferred from the dipolar spin-down law ($B$), Dispersion Measure (DM), observatory name where the observation took place (Telescope), and the data availability period (in MJD).}
\renewcommand\arraystretch{1.8}
\setlength{\tabcolsep}{6pt}
\begin{tabular}{cccccccc}
\hline
\hline
PSR JName & P & $\dot{P}$ ($10^{15}$) & $\tau$ & B & DM & Telescope & Available Data\\
 & (s) & (dimensionless) &  (kyr) & ($10^{12}$ G) & (pc cm$^{-3}$) &  & (MJD)\\ 
\hline
\hline
\centering
J0358$+$5413 & 0.15638412 & 4.39 & 563.77 & 0.84 & 57.14 & ORT & 57814--60156\\
\hline
J0525$+$1115 & 0.35443759 & 0.07 & 76295.26 & 0.16 & 79.42 & ORT & 58899--60137\\
\hline
J0528$+$2200 & 3.74553925 & 40.1 & 1481.65 & 12.39 & 50.87 & ORT & 57818--60243\\
\hline
J0534$+$2200 & 0.03339241 & 421 & 1.26 & 3.79 & 56.77 & ORT \& uGMRT & 56729--60248\\
\hline
J0659$+$1414 & 0.38492862 & 54.9 & 111.01 & 4.65 & 13.94 & ORT & 58900--60134\\
\hline
J0729--1448 & 0.25165871 & 113 & 35.20 & 5.40 & 91.89 & uGMRT & 58080--60245 \\
\hline
J0729--1836  & 0.51016034 & 19.0 & 426.38 & 3.15 & 61.29 & uGMRT & 58080--60245 \\
\hline
J0742--2822 & 0.16676229 & 16.8 & 157.08 & 1.69 & 73.73 & ORT & 56729--60248\\
\hline
J0835--4510 & 0.08932839 & 125 & 11.32 & 3.38 & 67.77 & ORT \& uGMRT & 56729--60248 \\
\hline
J0922$+$0638 & 0.43062710 & 13.7 & 496.95 & 2.46 & 27.30 & ORT & 58898--60149 \\
\hline
J1532$+$2745 & 1.12483574 & 0.78 & 22861.49 & 0.95 & 14.69 & ORT & 58899--60150\\
\hline
J1720--1633 & 1.56560115 & 5.80 & 4278.30 & 3.05 & 44.83 & ORT & 58900--60141\\
\hline
J1731--4744 & 0.82982879 & 164 & 80.35 & 11.79 & 123.06 & ORT \& uGMRT & 57856--60150 \\
\hline
J1740--3015 & 0.60688662 & 466 & 20.63 & 17.02 & 151.96 & uGMRT & 58070--60262 \\
\hline
J1803--2137 & 0.13366692 & 134 & 15.76 & 4.29 & 233.99 & uGMRT & 59163--60262\\
\hline
J1825--0935 & 0.76902100 & 52.4 & 232.72 & 6.42 & 19.38 & ORT \& uGMRT & 57820--60247 \\
\hline
J1826--1334 & 0.10148679 & 75.3 & 21.37 & 2.80 & 231.0 & uGMRT & 59163--60262 \\
\hline
J1847--0402 & 0.59780875 & 51.7 & 183.24 & 5.63 & 141.98 & uGMRT & 58089--60262 \\
\hline
J1909$+$0007 & 1.01694836 & 5.52 & 2919.71 & 2.40 & 112.79 & ORT & 58901--60141 \\
\hline
J1910--0309 & 0.50460606 & 2.19 & 3646.01 & 1.06 & 205.53 & ORT & 57851--60224 \\
\hline
J1919$+$0021 & 1.27226037 & 7.67 & 2628.20 & 3.16 & 90.32 & ORT & 58900--60143 \\
\hline
J2022$+$2854 & 0.34340216 & 1.89 & 2872.40 & 0.82 & 24.63 & ORT & 58900--60140 \\
\hline
J2219$+$4754 & 0.53846882 & 2.77 & 3085.31 & 1.23 & 43.50 & ORT & 58900--60144 \\
\hline
J2346--0609 & 1.18146338 & 1.36 & 13733.26 & 1.28 & 22.50 & ORT & 58899--60140 \\
\hline
\hline
\end{tabular}
\label{sample}
\end{table*}

We use India's two largest radio telescopes to observe the listed pulsars. The first telescope used was the ORT \citep{swarup1971ort}, a cylindrical paraboloid with an aperture of 530m $\times$ 30m, i.e., 530m long in the north-south direction and 30m wide in the east-west direction, covering a declination range of --60\textdegree \space to +60\textdegree. It was built on a hill with the same latitude as the geographic latitude (11\textdegree), making it an equatorially mounted telescope and allowing it to track sources for about 8 hours. The telescope has a total collecting area of around 16000 m$^2$ and an effective collecting area of around 8600 m$^2$. It provides an angular resolution of 2.3deg (in right ascension) and 5.5sec (declination) arcminute for the total power system. The primary reflector is made up of a network of around 1500 stainless steel wires running along the north-south direction. The cylindrical reflector directs the radio signals to 1056 dipoles located at the focal line in the north-south direction. Each set of 48 dipoles is arranged into a subarray called a module. The telescope has 22 modules, which are phased to form a module beam. ORT has a beam-forming network consisting of 12 beams. Observations in the ORT are conducted at a central frequency of 326.5 MHz. The Pulsar Ooty Radio Telescope New Digital Efficient Receiver, PONDER \citep{Naidu_2015}, which supports four modes for observations, provides us with real-time coherent dedispersed time series data. Dedispersion is performed with respect to the highest frequency in the band, which is 334.5 MHz. The DSPSR software \citep{DSPSR} was used for folding the raw data using the ephemeris created in our initial observations. It generates PSRFITS \citep{PSRCHIVE_and_PSRFITS2004} files. The data generally have 128 phase bins and at least three sub-integrations, used to provide reliable preliminary confirmations of glitches, even with one post-glitch epoch. Also, using three or more sub-integrations is beneficial for automated glitch detections \citep{Singha_2021_agdp}. At the ORT, bright glitching pulsars with low Dispersion Measure (DM), preferably less than 100 pc cm$^{-3}$, are being observed with a cadence of 1--14 days.

The second telescope used is the uGMRT \citep{Swarup1991GMRT, Ananthakrishnan1995GMRT}, an array of thirty 45-meter fully steerable parabolic antennas extended over a baseline of 25km, covering a declination range of -53\textdegree \space to +90\textdegree. The uGMRT \citep{ugmrt2017} supports observations in four bands: (i) Band 2: 125 to 250 MHz, (ii) Band 3: 250 to 500 MHz, (iii) Band 4: 550 to 850 MHz, and (vi) Band 5: 1000 to 1460 MHz. We have utilized Band 3, Band 4, and Band 5 for our pulsar observations. The phased array mode with an antenna configuration, including all the central square antennas and the first arm antennas, was used for observations. The data was recorded with 1024 and 2048 channels over a bandwidth of 200 MHz and 400 MHz, respectively. To reduce the raw data, we have used the \texttt{PINTA} pipeline \citep{pinta}, which produces PSRFITS files. The sources with high DM experience scatter broadening in the ORT and hence are chosen to be observed with the uGMRT. Additional criteria are that the glitching pulsars of variable glitch amplitudes and optimum observation time, in our case, less than 20 minutes each, are selected for observations. The cadence of the observations at uGMRT is 15-30 days. A few sources are common in both uGMRT and ORT either because they show exceptional post-glitch behaviour or are known to show radiative changes correlating with glitches, which is hard to show with the ORT, as it is a single polarisation telescope and can be better shown with uGMRT.
We prefer to observe pulsars with low DM at lower frequencies, where they require less integration time, as pulsars are intrinsically bright at lower radio frequencies. On the other hand, the pulsars with high DM were observed using higher frequencies, primarily at the uGMRT, to mitigate the impact of scatter broadening at lower frequencies. The data typically have 128 phase bins and at least three sub-integrations. At least three sub-integrations were used to provide reliable preliminary confirmations of glitches, even with one post-glitch epoch.

\section{Analysis}\label{Analysis}

\subsection{Timing Analysis}\label{TimingAnalysis}
The data obtained from observations using the uGMRT and the ORT were in PSRFITS format. Initial analysis was performed using \texttt{PSRCHIVE} \citep{PSRCHIVE_and_PSRFITS2004, PSRCHIVE2012}. The \texttt{psrstat} and \texttt{psrplot} functions were used for information extraction and visualization purposes. The RFI mitigation, including removal of bad subints, channels, or bins, was executed using \texttt{pazi}. The \texttt{pam} command was used
to collapse the archives in frequency and/or in time.
The command \texttt{paas} was utilized to generate a noise-free template profile from high signal-to-noise ratio profiles observed during observations. The template profile was cross-correlated with all other observed profiles using the \texttt{pat} command, which allows a frequency domain cross-correlation method to obtain Time of Arrivals (ToAs). A timing model, which contains the pulsar’s intrinsic spin evolution (spin frequency and its time derivatives), astrometric terms (position and proper motion), parallax, DM, and frame-of-reference terms, is utilized in the generation of the timing solution. A good timing solution yields white and minimized residuals obtained by fitting ToAs according to the timing model with a weighted least-squares algorithm \citep{BCJ2018}. The pulsar timing software \texttt{TEMPO2} \citep{Hobbs2006Tempo2, Edwards2006Tempo2} was used to perform the timing analysis of the pulsars. This traditional timing technique is highly successful, can be referred to as the foundation of observational studies in pulsar astronomy, and is widely used for various topics, including the detection of glitches. However, it does not consider the effects of timing noise in glitch verification methodology. In the traditional glitch analysis technique, glitch identification is conducted through visual inspection of the residuals, and a glitch event is defined by an abrupt positive change in spin frequency, $\Delta\nu$, or a negative change in frequency spin-down rate, $\Delta\dot{\nu}$. These two rapid changes simultaneously distinguish glitches from timing noise \citep{Espinoza_2014}. In case of small glitches, many times, no sudden discrete negative change in $\Delta\dot{\nu}$ is observed. Hence, it is necessary to take timing noise into account to verify if such glitches are real or a manifestation of timing noise while analyzing the residuals. We present a technique to estimate the timing noise and to differentiate glitches from timing noise in Sec.~\ref{TN_Analysis} and \ref{GP_realization}, respectively.

\begin{figure}
\centering
    \includegraphics[width=\linewidth]{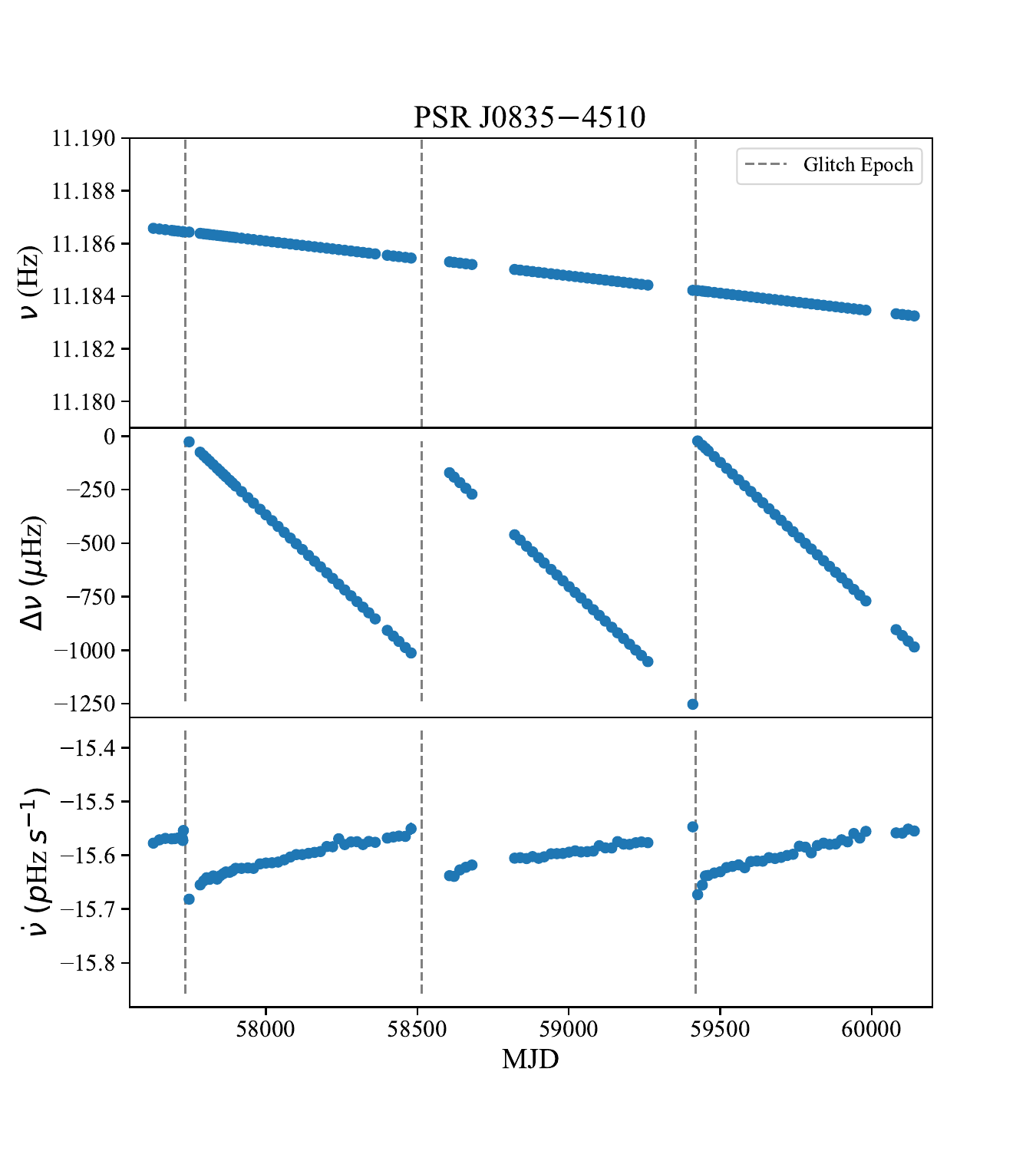}
    \caption{The rotational evolution of the Vela pulsar (PSR J0835--4510). The top panel represents the spin evolution. The middle panel shows the frequency residuals $\Delta \nu$, estimated by subtracting the pre-glitch timing solution, and the bottom panel represents the evolution of the spin-down rate. The vertical lines indicate the glitch epoch.}
    \label{J0835-4510_all}
\end{figure}

The rotational evolution of a pulsar during a glitch can be expressed as a Taylor series expansion \citep{erbil2022},

\begin{equation}
\begin{aligned}[c]
    \label{rotationmodel}
        \nu(t)  = & \nu_0 + \dot{\nu}_0(t-t_0) + \frac{1}{2} \Ddot{\nu}_0(t-t_0)^2 + \Delta \nu_g \noindent \\ & + \Delta\dot{\nu}_g(t-t_g) + \Delta \nu_d \exp(-(t-t_g)/\tau_d) \ ,
\end{aligned}
\end{equation}
\noindent
where, $\nu_0=1/P$ is the spin frequency, and $\dot{\nu}_0$ and $\Ddot{\nu}_0$ are the first and second-time derivatives of frequency at epoch $t_0$, respectively. The change in the spin frequency and its first-time derivative due to a glitch at epoch $t_g$ are represented by $\Delta \nu_g$ and $\Delta \dot{\nu}_g$, respectively. $\Delta \nu_d$ and $\tau_d$ are the amplitude and decay timescale of the exponential relaxation part of a glitch. The procedure followed to obtain the parameters of the model, given in Equation~\ref{rotationmodel}, is described in Sec.~\ref{parameter_estimation}. The long-term spin evolution for one of the pulsars in our sample, including several glitches, is given in Fig.~\ref{J0835-4510_all}.

\subsection{Bayesian Analysis} \label{Bayesian_analysis}
Bayesian analysis is a method of statistical analysis that uses Bayes' theorem to update and revise the probability for the hypothesis of an event based on new evidence or data. Bayes' theorem is the basis of Bayesian statistics, and the foundation of Bayes’ theorem is conditional probability. We provide a succinct description of the methods used, and a more exhaustive discussion of Bayesian regression and model comparison techniques can be found in \citet{Trotta2008, Sanjib, Kerscher} and \citet{Krishak} and references therein. The posterior probability distribution of a set of parameters $\theta$ for data $D$ and a hypothesis/model $H$ is given as:

\begin{equation}
    P(\theta|D,H) = \frac{P(D|\theta,H)P(\theta|H)}{P(D|H)},
\end{equation}
\noindent
where,
\begin{itemize}
    \item $P(\theta|H)$ is the prior, i.e., pre-knowledge or the probability distribution of parameter $\theta$ assuming a model.
    \item $P(D|H)$ is the Bayesian evidence - the probability of data or a normalization constant,
    \item $P(D|\theta,H)$ is called the likelihood - how likely the data is given the model or parameters and,
    \item  $P(\theta|D,H)$ is the Posterior, i.e., the probability that the hypothesis is true for a given data.
\end{itemize}

We have utilized the power of Bayesian analysis for model comparison and parameter estimation of various timing noise factors, which are described in detail in Section~\ref{TN_Analysis}. To select the most probable model, one has to estimate Bayesian Evidence, which is defined as:
\begin{equation}
    P(D|H) =  \int P(D|\theta,H)P(\theta|H) d\theta. \
\end{equation}

To decide the probable hypothesis between two models (say $H_{A}$ and $H_{B}$), we calculate the Bayes Factor (BF), which is defined as the ratio of evidence and can be written as:

\begin{equation}
    BF_{AB} = \frac{P(D|H_{A})}{P(D|H_{B})} = \frac{\int P(D|\theta_{A},H_{A})P(\theta_{A}|H_{A}) d\theta_{A} \ }{\int P(D|\theta_{B},H_{B})P(\theta_{B}|H_{B}) d\theta_{B} \ }.
\end{equation}

Jeffrey's scale \citep{Trotta2008} is used to examine the strength of evidence in favour of one hypothesis or model over another. The scale classifies the evidence into various levels, ranging from ``negative evidence" to ``decisive evidence". The interpretation of the scale is as follows: An evidence ratio less than 1 indicates negative evidence. A BF between 1 and 10 infers no substantial evidence. A BF between 10 and 100 suggests strong to very strong evidence, and if the BF is greater than 100, it indicates decisive evidence in favour of one hypothesis over another. We have utilized Jeffrey's scale for the model selection. Complex models are given preference only if they have decisive evidence. However, if the evidence is not decisive, then we followed Occam’s razor and selected the simplest model with fewer free parameters. We utilized Bayesian analysis to determine the timing noise parameters, discussed in the next section.

\subsection{Timing Noise Analysis}\label{TN_Analysis}
The irregularities in the spin create a time-correlated stochastic function in the TOAs, known as achromatic red noise, red spin noise, or timing noise. The timing noise in our sample of pulsars has been modelled as a Gaussian stationary random process using Fourier basis functions \citep{Lentati2013, Haasteren2014}. Unlike white noise, which has a constant power spectral density across all frequencies, red noise exhibits a power spectrum where the power decreases with increasing frequency. This results in a correlation structure where fluctuations at nearby time points are more strongly correlated than those further apart. The power spectral density of timing noise \citep{van_Haasteren_2013, Haasteren2014,lentati_16, parthasarathy2019_TimingI, lower2020_utmost2} for a given observing frequency $f$, can be written as,
  \begin{equation}
   \label{tn}
   \centering
   \mathcal{P}(f) = \frac{A_{\rm red}^2}{12\pi^2} \left(\frac{f}{f_{\text{yr}}}\right)^{-\gamma} \text{yr}^3\, ,
   \end{equation}
 \noindent
 where, $A_{\rm red}$ is the red-noise amplitude, in units of $\rm yr^{3/2}$, $\gamma$ is the power law index, and $f_{\rm yr}$ = 1.0 / yr. 

Our timing noise analysis model differs from \citet{Melatos2020_HMM} in terms that they have incorporated timing noise as a red noise in the spin-down rate. Other sources of uncertainties, such as radiometer noise and pulse jitter, can result in white noise features in the residuals \citep{Haasteren2014}. 
 These can be accounted by scaling the ToA uncertainties as given below,
  \begin{equation}
   \label{wn}
  \centering
   \sigma^2 = F^2 (\sigma_r^2 + \sigma_Q^2) .
   \end{equation}
Here, $\sigma_r$ is the ToA uncertainty provided by `pat' in \texttt{PSRCHIVE} and $F$ is known as EFAC. The EFAC represents the radiometer noise, which is affected by Radio Frequency Interference (RFI), etc, and hence, its value can vary. However, EFAC is preferably close to unity. The term $\sigma_Q$, known as EQUAD, is the error added in quadrature, constant for all the data points, and describes the stochastic fluctuation in the pulse profile, which can contribute additional noise to the timing data \citep{ Osłowski_2011, Haasteren2014, Kikunaga2024}. While EFAC could be epoch-dependent due to RFI or variations in telescope parameters,  EQUAD represents a time-independent white noise required for accurate error estimation. In timing noise analysis, we use four different combinations of noise models as listed below:
\begin{enumerate}
    \item F: Model consisting of only the white noise scaling  factor (EFAC).
    \item WN: Model consisting of the white noise scaling, EFAC, and the white noise error in quadrature (EQUAD).
    \item F+RN: Model consisting of the white noise scaling factor (EFAC) along with the intrinsic pulsar red noise.
    \item WN+RN: Model consisting of the white noise scaling factor (EFAC), white noise error in quadrature (EQUAD), and the intrinsic pulsar red noise.
\end{enumerate}

Our noise analysis is based on the Bayesian analysis method described in Section~\ref{Bayesian_analysis}. The prior used for our analysis are presented in Table~\ref{priors}. The prior for red noise parameters are chosen based on those used previously \citep{parthasarathy2019_TimingI, lower2020_utmost2} and further constrained during the analysis. The value of EFAC is observatory dependent, while EQUAD depends on the pulsar and observation frequency. The initial prior range of EFAC for uGMRT sources was chosen from previous uGMRT noise analysis \citep{Srivastava2023} as 0.1 to 5. During the initial analysis, we noticed that a range from 0.1 to 3 was also suitable as a prior range for our pulsars, so we used this range for the final analysis. The same range (0.1 to 3) for EFAC was assumed for the sources at the ORT and was found to be a convenient choice. The initial prior range for EQUAD was chosen as -3 to -10 \citep{parthasarathy2019_TimingI}, which includes the prior ranges used before \citep{Srivastava2023} for the uGMRT. During the initial analysis, we observed that the prior range used in \cite{Srivastava2023} and \cite{parthasarathy2019_TimingI} alone was insufficient to model the posterior distribution of several pulsars. Therefore, we expanded our range and found that the range of -1 to -10 \citep{lower2020_utmost2} was a suitable choice for all of our uGMRT and ORT pulsars. The timing noise is modelled using Gaussian likelihood \citep{Haasteren2009}, which is determined with \texttt{Enterprise} \citep{ENTERPRISE_Ellis2019} (Enhanced Numerical Toolbox Enabling a Robust PulsaR Inference SuitE). \texttt{Enterprise} is a pulsar timing analysis code that can perform noise analysis, gravitational-wave searches, and timing model analysis. 

\begin{table}
 \caption{Prior ranges used for the noise analysis of our pulsars. The first column represents the parameter name, followed by the prior range and the prior distribution.}
 \renewcommand\arraystretch{2.0}
 \setlength{\tabcolsep}{6pt}
\centering
\begin{tabular}{ lcr }
\hline \hline
 Parameter & Prior range & Prior distribution \\ \hline\hline
 EFAC (F) & (0.1,3) & Uniform \\
 EQUAD ($\sigma_Q)$ & (--10,--1) & log-Uniform \\ \hline
 Red noise Amplitude ($A_{red}$) & (--18,--5) & log-Uniform \\
 Red noise Spectral index ($\gamma$) & (0,9) & Uniform \\ \hline \hline
\end{tabular}
\label{priors}
\end{table}

\subsection{Fourier Modes Selection}
It is important to select the optimum number of Fourier modes ($M$), as the noise models are sensitive to the number of Fourier bins for each pulsar \citep{Chalumeau2021}. Young pulsars experience strong red noise because of highly dynamic internal and magnetospheric activities. Therefore, it is vital to have higher modes. If the data span is $T_{span}$ (in years), the lowest frequency mode is given as 1/$T_{span}$, and the highest mode is M/$T_{span}$. Therefore, in Fourier mode selection, the sample space is 1/$T_{span}$, 2/$T_{span}$, $\dots$, M/$T_{span}$. The value of modes can be decided by following the Nyquist criteria. Ideally, for daily cadence, the highest frequency is given as $2/$day. Therefore, the value of the highest number of Fourier modes for uniform daily observation is 730. However, our data are not uniform and has gaps between the observations. Additionally, the cadence is not strictly 1 day; for the uGMRT, it is 15--30 days, and for the ORT, 1--14 days. Accordingly, the choice for the highest mode frequency 48/$T_{span}$ signifies a once-a-weak frequency, while the lower component with a frequency of 3/$T_{span}$, four months is an adequate choice for Fourier modes sample space. In principle, we should use all the modes (from 1 to M) and select the model with the largest Bayes factor. However, it can be noted that the changes in the Bayes factor for very nearby modes may not be significant. Furthermore, it may be computationally expensive to use all the modes. Similar to the previous study on modes selection~\citep{Srivastava2023}, we are using a set of five Fourier modes defined as $M = 3, 6, 12, 24, 48$ (1/$T_{span}$), where $T_{span}$ is the time span of the data in years for the optimum mode selection. The Fourier modes selection has been done simultaneously with the model selection. We estimated the Bayesian evidence for all the models with our chosen five sets of Fourier modes. The number of Fourier modes has been selected for GP realization through a visual inspection of observed data and the realization plot. The GP realization includes observations near the glitch epochs; hence, a large number of modes, say around 100, is preferable for obtaining accurate realization plots. However, having a very high value of components does not add any extra advantage.

\subsection{Parameter Estimation}\label{parameter_estimation}
The glitch parameters have been obtained by the traditional timing analysis using \texttt{TEMPO2}. First, a pre-glitch timing solution was obtained, considering only the pre-glitch ToAs. Likewise, a post-glitch timing solution was obtained. Then, the glitch epoch is determined by identifying the x-coordinate of the intersection point (in MJD) of the fitted straight line in the pre-glitch and straight/quadratic curve in post-glitch residuals. Finally, the pre-glitch and post-glitch solutions were obtained with the glitch epoch as the reference. This provides the ratio of change in the rotation frequency corresponding to the post-glitch and pre-glitch solution to the pre-glitch rotational frequency. This fractional change in rotational frequency characterizes as the glitch amplitude. 
We utilized \texttt{Enterprise} to model the noise in pulsars, and \texttt{DYNESTY} (A Dynamic Nested Sampling Package for Estimating Bayesian Posteriors and Evidences) \citep{DYNESTY2020} has been utilized to estimate log evidence for the model and Fourier mode selection. After obtaining the preferred model and an optimum number of modes, we used \texttt{PTMCMCSampler} \citep{ptmcmc} for sampling and parameter estimation. The Python library \texttt{Corner} \citep{corner} has been utilized to plot the posteriors.

\section{The Gaussian Process Realization}\label{GP_realization}
Are all glitches real, or are some of them manifestations of timing noise? How to differentiate between a real glitch and a glitch that appeared because of strong timing noise? We have developed a novel approach to answer these questions. It is a five-step process: 
\begin{enumerate}
    \item Use the traditional glitch analysis process, i.e., using the pulsar timing technique to detect glitch-like events.
    \item Model the timing noise as a red noise process separately for the pre-glitch and post-glitch sections using the Bayesian inference.
    \item Utilize the results of Bayesian noise analysis to reconstruct a GP realization for pre-glitch and post-glitch parts.
    \item Subtract the median GP realization signals from the observed signals, which should result in residuals distributed as white noise.
    \item Use the whitened residuals to verify if the glitch is real or not. A real glitch is characterized by a jump soon after the glitch epoch in the whitened residuals.
\end{enumerate}

To plot the GP realization curve, each value of red noise amplitude $A_{\rm red}$ and spectral index $\gamma$, i.e., chains obtained from the timing noise analysis, are used. The Gaussian process is the family of curves obtained in the timing noise analysis, and the median value of the posterior for $A_{red}$ and $\gamma$ will give the median curve; if one subtracts the median curve from the residuals, then we obtain whitened residuals. For our analysis, we utilized the Python \texttt{LA-FORGE} \citep{La_Forge} package for the GP realization. 


The demonstration of the process is given in Section~\ref{GPR_results}, where we used a large glitch event (in PSR J0835--4510) and a small glitch (in pulsar J1847--0402), which appears like a glitch, but it turned out to be a manifestation of the timing noise. The GP realization helped us to verify these events. A step increase in the rotational frequency due to strong timing noise or lack of phase connection (due to large observation gaps) can lead to a small glitch. Hence, we recommend utilizing this technique to test all glitches.

\section{Results and Discussions}\label{ResultsandDiscussions}
We started the glitch monitoring program using the ORT and the uGMRT in 2014. The first results of this monitoring program, with 11 glitches in 8 pulsars, were reported in \citet{Basu2020}. Here, we present the detection of five glitches in four pulsars. Additionally, we present timing noise results for 20 pulsars, including three glitching pulsars, where the timing noise has been estimated for the pre-glitch and the post-glitch regions. Furthermore, we demonstrate our novel method to differentiate glitches from strong timing noise using the GP realization.

\subsection{Glitches}
\begin{table*}
\centering
\caption{The parameters of all glitches presented in this work. The J name of the pulsars, the epoch of the glitch, the pre-glitch rotation frequency, and the rotation frequency derivative at the glitch epoch are listed in the first four columns, respectively. The last two columns present the fractional change in the rotational frequency and its time-derivative, respectively. Pulsar with $*$ in JName represents the glitch-like event.}
\renewcommand\arraystretch{2}
\setlength{\tabcolsep}{8pt}
\begin{tabular}{ccccccccc}
\hline
\hline
PSR JName & Glitch Epoch & Pre-glitch &  Pre-glitch & $\frac{\Delta \nu}{\nu}$ &  $\frac{\Delta \dot\nu}{\dot\nu}$\\
\vspace{0.1mm}
 &  (MJD) & $\nu$ (Hz) & $\dot\nu$ & ($\times 10^{-9}$) & ($\times 10^{-3}$)\\
\hline
\hline
\centering
J0534+2200 & 58686.4 $\pm$ 0.8 & 29.6169020897(8) & --3.683384(6)$\times 10^{-10}$ &  26.6(3) & 0.6(1)\\
\hline
J0742--2822 & 59839.8 $\pm$ 0.5 & 5.9960110833(7) & --6.044(5)$\times 10^{-13}$ & 4299(10) & 90(8) \\
\hline
J0835--4510 & 58517 $\pm$ 7 & 11.1853972417(3) & --1.555659(4)$\times 10^{-11}$ & 2471(6) & 6(2) \\
& 59417.6 $\pm$ 0.1 & 11.184208478(5) & --1.5550(8)$\times 10^{-11}$ & 1235(5) & 8.0(7) \\
\hline
J1740--3015 & 59934.8 $\pm$ 0.5 & 1.6471971700(6) & --1.263924(7)$\times 10^{-12}$ & 327(1) & 1.3(3)  \\
\hline
J1825--0935$^*$ & 58025.6 $\pm$ 0.4 & 1.30035136047(3) & --8.8445(4)$\times 10^{-14}$ & 4.5(2) & -- \\
\hline
\hline
\end{tabular}
\label{glitch_table}
\end{table*}

We report the detection of 5 glitches in our 4 pulsars and a glitch-like event in PSR J1825--0935. The results are tabulated in Table~\ref{glitch_table}. The time evolution of two glitches, in PSR J0742--2822 and PSR J1740--3015, is being reported for the first time. 
Some of the glitches have also been reported by us earlier as Astronomer's Telegram\footnote{\label{note3}\url{https://astronomerstelegram.org/}} \citep{Atel_Singha_Vela_2021, ATel_Grover_J0742-2822, ATel_Grover_J1740-3015} as well as by other pulsar-monitoring programs \citep{basu2022, Lower_2021, Zubieta2024}. Most of the glitches reported here are large glitches. In addition to these large glitches, we detected around 8 glitch-like events. However, our new analysis technique implied that seven of these are unlikely to be glitches as these appear to be consistent with timing noise in these pulsars. In Section~\ref{GPR_results}, we critically examine a small glitch in PSR J1847--0402 with our new technique and show that this glitch is consistent with timing noise. We find that often such an event looks like a glitch either due to low cadence observation or big gaps between the observations. Therefore, we decided not to report these as glitches. We are only reporting one glitch-like event in PSR J1825--0935, which appears to be a slow glitch. More description of the event is provided in Section~\ref{slowglitch_J1825}. Now, we concisely describe glitches in each pulsar. 

\subsubsection{PSR J0534+2200}
\begin{figure}
    \centering
    \includegraphics[width=\linewidth]{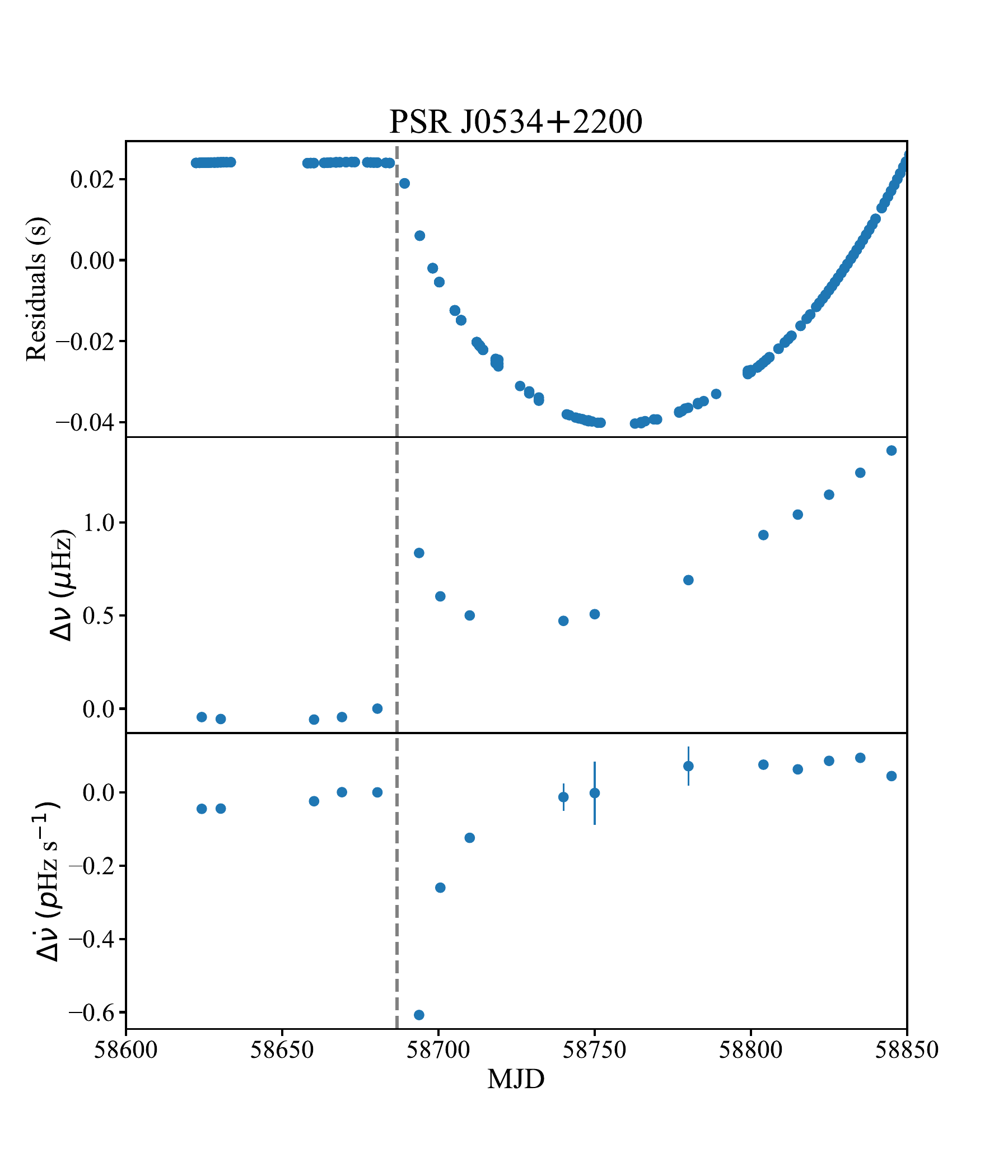}
    \caption{A glitch seen in PSR J0534+2200 on MJD 58686. The top panel shows the residuals. The middle panel displays the $\Delta \nu$ evolution, and the bottom panel presents the evolution of $\Delta \dot\nu$. The glitch epoch is shown by the vertical line.}
    \label{Glitch_J0534+2200_58686}
\end{figure}

The Crab pulsar (PSR J0534+2200) is one of the frequently glitching pulsars, with about 30 glitches observed since its first glitch. This pulsar is monitored regularly with a cadence of 1-3 days using the ORT and a biweekly cadence using the uGMRT. In our monitoring program, we have detected multiple glitches, including the largest ever glitch seen in this pulsar \citep{Basu2020}. In this work, we present the glitch parameters of the 30th glitch reported for the Crab pulsar \citep{Shaw2021}, estimated to occur at MJD 58686.4(8). We calculated the fractional increase in the spin to be $26.6(3)\times10^{-9}$, and the time derivative of spin to be $0.6(1)\times10^{-3}$. The timing residuals and time evolution of the frequency and its derivative are presented in Fig.~\ref{Glitch_J0534+2200_58686}. The previous glitches from our monitoring program, which were published in the previous paper \citep{Basu2020} are not included in the present work. 

\subsubsection{PSR J0742--2822}
\begin{figure}
\centering
    \includegraphics[width=\linewidth]{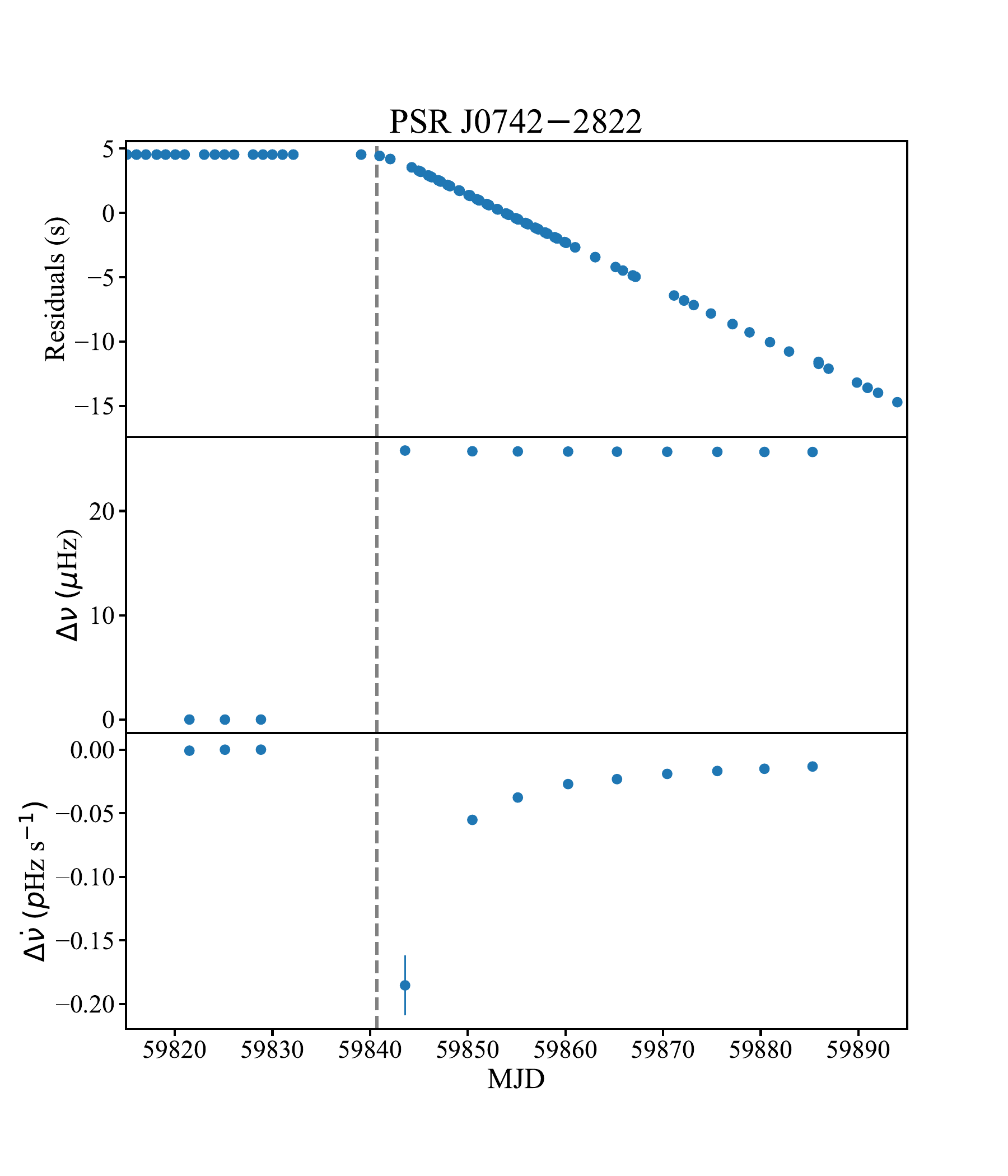}
    \caption{A glitch seen in PSR J0742--2822. The top panel shows the residuals. The middle and the bottom panel show the evolution of  $\Delta \nu$ and $\Delta \dot\nu$. The vertical line indicates the glitch epoch.}
    \label{Glitch_J0742-2822_59839}
\end{figure}

With an age of 157 Kyr, PSR J0742--2822 is an intriguing old glitching pulsar, which has displayed many glitches of various amplitudes. It is also known to show frequent changes in the magnetospheric state that may occur due to a glitch \citep{Keith2013}, making it an excellent candidate to investigate. It is currently monitored at the ORT with a cadence of 1-3 days. We observed the largest glitch detected in this pulsar on MJD 59839.8(5), which was detected by the Automated Glitch Detection Pipeline implemented at the ORT \citep{Singha_2021_agdp} and reported by us in the Astronomer's Telegram \citep{ATel_Grover_J0742-2822}. The glitch was also reported by \citet{ATel_shaw_J0742}, further confirmed by \citet{ATel_Dunn_J0742}, and \citet{ATel_Zubieta_J0742} on the Astronomer's Telegram. This article presents a more complete analysis of all our data and final updates to our preliminary results reported in the Astronomer's Telegram \citep{ATel_Grover_J0742-2822}. The fractional rise in the spin is observed as $4299(10)\times10^{-9}$, and the time derivative of spin is estimated to be $90(8)\times10^{-3}$. We also note around 35-day recovery in the glitch, which is being reported for the first time and matches with recent reports \citep{Zubieta2024}. The glitch parameters are broadly consistent with preliminary estimates reported by other groups in the Astronomer's Telegram\citep{ATel_shaw_J0742,ATel_Dunn_J0742,ATel_Zubieta_J0742}, while the timing residuals, frequency, and 
frequency derivative evolution, shown in Fig.~\ref{Glitch_J0742-2822_59839}, are being reported for the first time.

\subsubsection{PSR J0835--4510}
\begin{figure}
\centering
\includegraphics[width=\linewidth]{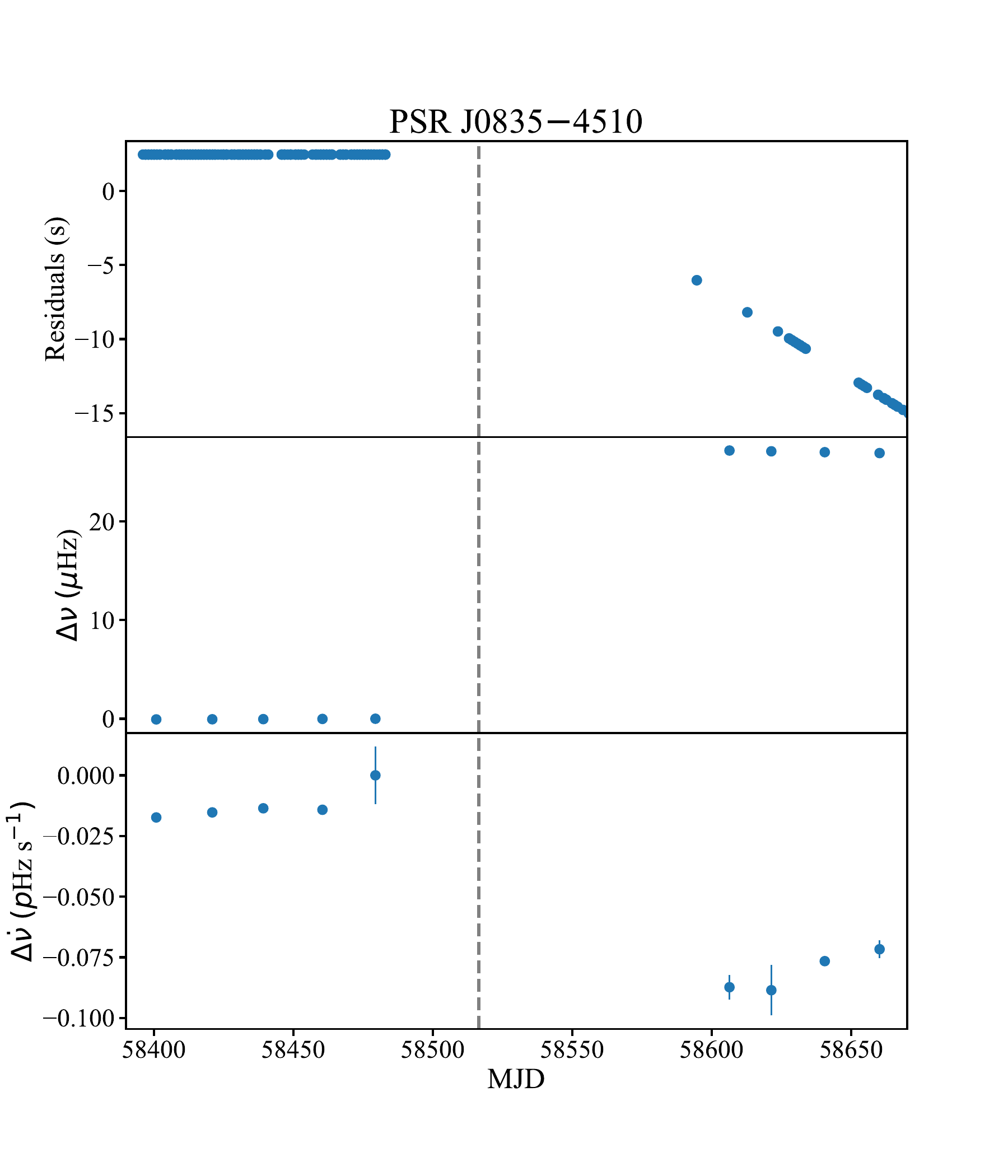}
\includegraphics[width=\linewidth]{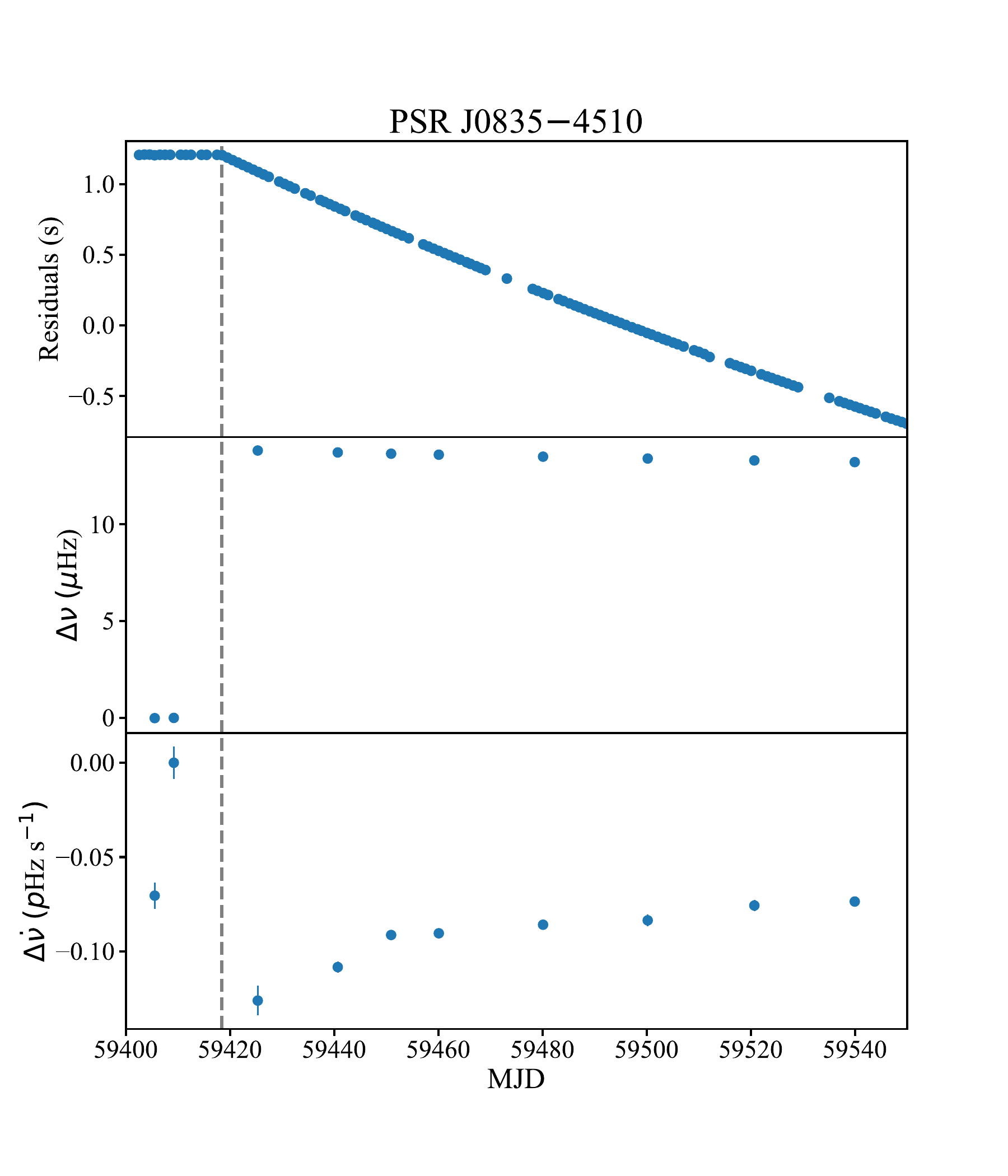}
\caption{Two glitches observed in PSR J0835--4510 on MJD 58517 and 59417 respectively. The top panels represent the residuals. The middle and the bottom panels display the evolution of  $\Delta \nu$ and $\Delta \dot\nu$. The vertical line indicates the glitch epoch.}
\label{J0835_gliches}
\end{figure}

The Vela Pulsar, PSR J0835--4510, is a fantastic pulsar for glitch observations, as it shows regular glitches of similar amplitudes (large glitches with fractional sizes $\gtrsim10^{-6}$). We have been regularly monitoring this pulsar using the ORT with a cadence of 1-3 days and a biweekly cadence using the uGMRT. We present two recent glitches observed in the Vela pulsar using the ORT. Both these glitches are large glitches, which are the characteristic features of the glitches seen in this pulsar. The timing residuals, the evolution of the frequency, and its time derivatives for both the glitches are shown in Fig.~\ref{J0835_gliches}. We detected its twenty-third (G23) \citep{ATelVelaG23_Kerr, lower2020_utmost2, erbil2022} and twenty-fourth (G24) glitches \citep{ATel_Sosa_G24_Vela, Zubieta_2023}. We estimated the glitch epoch for G23 as MJD 58517(7), the fractional change in frequency is $2471(6)\times10^{-9}$, and the fractional increase in the time derivative of frequency is $6(2)\times10^{-3}$; however, the reported values in the literature for the fractional increases are slightly higher. This is probably due to the number of pre-glitch and post-glitch epochs used for estimations and uncertainty due to slight differences in epoch estimation. For glitch G24, the glitch epoch is estimated as MJD 58417.6(1), the fractional change in frequency is $1235(5)\times10^{-9}$, and the time derivative of frequency is $8.0(7)\times10^{-3}$. The fractional change in frequency is again slightly lower than the value reported on the Jodrell Bank Observatory catalogue; again, this could be because of the number of pre-glitch and post-glitch epochs used for estimations and uncertainty due to slight differences in epoch estimation.

\subsubsection{PSR J1740--3015}
\begin{figure}
\centering
    \includegraphics[width=\linewidth]{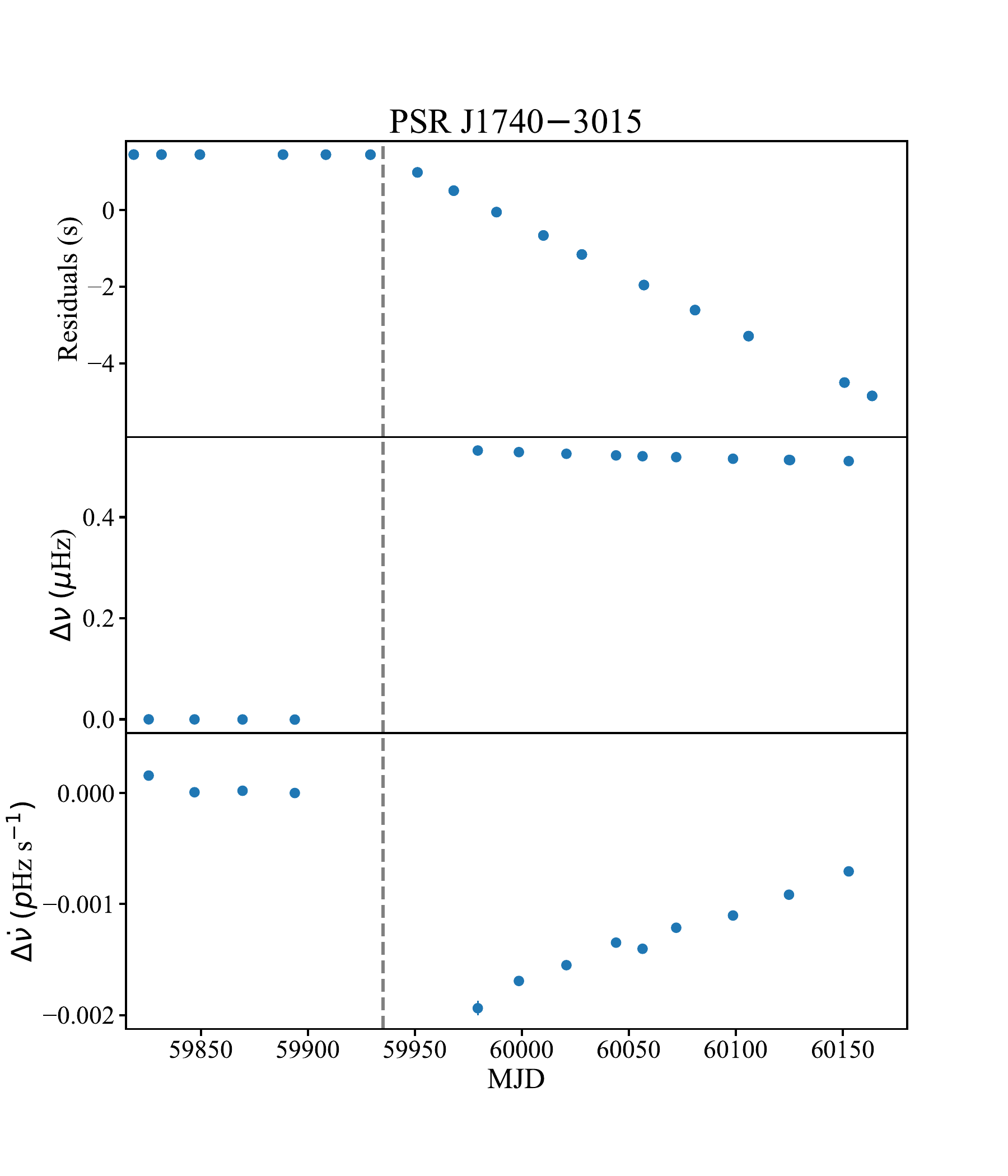}
    \caption{A glitch detected in PSR J1740--3015. The top panel shows the residuals. The middle panel displays the $\Delta \nu$ evolution, and the bottom panel presents the evolution of $\Delta \dot\nu$. The vertical line represents the glitch epoch}
    \label{Glitch_J1740-3015}
\end{figure}
PSR J1740--3015 is one of the most frequently glitching pulsars with 37 glitches. It is presently in the second position in glitch count, which makes it a remarkable candidate for studying timing irregularities. It is currently monitored at uGMRT with a cadence of 15--20 days. We report the detection of a large glitch in this pulsar observed on MJD 59934.8(5), which was initially announced in the Astronomer's Telegram by \citep{ATel_Zubieta_J1740} and subsequently confirmed in the Astronomer's Telegram by \citep{ATel_Dunn_J1740} and \citep{ATel_Grover_J1740-3015}. We update our preliminary results reported on Astronomer’s Telegram \citep{ATel_Grover_J1740-3015} by augmenting subsequent observations. We calculated the fractional increase in the spin as $327(1)\times10^{-9}$, and the time derivative of spin is $1.3(3)\times10^{-3}$. While we believe these are more accurate estimated parameters, they are broadly in line with the parameters reported earlier. The timing residuals, spin, and spin derivative evolution are given in Fig.~\ref{Glitch_J1740-3015} and are being presented for the first time for this new glitch.

\subsubsection{Glitch-like event in J1825--0935}
\label{slowglitch_J1825}
\begin{figure}
\centering
    \includegraphics[width=\linewidth]{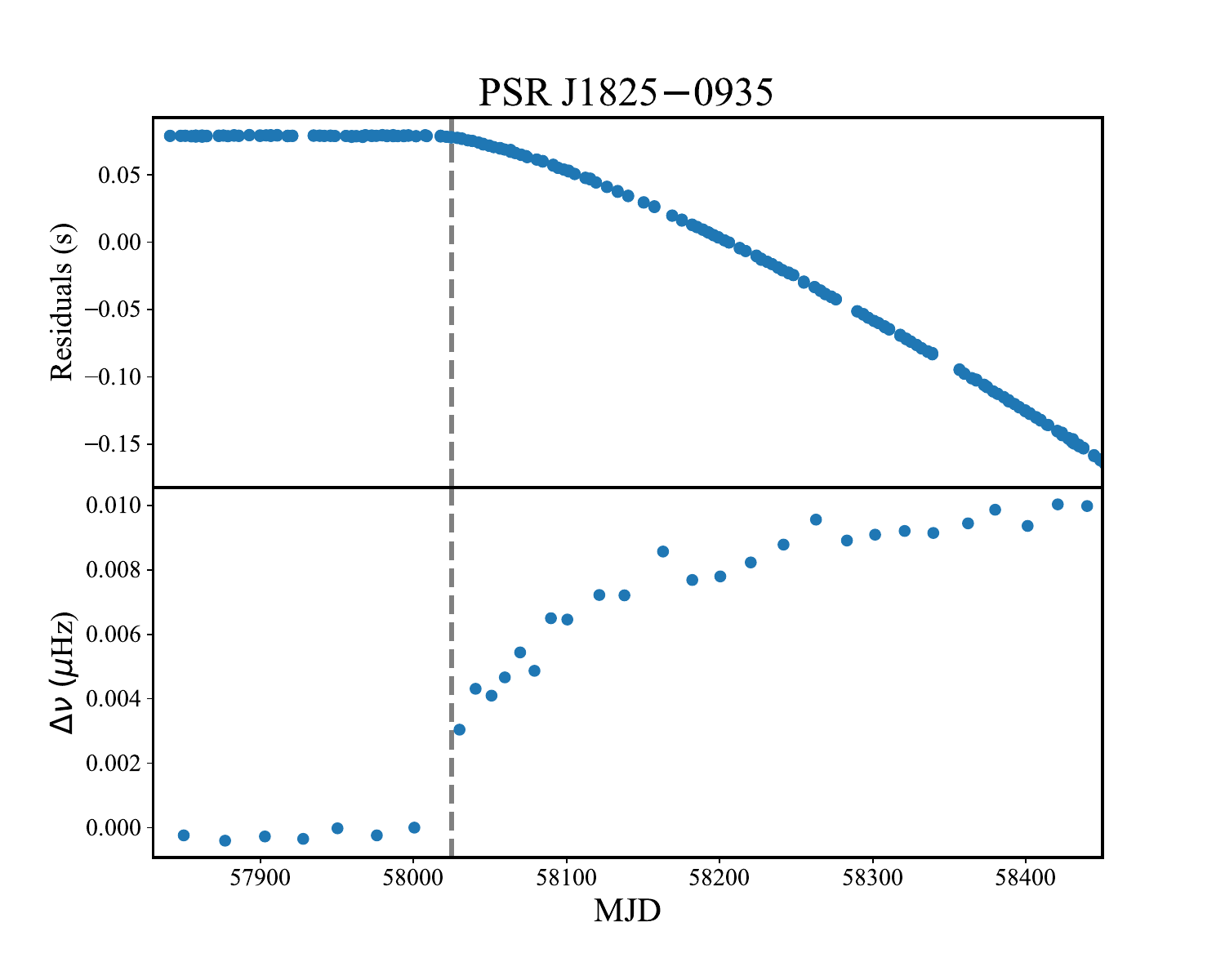}
    \caption{A glitch seen in PSR J1825--0935. The top panel presents the residuals. The middle panel displays the $\Delta \nu$ evolution. The vertical line indicates the glitch epoch.}
    \label{J1825_glitch}
\end{figure}
PSR J1825--0935 is one of the few pulsars which shows slow glitches \citep{Zou2004, Shabanova1998_slowglitch, Shabanova2005, Zhou2019_slowglitch} along with conventional glitches. Slow glitches are characterized by an extended rise in their frequency evolution, with a typical rise timescale ranging from months to years. We have been monitoring this pulsar regularly using the uGMRT and the ORT, and we detected a slow glitch-like event also reported in \citet{lower2020_utmost2} and \citet{Singha_2022_SKA_Review}. The residual and spin evolution plots are given in Fig.~\ref{J1825_glitch}. 

To date, only $\sim$ 32 slow glitches have been reported \citep{Zhou2019_slowglitch} and the mechanisms behind slow glitches are not yet fully understood. However, there are several theoretical models that suggest the occurrence of slow glitches. One such model suggests that the cause of slow glitches is the oscillation between two phases of an anisotropic superfluid \citep{peng2018}. Others proposed that a slow glitch occurs due to vortex bending oscillation in the spin-down rate, generated either by magnetospheric changes or crustquakes \citep{Erbil_2023}. Slow glitches are dominantly observed in old pulsars with low spin-down rates. As in such systems, superfluid recoupling timescale would be longer, and this, in turn, affects the migration rate of vortex lines once they become unpinned. When transported radially inward driven by a crustquake, completion of the angular momentum transfer process and achieving the equilibrium configuration for vortex lines take a longer time due to the old age of a pulsar \citep{Gugercinoglu_2019}. The occurrence of slow glitches in PSR B0919+06 concurrent with cyclic changes in its spin-down rate is a firm demonstration \citep{Shabanova_2010}. 

PSR J1825--0935 is a very interesting pulsar because it also exhibits profile changes concurrent with its glitches. It is known to display interpulse, drifting subpulses, and mode switching between two emission states: Burst/B-mode and Quiet/Q-mode \citep{Fowler1981, morris1981, Fowler1982, Gil1994}. In the burst mode, an additional faint precursor unit adjacent to the intense main pulse is visible, whereas in the quiet mode, the precursor unit exhibits minimal emission, and the interpulse appears brighter. The timing study of this pulsar has complications because of mode switching. According to \cite{lyne2010}, the slow glitches are actually not related to the glitch phenomenon and are consequences of the switching between the two magnetospheric spin-down modes. The observed oscillations in the spin-down rate,   correlated with pulse profile changes or induced after glitches \citep{Shaw_2022}, support the view that either magnetospheric processes or crustquakes lead to a change in the strength of coupling of the interior superfluid torque acting on the neutron star surface \citep{Erbil_2017, Erbil_2023}.

It has also been suggested that slow glitches are not unique and are a manifestation of the timing noise \citep{Hobbs_2010}. However, if slow glitches were manifestations of merely timing noise in pulsars, they should not be rare \citep{Zhou2019_slowglitch}. Our novel glitch verification methodology is capable of differentiating a real glitch from timing noise. However, the observed data using ORT and uGRMT do not have sufficient pre-glitch observations for timing noise analysis. Hence, the GP realization has not been utilized for this pulsar. Additionally, due to the complications in the timing from mode switching, PSR J1825--0935 is not a good example to answer if slow glitches are real. A rigorous analysis accounting for all the reported slow glitches is required to answer whether they are real glitches or a special case of timing noise.

\subsection{Timing Noise}\label{timingnoise}
\begin{figure*}
    \centering
    \begin{subfigure}{.99\textwidth}
  \centering
  \includegraphics[width=\linewidth]{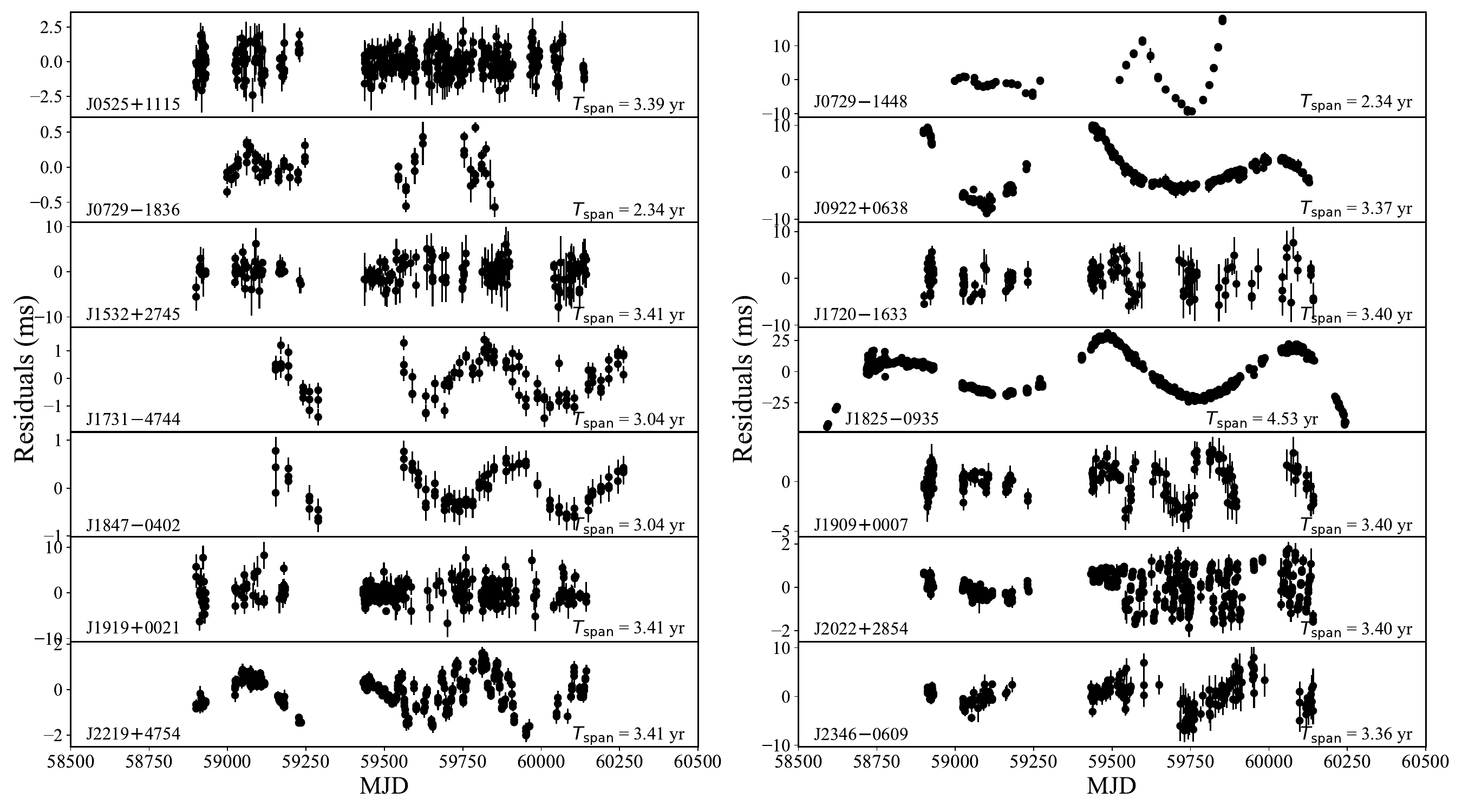}  
\end{subfigure}
\begin{subfigure}{.9\textwidth}
  \centering
  \includegraphics[width=.8\linewidth]{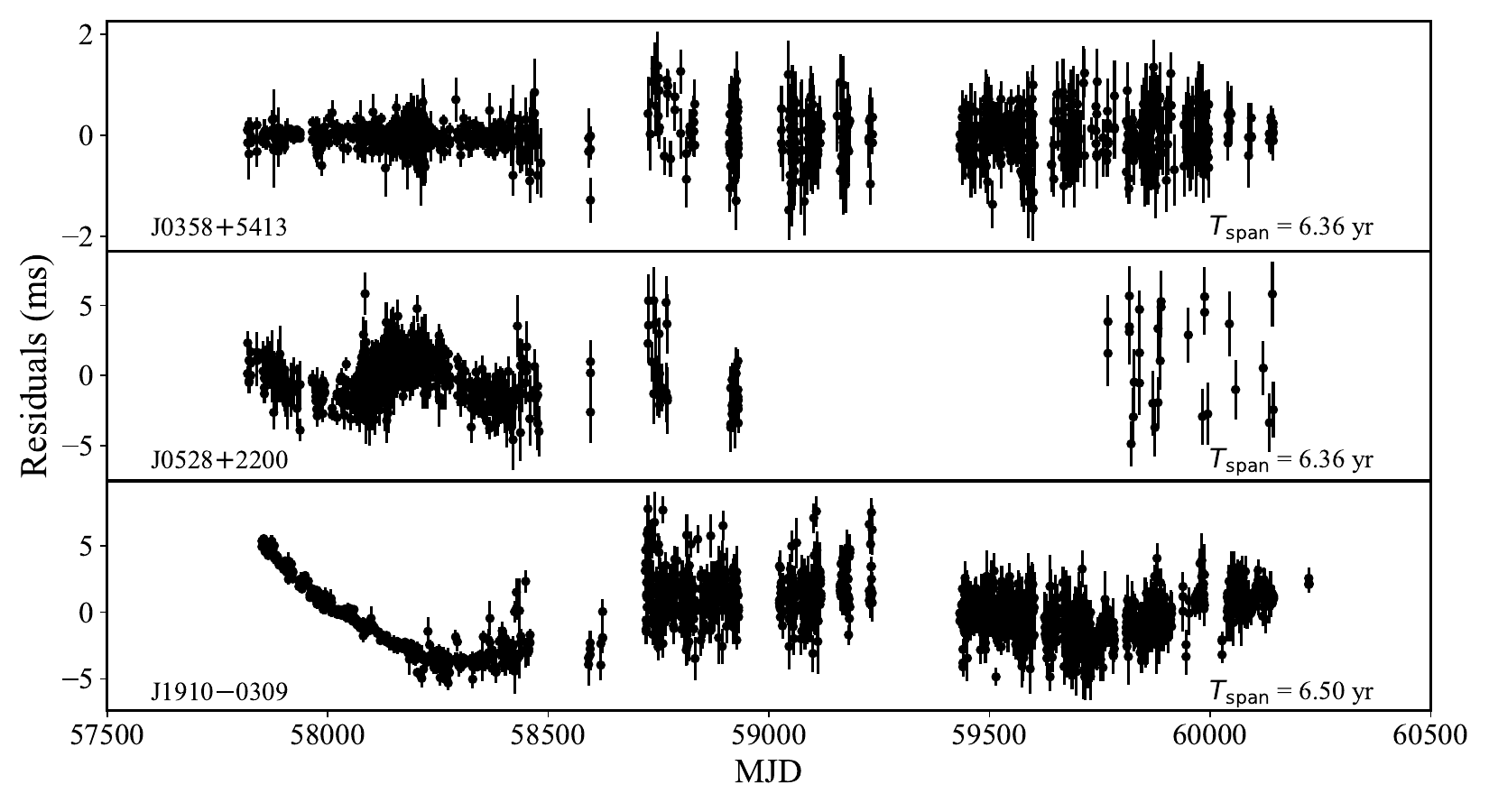}  
    \end{subfigure}   
\caption{Timing residuals plots for 17 pulsars used for single pulsar noise analysis presented in this work.}
\label{Residuals}
\end{figure*}

\begin{table*}
\centering
\caption{The values of $\ln(\text{BF})$ with respect to the simplest model, in terms of parameters (F) for the 20 pulsars in our sample. Bold values against a model indicate it is the preferred model for the corresponding pulsar. This preferred model has been selected based on the values of the BF and the number of free parameters in the model. The pulsars marked with $*$ have experienced glitches. The model comparison results correspond to the pre-glitch region, and for the post-glitch region, the preferred model remains the same; however, the value of $\ln(\text{BF})$ may vary.}
\label{model_select}
\renewcommand\arraystretch{2.0}
\setlength{\tabcolsep}{13pt}
\centering
\begin{tabular}{ccccccc}
\hline
\hline
PSR JName & $T_{span}$ (yr) & F & WN & F+RN & WN+RN & Fourier
Modes  \\ \hline \hline
J0358+5413  & 6.36 & 0 & 0.2  & \textbf{10.8} & 10.8 & 38 \\ 
J0525+1115  & 3.39 & 0 & \textbf{9.7}  &  0.5  & 9.7 & 81 \\ 
J0528+2200  & 6.36 & 0 & 72.1  & 297.4   & \textbf{324.4} & 19 \\
J0729--1448 & 2.34 & 0 & 1353.9  & \textbf{1584.3} & 1584.2 & 14 \\ 
J0729--1836 & 2.34 & 0 & 273.0   & \textbf{509.0} & 508.7 & 14 \\ 
J0742--2822$^*$ & 4.37 & 0 & 333589.3 & 338217.8 & \textbf{338233.9} & 210 \\
J0835--4510$^*$ & 1.27 & 0 & 1677.0 & 2932.6 & \textbf{2957.7} & 61 \\
J0922+0638  & 3.37 & 0 & 226.8 & \textbf{1023.9} & 1023.4 & 162 \\
J1532+2745 & 3.41 & 0 & 1.0 & \textbf{4.3} & 4.5 & 10 \\
J1720--1633 & 3.40 & 0 & \textbf{15.2} & 0.1 & 15.1 & 82 \\
J1731--4744 & 3.04 & 0 & -- 0.3 & \textbf{47.0} & 46.7 & 36 \\
J1740--3015$^*$ & 1 & 0 & -- 0.1 & \textbf{28.4} & 28.4 & 24 \\
J1825--0935 & 4.53 & 0  & 10459.3 & 13150.6 & \textbf{13370.9} & 218 \\ 
J1847--0402 & 3.04 & 0 & -- 0.2  & \textbf{82.7}  & 82.0  & 9 \\ 
J1909+0007  & 3.40 & 0 & -- 0.1  & \textbf{60.5}  & 60.9  & 41  \\
J1910--0309 & 6.50 & 0 & 310.5  & \textbf{1504.0}  & 1503.3 & 78   \\ 
J1919+0021 & 3.41 & 0 & \textbf{7.3} & -- 0.3 & 6.6 & 20  \\ 
J2022+2854 & 3.40 & 0 & 32.0 & \textbf{566.8} & 566.2 & 82 \\
J2219+4754 & 3.41 & 0 & 27.1 & \textbf{487.1} & 487.6 & 41 \\
J2346--0609 & 3.36 & 0 & 3 & \textbf{144.4} & 144.2 & 10 \\   \hline\hline
\end{tabular}
\end{table*}

\begin{table*}
\centering
\caption{The noise parameters obtained for our sample of pulsars that have not experienced any glitch. The 1st column represents the pulsar J name, followed by the time span of observation, the 4th column represents the most preferred model for the pulsar, the 5th and 6th columns present the white noise parameters, and the 7th and 8th columns are the red noise parameters. The last column corresponds to the number of glitch-like events observed in the respective pulsar, which are unlikely to be real glitches as they appear consistent with timing noise in these pulsars.}
\label{TNresultstable}
\renewcommand\arraystretch{2.2}
\setlength{\tabcolsep}{6pt}
\begin{tabular}{ccccccccc}
\hline
\hline
PSR JName & MJD &  T & Model  & EFAC (F) & $\log_{10}(\sigma_Q)$ &  $\log_{10}(A_{\rm red})$  & $\gamma$ & Glitch-like events\\ 
     & (days) &  (yr) &  &   &   &  (yr$^{(3/2)}$) & &    \\ \hline \hline
    J0358+5413 & 57818--60143 & 6.36 & F+RN & $1.02^{+0.02}_{-0.02}$  & $---$ & $-10.84^{+0.14}_{-0.15}$ & $1.93^{+0.53}_{-0.57}$ & 0\\ 
    \hline
    J0525+1115 & 58899--60137 & 3.39 & WN &   $0.90^{+0.09}_{-0.09}$  & $-3.25^{+0.10}_{-0.10}$ & $---$  &  $---$ & 0\\ 
    \hline
    J0528+2200 & 57818--60143 & 6.36 & WN+RN & $1.02^{+0.05}_{-0.05}$  & $-3.22^{+0.05}_{-0.05}$ & $-9.75^{+0.09}_{-0.08}$ &  $3.40^{+0.64}_{-0.60}$ & 0\\ 
    \hline
    J0729--1448 & 58998--59852 & 2.34 & F+RN & $1.14^{+0.11}_{-0.10}$  & $---$ & $-8.58^{+0.13}_{-0.11}$ & $3.97^{+0.46}_{-0.44}$ & 1 \\ 
    \hline
    J0729--1836 & 58998--59852 & 2.34 & F+RN & $1.59^{+0.19}_{-0.16}$  & $---$ & $-10.08^{+0.22}_{-0.22}$ & $2.89^{+1.12}_{-0.90}$ & 0\\ 
    \hline
    J0922+0638 & 58898--60128 & 3.37 & F+RN  & $0.93^{+0.04}_{-0.04}$  & $---$ & $-9.11^{+0.10}_{-0.09}$ & $3.13^{+0.28}_{-0.26}$ & 1 \\ 
    \hline
    J1532+2745 & 58899-60143 & 3.41 & F+RN & $0.95^{+0.05}_{-0.05}$  & $---$ & $-9.74^{+0.18}_{-0.20}$ & $3.14^{+1.32}_{-1.08}$ & 0 \\
    \hline
    J1720--1633 & 58900--60141 & 3.40 & WN &  $0.82^{+0.20}_{-0.21}$  & $-2.57^{+0.17}_{-0.14}$ & $---$ & $---$  & 0\\
    \hline
    J1731--4744 & 59153--60262 & 3.04 & F+RN &  $1.25^{+0.11}_{-0.10}$  & $---$ & $-9.75^{+0.13}_{-0.12}$ & $3.09^{+0.93}_{-0.63}$ & 1 \\
    \hline    
    J1825--0935 & 58591--60244 & 4.53 & WN+RN &  $0.91^{+0.11}_{-0.11}$ & $-2.82^{+0.07}_{-0.06}$ & $-8.78^{+0.07}_{-0.06}$ & $3.64^{+0.26}_{-0.23}$ & 1 \\
    \hline
    J1847--0402 & 59153--60262 & 3.04 & F+RN &  $0.50^{+0.05}_{-0.04}$  & $---$ & $-9.99^{+0.11}_{-0.10}$ & $4.23^{+1.14}_{-0.89}$ & 2 \\
    \hline
    J1909+0007 & 58901--60141 & 3.40 & F+RN &  $0.88^{+0.05}_{-0.05}$   & $---$ & $-9.36^{+0.14}_{-0.13}$ & $1.77^{+0.52}_{-0.48}$ & 0\\
    \hline
    J1910--0309 & 57851--60224 & 6.50 & F+RN &  $1.34^{+0.02}_{-0.02}$ & $---$ & $-9.68^{+0.05}_{-0.05}$ & $1.75^{+0.15}_{-0.14}$ & 0\\
    \hline
    J1919+0021 & 58900--60143 & 3.41 & WN & $1.09^{+0.10}_{-0.11}$ & $-2.96^{+0.10}_{-0.11}$ & $---$ & $---$ & 0\\
    \hline
    J2022+2854 & 58900--60140 & 3.40 & F+RN & $1.14^{+0.04}_{-0.04}$ &  $---$   & $-10.00^{+0.09}_{-0.08}$ & $0.44^{+0.17}_{-0.16}$ & 0 \\
    \hline
    J2219+4754 & 58900--60144 & 3.41 & F+RN  & $0.70^{+0.03}_{-0.03}$ &  $---$  & $-9.60^{+0.12}_{-0.11}$ & $1.83^{+0.25}_{-0.23}$ & 0\\
    \hline
    J2346--0609 & 58912--60140 & 3.36 & F+RN  & $0.96^{+0.05}_{-0.04}$ &  $---$  & $-9.26^{+0.11}_{-0.10}$ & $3.08^{+1.12}_{-0.94}$ & 0\\
    \hline\hline
\end{tabular}
\end{table*}

\begin{table*}
\centering
\caption{The noise parameters obtained for our sample of pulsars that have experienced glitches. The 1st column represents the pulsar J name, followed by the time span of observation, the 4th column represents the most preferred model for the pulsar, the 5th and 6th columns present the white noise parameters, and the 7th and 8th columns are the red noise parameters. The last column corresponds to the number of glitch-like events observed in the respective pulsar, which are unlikely to be real glitches as they appear consistent with timing noise in these pulsars.}
\label{Glitch_TNresultstable}
\renewcommand\arraystretch{2}
\setlength{\tabcolsep}{6pt}
\begin{tabular}{ccccccccc}
\hline
\hline
PSR JName & MJD &  T & Model  & EFAC(F) & $\log_{10}(\sigma_Q)$ &  $\log_{10}(A_{\rm red})$  & $\gamma$ & Glitch-like \\ 
     & (days) &  (yr) &  &   &   &  (yr$^{(3/2)}$) &  & events  \\ \hline \hline
    J0742--2822 & 58237-59832 & 4.37 & WN+RN & $1.20^{+0.10}_{-0.09}$  & $-4.24^{+0.06}_{-0.07}$ & $-9.12^{+0.08}_{-0.07}$ & $4.18^{+0.29}_{-0.24}$ & 0\\ 
     & 59841-60248 & 1.12 & WN+RN & $0.81^{+0.07}_{-0.07}$  & $-3.92^{+0.06}_{-0.06}$ & $-7.47^{+0.17}_{-0.16}$ & $6.16^{+0.38}_{-0.36}$ & -- \\ 
    \hline
    J0835--4510 & 58018-58482 & 1.27 & WN+RN & $0.63^{+0.58}_{-0.38}$  & $-3.73^{+0.41}_{-0.30}$ & $-9.33^{+0.16}_{-0.14}$ & $4.14^{+0.42}_{-0.36}$ & 0\\ 
     & 58611-58999 & 1.06 & WN+RN & $1.11^{+0.19}_{-0.17}$  & $-4.18^{+0.10}_{-0.10}$ & $-9.02^{+0.14}_{-0.12}$ & $3.83^{+0.29}_{-0.27}$ & -- \\ 
    \hline
    J0835--4510 & 58803-59271 & 1.28 & WN+RN & $0.82^{+0.15}_{-0.16}$  & $-4.06^{+0.11}_{-0.10}$ & $-9.57^{+0.10}_{-0.09}$ & $2.74^{+0.18}_{-0.17}$ & 0\\ 
     & 59418-60248 & 2.27 & WN+RN & $1.22^{+0.18}_{-0.17}$  & $-3.88^{+0.07}_{-0.07}$ & $-8.68^{+0.09}_{-0.08}$ & $5.62^{+0.44}_{-0.39}$ & -- \\ 
    \hline
    J1740--3015 & 59562-59929 & 1.00 & F+RN & $0.56^{+0.07}_{-0.06}$  & $---$ & $-9.34^{+0.30}_{-0.26}$ & $4.04^{+1.14}_{-0.98}$ & 2 \\ 
     & 59951-60262 & 0.85 & F+RN & $0.48^{+0.08}_{-0.06}$  & $---$ & $-9.77^{+0.46}_{-0.48}$ & $3.57^{+1.86}_{-1.58}$ & -- \\ 
    \hline
    \hline
\end{tabular}
\end{table*}

We now present the noise analysis of 20 pulsars in our sample, including 3 glitching pulsars. The timing residuals of 17 pulsars, with no glitch observed using the uGMRT and the ORT are shown in Fig.~\ref{Residuals}. The pulsar timing package, \texttt{Enterprise}, was used to obtain the timing noise parameters, and \texttt{PTMCMCSampler} has been used for sampling.
A description of the four different combinations of noise models and various model parameters is given in Sec.~\ref{TN_Analysis}. We performed the Bayesian model selection using these models. In Table~\ref{model_select}, we present the $\ln$(BF) of the various models with respect to the simplest model in terms of the number of parameters. The best model was selected for each pulsar based on the values of the Bayes Factor. The bold value of the number against a model indicates that it is the most preferred model. This preferred model was selected based on the values of the BF and for its simplicity. In cases where the BF of two or more models are similar, we selected the model containing the least number of parameters (Occam's razor). The preferred models and their corresponding parameters for the 20 pulsars are tabulated in Table~\ref{TNresultstable} and Table~\ref{Glitch_TNresultstable}. The posterior distribution for the parameters of the most preferred noise models for our pulsars are given in~\ref{appendix_TN_posterior}. The posterior distributions are plotted with 68\% and 95\% credible intervals for all the cases.

Three of our pulsars, namely, PSR J0525+1115, J1720--1633 and J1919+0021 show evidence of only white noise (EFAC + EQUAD) in the timing residuals. The following pulsars, J0358+5413, J0729--1448, J0729--1836, J0922+0638, J1532+2745, J1731--4744, J1740--3015, J1847--0402, J1909+ 0007, J1910--0309, J2022+2854, J2219+4754, J2346--0609 show evidence for red noise and do not prefer the inclusion of EQUAD in the noise model. On the other hand, PSRs J0528+ 2200, J0742--2822, J0835--4510, and J1825--0935 preferred the full white noise (EFAC + EQUAD) along with the red noise model. Some pulsars prefer the inclusion of EQUAD while others do not because, for some pulsars, the observed timing residuals contain extra noise contributions from epoch to epoch pulse shape changes or jitter \citep{Osłowski_2011} that are not well-characterized by the formal measurement uncertainties. Adding an EQUAD term helps to account for this extra noise, leading to more accurate parameter estimates and uncertainties.
The noise results for PSRs J0525+1115, J0729--1836, J0742--2822, J0835--4510, J0922+0638, J1720--1633, J1731--4744, J1740--3015, J1825--0935, J1847--0402, J1909+0007, J1910 --0309, J1919+0021, J2346--0609 have been reported in literature \citep{lower2020_utmost2, parthasarathy2019_TimingI}. 

The white noise model for PSRs J0525+1115, J1720--1633, J1919+0021 and the red noise model for PSRs J0729--1836, J0922+0638, J1731--4744, J1825--0935, J1847--0402, J1909+ 0007, J2346--0609 were also preferred in \citep{lower2020_utmost2}. The only difference was in pulsar J1910--0309, which preferred the white noise model in their results and showed evidence of red noise in ours. This is probably due to the availability of a larger data span in our case. The reported red noise parameters for most of the pulsars, for e.g., J0729--1836, J0922+0638, J1731--4744, J1825--0935, J1847--0402, were within the error bars of the parameters reported by them. However, the timing noise parameters are different for a few pulsars, such as PSRs J1909+0007 and J2346--0609.
This inconsistency can be due to the unmodelled interstellar medium effects in both ours and \cite{lower2020_utmost2} analyses. The timing noise (or red noise) parameters for pulsars J0358+5413, J0528+2200, J0729--1448, J1532+2745, J1910--0309, J2022+2854, and J2219+4754 were not reported. Therefore, we present the new timing noise results for at least seven pulsars. 

We have also analyzed the timing noise in glitching pulsars in the pre-glitch and the post-glitch regions for three pulsars for the first time to the best of our knowledge. The preferred models and their corresponding parameters for three glitching pulsars are listed in Table~\ref{Glitch_TNresultstable}, and the timing noise posteriors of the most preferred noise models are given in~\ref{appendix_TN_posterior}. A significant variation in 
timing noise parameters has been observed before and after the MJD 59839.8 glitch in 
PSR J0742--2822 and the MJD 59417.6 glitch in PSR J0835--4510. This is probably due to the consideration of the exponential post-glitch recovery region in the noise analysis. However, as no exponential recovery was present for the MJD 58517 glitch in PSR J0835--4510, the noise results in the pre-glitch and the post-glitch regions are consistent.

\subsection{Differentiating glitches from timing noise : GP realization}\label{GPR_results}
As mentioned in Section 4, we devise a novel method to distinguish real glitches from timing noise by taking into account the timing noise in glitching pulsars. We illustrate this process here with two extreme examples.

\subsubsection{PSR J0835--4510}
\begin{figure*}
    \centering
    \begin{subfigure}{0.9\textwidth}
        \includegraphics[width=\linewidth]{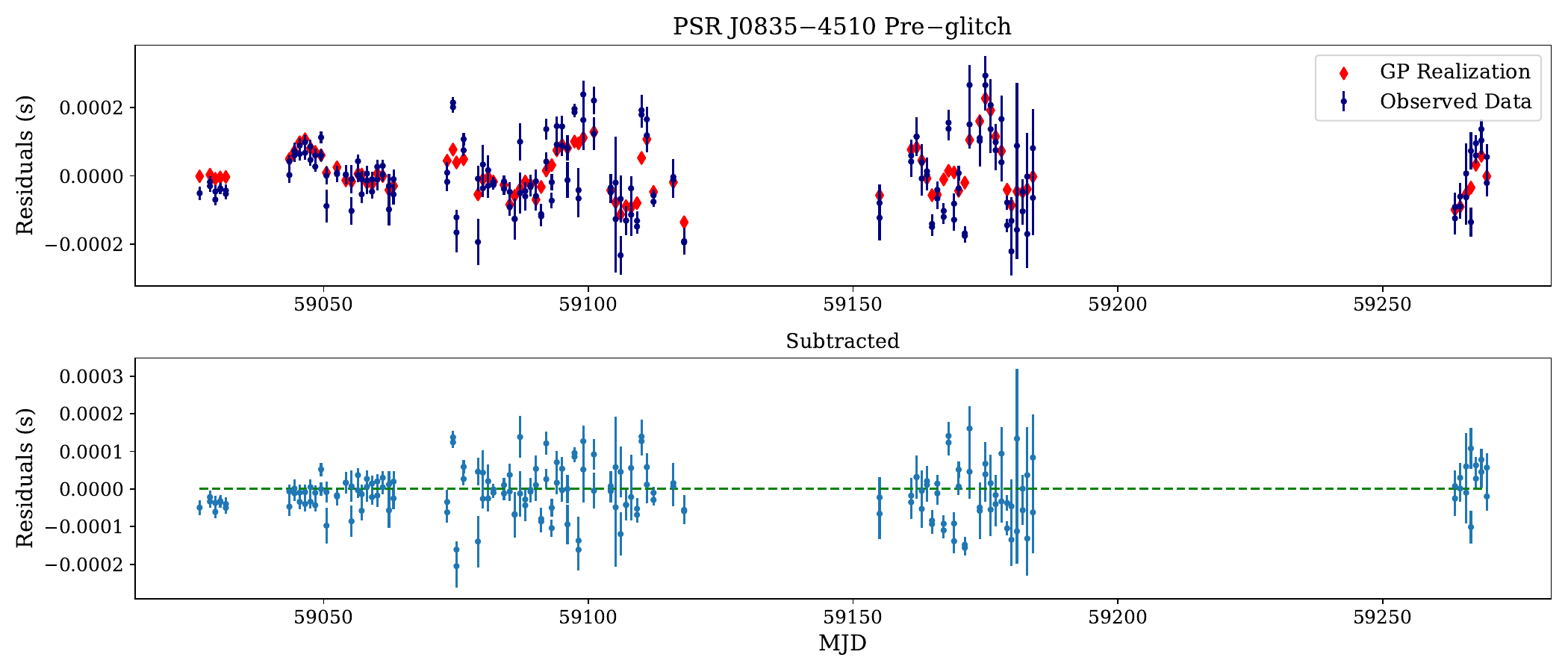}
        \caption{The pre-glitch GP realization of PSR J0835--4510.}
        \label{GPR_J0835-4510_59417_pregl}
    \end{subfigure}
    \begin{subfigure}{0.9\textwidth}
    \centering
        \includegraphics[width=\linewidth]{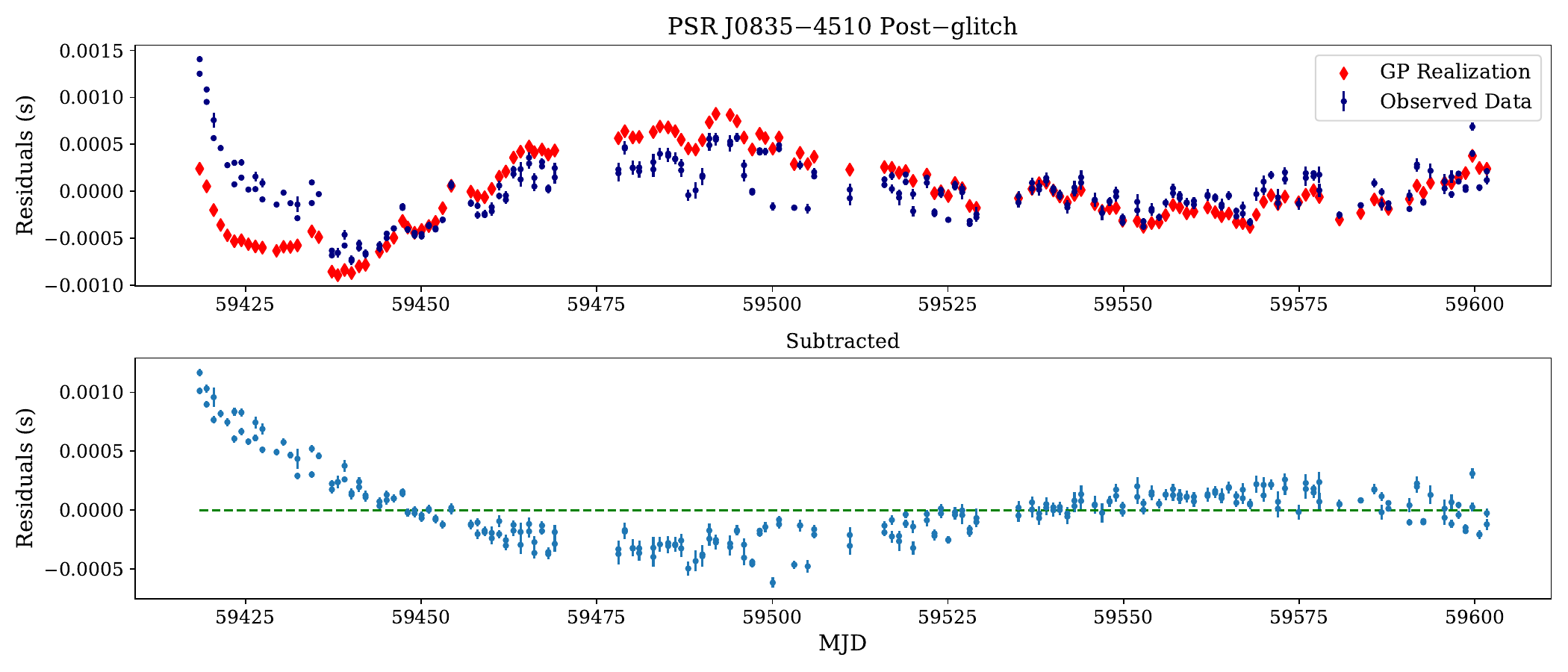}
        \caption{The post-glitch GP realization of PSR J0835--4510 with exponential recovery region.}
        \label{GPR_J0835-4510_59417_withexp}
    \end{subfigure}
    \begin{subfigure}{0.9\textwidth}
    \centering
        \includegraphics[width=\linewidth]{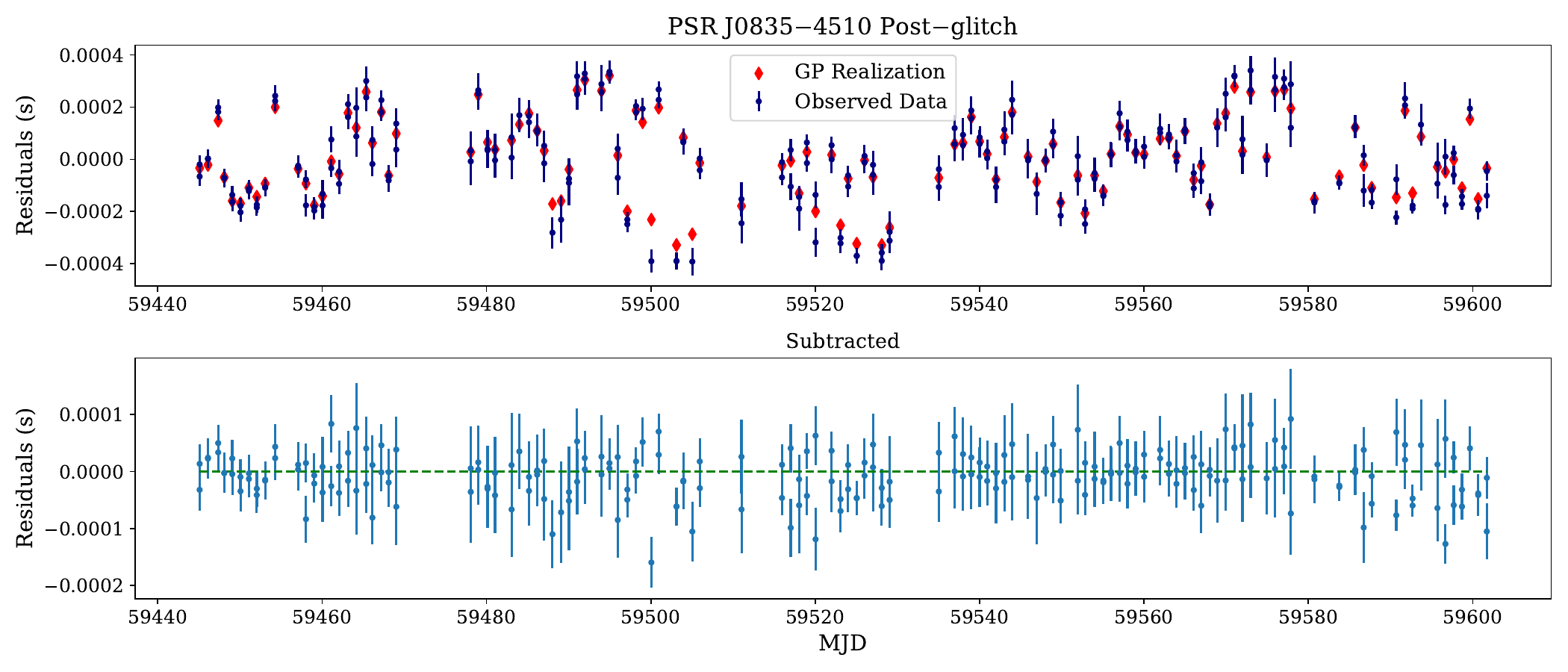}
        \caption{The post-glitch GP realization of PSR J0835--4510 without exponential recovery region.}
        \label{GPR_J0835-4510_59417_noexp}
    \end{subfigure}
\caption{Verification of glitch using our new technique employing GP realization for PSR J0835--4510.}
\label{GPR_J0835-4510_59417}
\end{figure*}

\begin{figure*}
\centering
    \includegraphics[width=\linewidth]{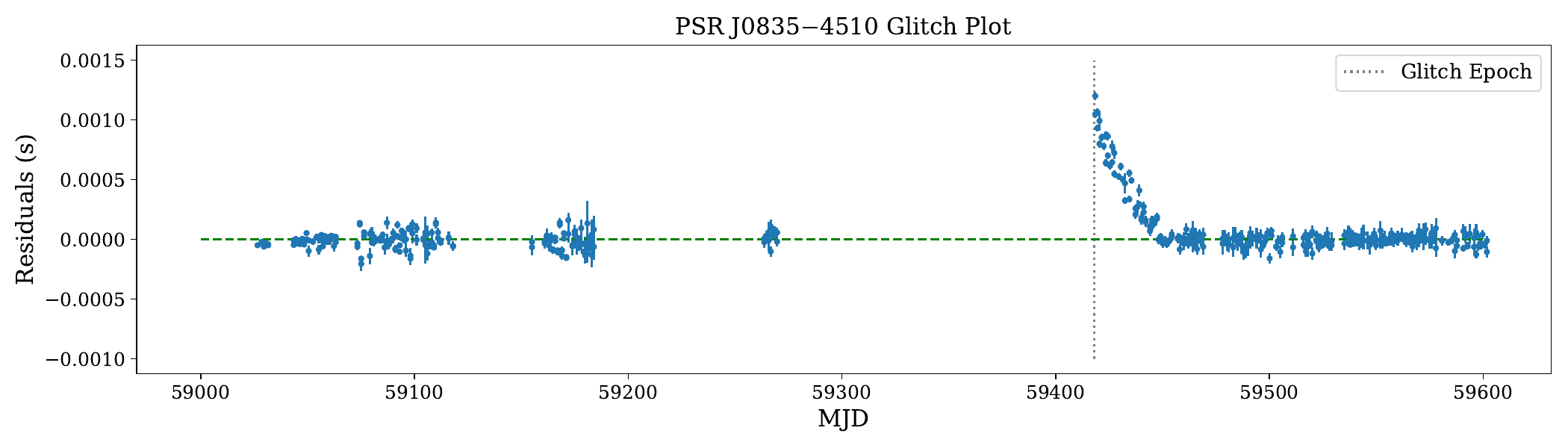}
\caption{The plot for PSR J0835--4510 residuals obtained after subtracting the GP realization from the timing residuals revealing the clear signature of a glitch.}
\label{GPR_J0835-4510_59417_glitchplot}
\end{figure*}

We begin with the glitch in the Vela Pulsar, which is a large glitch; hence, there is a high probability that it is a real glitch.  Additionally, we have a high cadence ($\sim$ 1-3 days) data for this glitch, as it is being observed at the ORT regularly. We started with the pre-glitch timing solution for this pulsar and utilized it to investigate the timing noise and obtain timing noise parameters/chains. Further, timing noise results were used to produce a GP realization. The results are illustrated in Fig.~\ref{GPR_J0835-4510_59417_pregl}. The top panel of the figure displays the observed data points alongside the median GP realization. The bottom plot illustrates the subtracted data obtained by subtracting the GP realization from the observed data points. The residual data appears consistent with zero mean white noise. Now, in a glitch event, we anticipate observing a jump in the post-glitch region after eliminating the timing noise. Fig.~\ref{GPR_J0835-4510_59417_withexp} presents the GP realization for the post-glitch region. The curvature observed in the bottom plot is attributed to the residual data being fitted in \texttt{tempo2}. This process generated a negative compensated component to counteract the glitch jump event, resulting in the curvature observed in the white noise signal. The curvature gives an initial hint that the glitch will probably be real. If we exclude the initial few weeks of observations, primarily representing the exponential recovery phase, and focus on the linear recovery region, we anticipate observing a flat line once more, as shown in Fig.~\ref{GPR_J0835-4510_59417_noexp}. Finally, we extrapolate the residuals described in Fig.~\ref{GPR_J0835-4510_59417_noexp} and subtract the extrapolated value for the initial observations or the exponential recovery phase. This yields the final glitch plot, providing evidence that the glitch is real. The final glitch plot is shown in Fig.~\ref{GPR_J0835-4510_59417_glitchplot}. 

\subsubsection{PSR J1847--0402}
\begin{figure*}
\centering
    \begin{subfigure}{\textwidth}
    \centering
    \includegraphics[width=\linewidth]{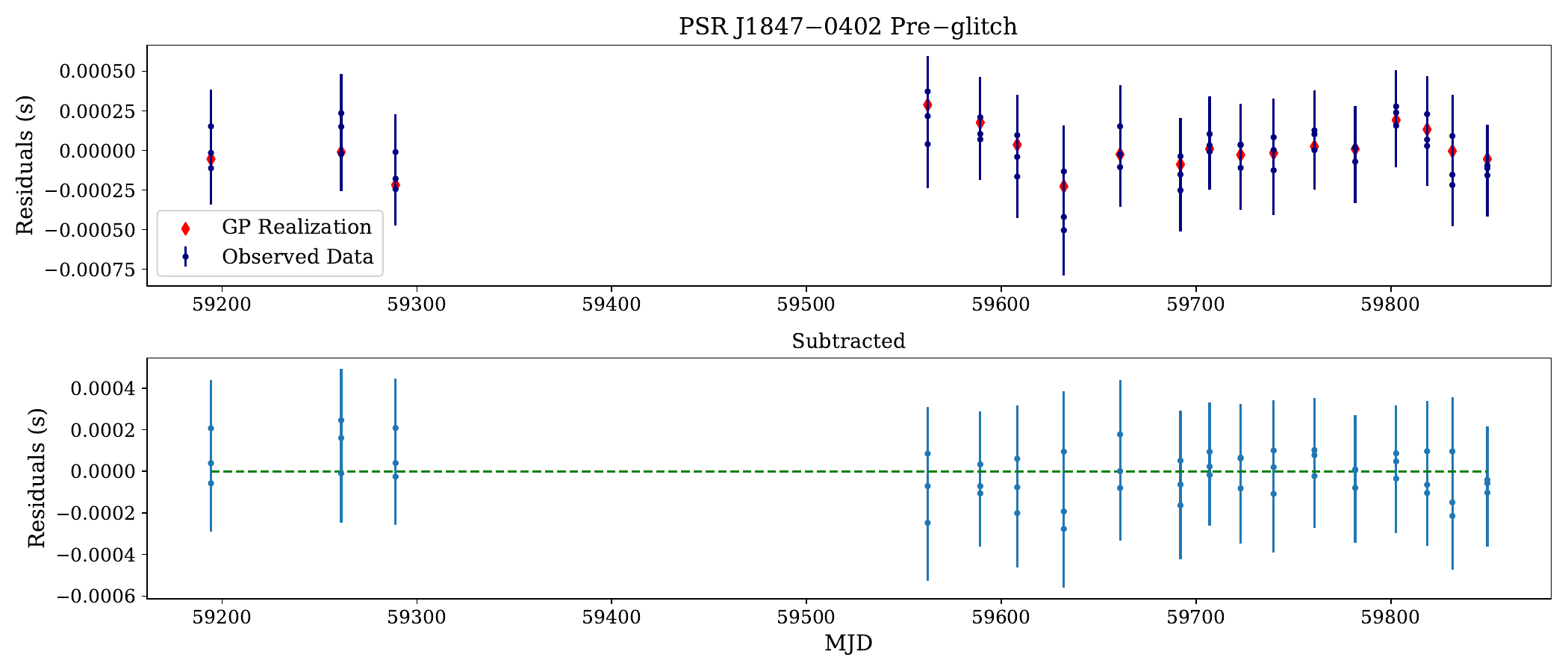}
    \caption{The pre-glitch GP realization of PSR J1847--0402.}
    \label{GPR_J1847-0402_pregl}
    \end{subfigure}
    \begin{subfigure}{\textwidth}
    \centering
    \includegraphics[width=\linewidth]{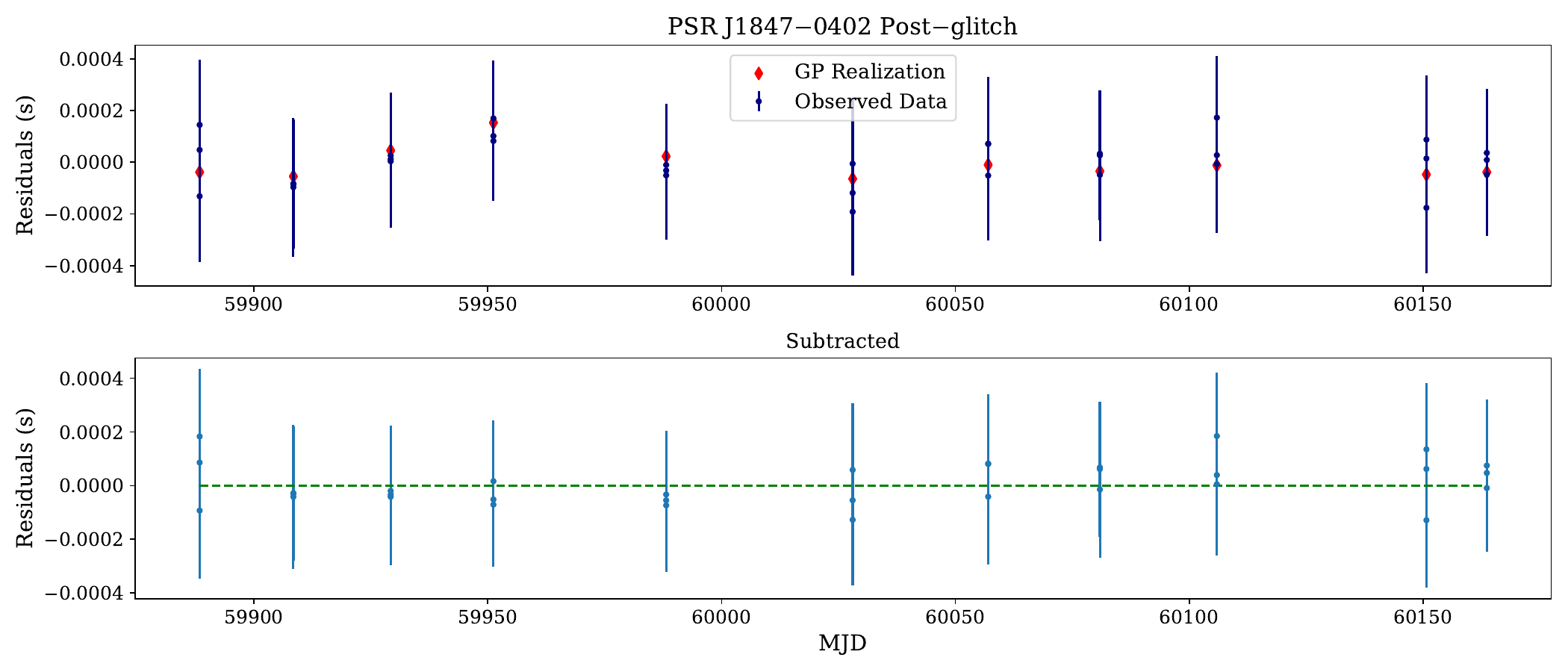}
    \caption{The post-glitch GP realization of PSR J1847--0402.}
    \label{GPR_J1847-0402_postgl}
    \end{subfigure}
    \begin{subfigure}{\textwidth}
    \centering
    \includegraphics[width=\linewidth]{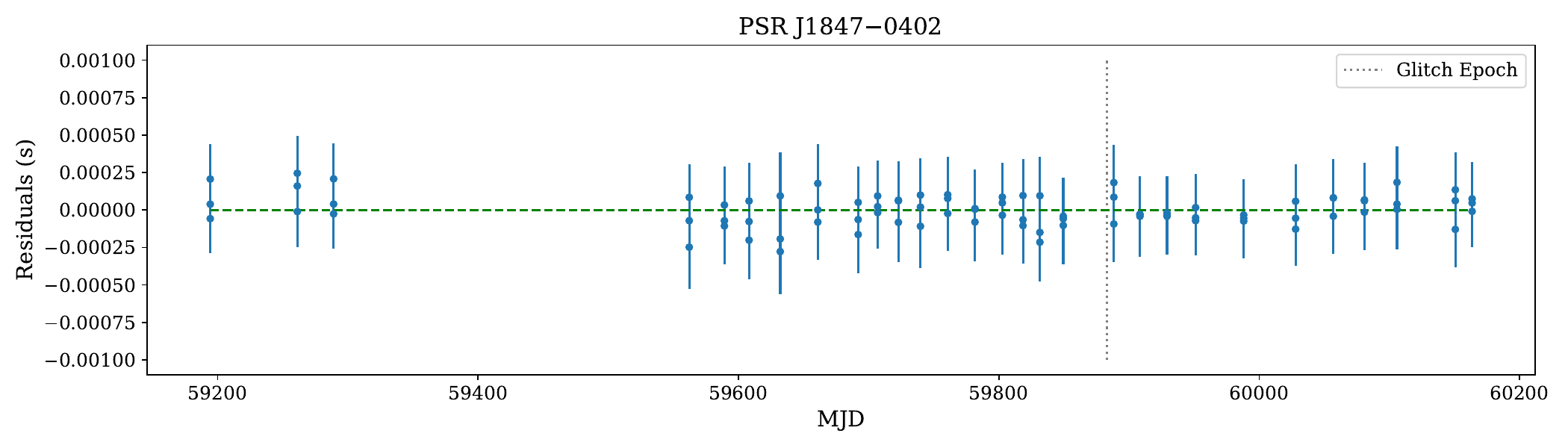}
    \caption{The plot for PSR J1847--0402 obtained after subtracting GP realization from the timing 
residuals showing an absence of glitch-like structure.}
    \label{GPR_J1847-0402_glitchplot}
    \end{subfigure}
\caption{Verification of glitch using our new technique employing GP realization for PSR J1847--0402.}
\label{GPR_J1847-0402}
\end{figure*}

Here, we discuss the case of PSR J1847--0402. Three small glitches (glitch amplitude $\sim 10^{-10}$) have previously been reported for this pulsar. We detected a glitch-like event on MJD 59883(6) using timing analysis with a glitch amplitude of $0.18(3)\times 10^{-9}$, which is close to the range reported previously. We utilized the GP realization to verify this glitch. The results are shown in Fig.~\ref{GPR_J1847-0402}, authenticating that this event was not a real glitch but a manifestation of the timing noise. It is possible that all other small glitches reported in this pulsar appeared because of strong timing noise. Hence,
it is important to utilize our new method of employing GP realizations to investigate small glitches, which are expected to appear frequently because of strong timing noise. Overall, we detected around 8 glitch-like events, seven of which appeared like glitches because of large observation gaps or strong timing noise, and the 8th one is a glitch-like event in J1825--0935, as discussed in Section~\ref{slowglitch_J1825}.

\section{Conclusions and Future work}\label{conclusion}
Pulsar glitches and timing noise probe into the dynamics of superfluid interiors and help us to understand the physical conditions prevailing inside neutron stars. We present the recent results of our glitch monitoring program using the uGMRT and the ORT. This includes the detection of five glitches in 4 pulsars. The time evolution of two of these glitches has been presented for the first time, showing significant recovery from the glitch. Most of the glitches presented in this work are large glitches. Additionally, we report noise analysis for 20 pulsars using Bayesian analysis techniques given in Table~\ref{TNresultstable} and Table~\ref{Glitch_TNresultstable}. The timing noise analyses for 14 of the pulsars from our sample were presented before \citep{lower2020_utmost2}. We are reporting new results for noise analysis of 6 pulsars and red noise results for J1909--0309 for the first time. Furthermore, we have investigated the timing noise in the pulsars that have shown glitches and presented the timing noise parameter values in pre-glitch and post-glitch regions. A significant variation in the pre-glitch and post-glitch red noise parameters has been observed (given in Table~\ref{Glitch_TNresultstable}) for the cases where exponential recovery has been considered for the noise analysis.

The continuous increment in the detected glitches may shed light on many correlations that are not well understood yet, for example, correlation with inter-glitch time, and may provide more information about the mechanism responsible for the occurrence of glitches. The reported glitch data sets are utilized to study statistical properties like differentiating the glitching pulsar population from the non-glitching population. Hence, it is important to report real glitches only. A discontinuity in observations or strong timing noise may manifest itself as a small glitch and result in a false detection, which should be taken care of. To tackle this problem, we have suggested a novel glitch verification method that can differentiate between a real glitch and a pseudo-glitch, which is a manifestation of any other phenomenon.

About $\sim$ 35\% of the total number of glitches reported have relatively small amplitudes. This suggests that many of the small glitches that have been reported could be due to strong timing noise in pulsars. Hence, it is very important to use our updated glitch verification methodology to make sure that any considered glitch is real and not a manifestation of timing noise.

The proposed GP realization technique for glitch analysis is capable of differentiating between a real glitch and a pseudo-glitch, which is demonstrated in Section~\ref{GPR_results} with basic noise models, consist combinations of white noise and achromatic red noise. These simple noise models serve as a good starting point. However, in future endeavors, it may be necessary to employ more comprehensive timing noise models by including several effects, such as the effect of the interstellar medium, which can play a significant role for some of the pulsars, e.g., the Crab pulsar. 

\begin{acknowledgement}
We acknowledge the support of staff at the Radio Astronomy Centre, Ooty, and the upgraded Giant Metrewave Radio Telescope during these observations. The ORT and the uGMRT are operated by the National Centre for Radio Astrophysics. 
We acknowledge the National Supercomputing Mission (NSM) for providing computing resources of ‘PARAM Ganga’ at the Indian Institute of Technology Roorkee, which is implemented by C-DAC and supported by the Ministry of Electronics and Information Technology (MeitY) and Department of Science and Technology (DST), Government of India. We would like to thank the anonymous reviewer for the insightful comments, which have significantly contributed to the improvement of the manuscript.
\end{acknowledgement}

\paragraph{Funding Statement}

BCJ acknowledges the support from Raja Ramanna Chair fellowship of the Department of Atomic Energy, Government of India (RRC - Track I Grant 3/3401 Atomic Energy Research 00 004 Research and Development 27 02 31 1002//2/2023/RRC/R\&D-II/13886). BCJ also acknowledges support from the Department of Atomic Energy Government of India, under project number 12-R\&D-TFR-5.02-0700.
JS acknowledges funding from the South African Research Chairs Initiative of the Depart of Science and Technology and the National Research Foundation of South Africa. 
EG is supported by the National Natural Science Foundation of China (NSFC) programme 11988101 under the foreign talents grant QN2023061004L.
PA acknowledges the support from SERB-DST, Govt.~of India, via project code \sloppy{CRG/2022/009359}.
JOC acknowledges financial support from the South African Department of Science and Innovation's National Research Foundation under the ISARP RADIOMAP Joint Research Scheme (DSI-NRF Grant Number 150551).



\paragraph{Data Availability Statement}

Data and codes will be provided on reasonable request.

\bibliography{main}

\begin{thebibliography}{}
\expandafter\ifx\csname natexlab\endcsname\relax\def\natexlab#1{#1}\fi

\bibitem[{{Alpar} {et~al.}(1984){Alpar}, {Pines}, {Anderson}, \& {Shaham}}]{AlparvortexcreepI1984}
{Alpar}, M.~A., {Pines}, D., {Anderson}, P.~W., \& {Shaham}, J. 1984, \apj, 276, 325

\bibitem[{{Ananthakrishnan}(1995)}]{Ananthakrishnan1995GMRT}
{Ananthakrishnan}, S. 1995, Journal of Astrophysics and Astronomy Supplement, 16, 427

\bibitem[{{Anderson} \& {Itoh}(1975)}]{AndersonItoh1975}
{Anderson}, P.~W., \& {Itoh}, N. 1975, \nat, 256, 25

\bibitem[{Antonelli {et~al.}(2022)Antonelli, Montoli, \& Pizzochero}]{Antonelli_2022}
Antonelli, M., Montoli, A., \& Pizzochero, P.~M. 2022, Insights Into the Physics of Neutron Star Interiors from Pulsar Glitches (WORLD SCIENTIFIC), 219–281

\bibitem[{{Antonopoulou} {et~al.}(2022){Antonopoulou}, {Haskell}, \& {Espinoza}}]{Antonopoulou_2022}
{Antonopoulou}, D., {Haskell}, B., \& {Espinoza}, C.~M. 2022, Reports on Progress in Physics, 85, 126901

\bibitem[{{Archibald} {et~al.}(2013){Archibald}, {Kaspi}, {Ng}, {Gourgouliatos}, {Tsang}, {Scholz}, {Beardmore}, {Gehrels}, \& {Kennea}}]{Archibald2013_anti-glitch}
{Archibald}, R.~F., {Kaspi}, V.~M., {Ng}, C.~Y., {et~al.} 2013, \nat, 497, 591

\bibitem[{{Arumugam} \& {Desai}(2023)}]{Arumugam_Desai_2023}
{Arumugam}, S., \& {Desai}, S. 2023, Journal of High Energy Astrophysics, 37, 46

\bibitem[{{Basu} {et~al.}(2020){Basu}, {Joshi}, {Krishnakumar}, {Bhattacharya}, {Nandi}, {Bandhopadhay}, {Char}, \& {Manoharan}}]{Basu2020}
{Basu}, A., {Joshi}, B.~C., {Krishnakumar}, M.~A., {et~al.} 2020, \mnras, 491, 3182

\bibitem[{{Basu} {et~al.}(2022){Basu}, {Shaw}, {Antonopoulou}, {Keith}, {Lyne}, {Mickaliger}, {Stappers}, {Weltevrede}, \& {Jordan}}]{basu2022}
{Basu}, A., {Shaw}, B., {Antonopoulou}, D., {et~al.} 2022, \mnras, 510, 4049

\bibitem[{{Celora} {et~al.}(2020){Celora}, {Khomenko}, {Antonelli}, \& {Haskell}}]{Celora2020}
{Celora}, T., {Khomenko}, V., {Antonelli}, M., \& {Haskell}, B. 2020, \mnras, 496, 5564

\bibitem[{{Chalumeau} {et~al.}(2022){Chalumeau}, {Babak}, {Petiteau}, {Chen}, {Samajdar}, {Caballero}, {Theureau}, {Guillemot}, {Desvignes}, {Parthasarathy}, {Liu}, {Shaifullah}, {Hu}, {van der Wateren}, {Antoniadis}, {Bak Nielsen}, {Bassa}, {Berthereau}, {Burgay}, {Champion}, {Cognard}, {Falxa}, {Ferdman}, {Freire}, {Gair}, {Graikou}, {Guo}, {Jang}, {Janssen}, {Karuppusamy}, {Keith}, {Kramer}, {Lee}, {Liu}, {Lyne}, {Main}, {McKee}, {Mickaliger}, {Perera}, {Perrodin}, {Porayko}, {Possenti}, {Sanidas}, {Sesana}, {Speri}, {Stappers}, {Tiburzi}, {Vecchio}, {Verbiest}, {Wang}, {Wang}, \& {Xu}}]{Chalumeau2021}
{Chalumeau}, A., {Babak}, S., {Petiteau}, A., {et~al.} 2022, \mnras, 509, 5538

\bibitem[{{Cordes} \& {Helfand}(1980)}]{Cordes1980}
{Cordes}, J.~M., \& {Helfand}, D.~J. 1980, \apj, 239, 640

\bibitem[{{{\c{S}}a{\c{s}}maz Mu{\c{s}}} {et~al.}(2014){{\c{S}}a{\c{s}}maz Mu{\c{s}}}, {Ayd{\i}n}, \& {G{\"o}{\u{g}}{\"u}{\c{s}}}}]{Sinem2014_anti-glitch}
{{\c{S}}a{\c{s}}maz Mu{\c{s}}}, S., {Ayd{\i}n}, B., \& {G{\"o}{\u{g}}{\"u}{\c{s}}}, E. 2014, \mnras, 440, 2916

\bibitem[{{D'Alessandro}(1996)}]{Alessandro_1996}
{D'Alessandro}, F. 1996, \apss, 246, 73

\bibitem[{{Downs}(1981)}]{Downs_1981}
{Downs}, G.~S. 1981, \apj, 249, 687

\bibitem[{{Dunn} {et~al.}(2023{\natexlab{a}}){Dunn}, {Melatos}, {Espinoza}, {Antonopoulou}, \& {Dodson}}]{Dunn2023_HMM_Vela}
{Dunn}, L., {Melatos}, A., {Espinoza}, C.~M., {Antonopoulou}, D., \& {Dodson}, R. 2023{\natexlab{a}}, \mnras, 522, 5469

\bibitem[{{Dunn} {et~al.}(2022{\natexlab{a}}){Dunn}, {Flynn}, {Bailes}, {Bateman}, {Campbell-Wilson}, {Deller}, {Green}, {Gupta}, {Jameson}, {Lee}, {Mandlik}, {Melatos}, \& {Urquhart}}]{ATel_Dunn_J0742}
{Dunn}, L., {Flynn}, C., {Bailes}, M., {et~al.} 2022{\natexlab{a}}, The Astronomer's Telegram, 15631, 1

\bibitem[{{Dunn} {et~al.}(2022{\natexlab{b}}){Dunn}, {Melatos}, {Suvorova}, {Moran}, {Evans}, {Os{\l}owski}, {Lower}, {Bailes}, {Flynn}, \& {Gupta}}]{Dunn2022_HMM_upper_limits}
{Dunn}, L., {Melatos}, A., {Suvorova}, S., {et~al.} 2022{\natexlab{b}}, \mnras, 512, 1469

\bibitem[{{Dunn} {et~al.}(2023{\natexlab{b}}){Dunn}, {Flynn}, {Bailes}, {Bateman}, {Campbell-Wilson}, {Deller}, {Green}, {Gupta}, {Jameson}, {Lee}, {Mandlik}, {Melatos}, \& {Urquhart}}]{ATel_Dunn_J1740}
{Dunn}, L., {Flynn}, C., {Bailes}, M., {et~al.} 2023{\natexlab{b}}, The Astronomer's Telegram, 15839, 1

\bibitem[{{Edwards} {et~al.}(2006){Edwards}, {Hobbs}, \& {Manchester}}]{Edwards2006Tempo2}
{Edwards}, R.~T., {Hobbs}, G.~B., \& {Manchester}, R.~N. 2006, \mnras, 372, 1549

\bibitem[{Ellis \& van Haasteren(2017)}]{ptmcmc}
Ellis, J., \& van Haasteren, R. 2017, jellis18/PTMCMCSampler: Official Release, doi:10.5281/zenodo.1037579

\bibitem[{{Ellis} {et~al.}(2019){Ellis}, {Vallisneri}, {Taylor}, \& {Baker}}]{ENTERPRISE_Ellis2019}
{Ellis}, J.~A., {Vallisneri}, M., {Taylor}, S.~R., \& {Baker}, P.~T. 2019, {ENTERPRISE: Enhanced Numerical Toolbox Enabling a Robust PulsaR Inference SuitE}, Astrophysics Source Code Library, record ascl:1912.015, ascl:1912.015

\bibitem[{{Espinoza} {et~al.}(2014){Espinoza}, {Antonopoulou}, {Stappers}, {Watts}, \& {Lyne}}]{Espinoza_2014}
{Espinoza}, C.~M., {Antonopoulou}, D., {Stappers}, B.~W., {Watts}, A., \& {Lyne}, A.~G. 2014, \mnras, 440, 2755

\bibitem[{{Eya} {et~al.}(2019){Eya}, {Urama}, \& {Chukwude}}]{Eya_2019}
{Eya}, I.~O., {Urama}, J.~O., \& {Chukwude}, A.~E. 2019, Research in Astronomy and Astrophysics, 19, 089

\bibitem[{Foreman-Mackey(2016)}]{corner}
Foreman-Mackey, D. 2016, The Journal of Open Source Software, 1, 24

\bibitem[{{Fowler} \& {Wright}(1982)}]{Fowler1982}
{Fowler}, L.~A., \& {Wright}, G.~A.~E. 1982, \aap, 109, 279

\bibitem[{{Fowler} {et~al.}(1981){Fowler}, {Wright}, \& {Morris}}]{Fowler1981}
{Fowler}, L.~A., {Wright}, G.~A.~E., \& {Morris}, D. 1981, \aap, 93, 54

\bibitem[{Fuentes {et~al.}(2017)Fuentes, Espinoza, Reisenegger, Shaw, Stappers, \& Lyne}]{Fuentes_2017}
Fuentes, J.~R., Espinoza, C.~M., Reisenegger, A., {et~al.} 2017, \aap, 608, A131

\bibitem[{{Gil} {et~al.}(1994){Gil}, {Jessner}, {Kijak}, {Kramer}, {Malofeev}, {Malov}, {Seiradakis}, {Sieber}, \& {Wielebinski}}]{Gil1994}
{Gil}, J.~A., {Jessner}, A., {Kijak}, J., {et~al.} 1994, Astronomy {\&} Astrophysics, 282, 45

\bibitem[{{Grover} {et~al.}(2022){Grover}, {Singha}, {Joshi}, \& {Arumugam}}]{ATel_Grover_J0742-2822}
{Grover}, H., {Singha}, J., {Joshi}, B.~C., \& {Arumugam}, P. 2022, The Astronomer's Telegram, 15629, 1

\bibitem[{{Grover} {et~al.}(2023){Grover}, {Singha}, {Joshi}, {Arumugam}, {Gugercinoglu}, {Bandyopadhyay}, {Desai}, {Banik}, {Eya}, {Chibueze}, {Urama}, \& {Kundu}}]{ATel_Grover_J1740-3015}
{Grover}, H., {Singha}, J., {Joshi}, B.~C., {et~al.} 2023, The Astronomer's Telegram, 15851, 1

\bibitem[{{G{\"u}gercino{\u{g}}lu} \& {Alpar}(2020)}]{Gugercinoglu_2020}
{G{\"u}gercino{\u{g}}lu}, E., \& {Alpar}, M.~A. 2020, \mnras, 496, 2506

\bibitem[{{G{\"u}gercino{\u{g}}lu} {et~al.}(2022){G{\"u}gercino{\u{g}}lu}, {Ge}, {Yuan}, \& {Zhou}}]{erbil2022}
{G{\"u}gercino{\u{g}}lu}, E., {Ge}, M.~Y., {Yuan}, J.~P., \& {Zhou}, S.~Q. 2022, \mnras, 511, 425

\bibitem[{{G{\"u}gercino{\u{g}}lu} {et~al.}(2023){G{\"u}gercino{\u{g}}lu}, {K{\"o}ksal}, \& {G{\"u}ver}}]{Erbil_2023}
{G{\"u}gercino{\u{g}}lu}, E., {K{\"o}ksal}, E., \& {G{\"u}ver}, T. 2023, \mnras, 518, 5734

\bibitem[{{G{\"u}gercino{\v{g}}lu} \& {Alpar}(2017)}]{Erbil_2017}
{G{\"u}gercino{\v{g}}lu}, E., \& {Alpar}, M.~A. 2017, \mnras, 471, 4827

\bibitem[{{G{\"u}gercino{\v{g}}lu} \& {Alpar}(2019)}]{Gugercinoglu_2019}
---. 2019, \mnras, 488, 2275

\bibitem[{{Gupta} {et~al.}(2017){Gupta}, {Ajithkumar}, {Kale}, {Nayak}, {Sabhapathy}, {Sureshkumar}, {Swami}, {Chengalur}, {Ghosh}, {Ishwara-Chandra}, {Joshi}, {Kanekar}, {Lal}, \& {Roy}}]{ugmrt2017}
{Gupta}, Y., {Ajithkumar}, B., {Kale}, H.~S., {et~al.} 2017, Current Science, 113, 707

\bibitem[{Gügercinoğlu \& Alpar(2016)}]{erbil2016}
Gügercinoğlu, E., \& Alpar, M.~A. 2016, \mnras, 462, 1453

\bibitem[{{Haskell} \& {Melatos}(2015)}]{Haskell_2015}
{Haskell}, B., \& {Melatos}, A. 2015, International Journal of Modern Physics D, 24, 1530008

\bibitem[{{Hobbs} {et~al.}(2010){Hobbs}, {Lyne}, \& {Kramer}}]{Hobbs_2010}
{Hobbs}, G., {Lyne}, A.~G., \& {Kramer}, M. 2010, \mnras, 402, 1027

\bibitem[{{Hobbs} {et~al.}(2006){Hobbs}, {Edwards}, \& {Manchester}}]{Hobbs2006Tempo2}
{Hobbs}, G.~B., {Edwards}, R.~T., \& {Manchester}, R.~N. 2006, \mnras, 369, 655

\bibitem[{{Hotan} {et~al.}(2004){Hotan}, {van Straten}, \& {Manchester}}]{PSRCHIVE_and_PSRFITS2004}
{Hotan}, A.~W., {van Straten}, W., \& {Manchester}, R.~N. 2004, \pasa, 21, 302

\bibitem[{{I{\c{c}}dem} {et~al.}(2012){I{\c{c}}dem}, {Baykal}, \& {Inam}}]{Icdem2012_anti-glitch}
{I{\c{c}}dem}, B., {Baykal}, A., \& {Inam}, S.~{\c{C}}. 2012, \mnras, 419, 3109

\bibitem[{{Joshi} {et~al.}(2018){Joshi}, {Arumugasamy}, {Bagchi}, {Bandyopadhyay}, {Basu}, {Dhanda Batra}, {Bethapudi}, {Choudhary}, {De}, {Dey}, {Gopakumar}, {Gupta}, {Krishnakumar}, {Maan}, {Manoharan}, {Naidu}, {Nandi}, {Pathak}, {Surnis}, \& {Susobhanan}}]{BCJ2018}
{Joshi}, B.~C., {Arumugasamy}, P., {Bagchi}, M., {et~al.} 2018, Journal of Astrophysics and Astronomy, 39, 51

\bibitem[{{Keith} {et~al.}(2013){Keith}, {Shannon}, \& {Johnston}}]{Keith2013}
{Keith}, M.~J., {Shannon}, R.~M., \& {Johnston}, S. 2013, \mnras, 432, 3080

\bibitem[{{Kerr}(2019)}]{ATelVelaG23_Kerr}
{Kerr}, M. 2019, The Astronomer's Telegram, 12481, 1

\bibitem[{{Kerscher} \& {Weller}(2019)}]{Kerscher}
{Kerscher}, M., \& {Weller}, J. 2019, SciPost Physics Lecture Notes, 9, arXiv:1901.07726

\bibitem[{{Kikunaga} {et~al.}(2024){Kikunaga}, {Hisano}, {Batra}, {Desai}, {Joshi}, {Bagchi}, {Prabu}, {Takahashi}, {Arumugam}, {Bathula}, {Dandapat}, {Deb}, {Dwivedi}, {Gupta}, {Jacob}, {Kareem}, {Nobleson}, {Mamidipaka}, {Paladi}, {Pandian}, {Rana}, {Singha}, {Srivastava}, {Surnis}, \& {Tarafdar}}]{Kikunaga2024}
{Kikunaga}, T., {Hisano}, S., {Batra}, N.~D., {et~al.} 2024, \pasa, 41, e036

\bibitem[{{Krishak} \& {Desai}(2020)}]{Krishak}
{Krishak}, A., \& {Desai}, S. 2020, JCAP, 2020, 006

\bibitem[{{Lentati} {et~al.}(2013){Lentati}, {Alexander}, {Hobson}, {Taylor}, {Gair}, {Balan}, \& {van Haasteren}}]{Lentati2013}
{Lentati}, L., {Alexander}, P., {Hobson}, M.~P., {et~al.} 2013, \prd, 87, 104021

\bibitem[{{Lentati} {et~al.}(2016){Lentati}, {Shannon}, {Coles}, {Verbiest}, {van Haasteren}, {Ellis}, {Caballero}, {Manchester}, {Arzoumanian}, {Babak}, {Bassa}, {Bhat}, {Brem}, {Burgay}, {Burke-Spolaor}, {Champion}, {Chatterjee}, {Cognard}, {Cordes}, {Dai}, {Demorest}, {Desvignes}, {Dolch}, {Ferdman}, {Fonseca}, {Gair}, {Gonzalez}, {Graikou}, {Guillemot}, {Hessels}, {Hobbs}, {Janssen}, {Jones}, {Karuppusamy}, {Keith}, {Kerr}, {Kramer}, {Lam}, {Lasky}, {Lassus}, {Lazarus}, {Lazio}, {Lee}, {Levin}, {Liu}, {Lynch}, {Madison}, {McKee}, {McLaughlin}, {McWilliams}, {Mingarelli}, {Nice}, {Os{\l}owski}, {Pennucci}, {Perera}, {Perrodin}, {Petiteau}, {Possenti}, {Ransom}, {Reardon}, {Rosado}, {Sanidas}, {Sesana}, {Shaifullah}, {Siemens}, {Smits}, {Stairs}, {Stappers}, {Stinebring}, {Stovall}, {Swiggum}, {Taylor}, {Theureau}, {Tiburzi}, {Toomey}, {Vallisneri}, {van Straten}, {Vecchio}, {Wang}, {Wang}, {You}, {Zhu}, \& {Zhu}}]{lentati_16}
{Lentati}, L., {Shannon}, R.~M., {Coles}, W.~A., {et~al.} 2016, \mnras, 458, 2161

\bibitem[{{Lower} {et~al.}(2020){Lower}, {Bailes}, {Shannon}, {Johnston}, {Flynn}, {Os{\l}owski}, {Gupta}, {Farah}, {Bateman}, {Green}, {Hunstead}, {Jameson}, {Jankowski}, {Parthasarathy}, {Price}, {Sutherland}, {Temby}, \& {Venkatraman Krishnan}}]{lower2020_utmost2}
{Lower}, M.~E., {Bailes}, M., {Shannon}, R.~M., {et~al.} 2020, \mnras, 494, 228

\bibitem[{{Lower} {et~al.}(2021){Lower}, {Johnston}, {Dunn}, {Shannon}, {Bailes}, {Dai}, {Kerr}, {Manchester}, {Melatos}, {Oswald}, {Parthasarathy}, {Sobey}, \& {Weltevrede}}]{Lower_2021}
{Lower}, M.~E., {Johnston}, S., {Dunn}, L., {et~al.} 2021, \mnras, 508, 3251

\bibitem[{{Lyne} {et~al.}(2010){Lyne}, {Hobbs}, {Kramer}, {Stairs}, \& {Stappers}}]{lyne2010}
{Lyne}, A., {Hobbs}, G., {Kramer}, M., {Stairs}, I., \& {Stappers}, B. 2010, Science, 329, 408

\bibitem[{{Lyne} {et~al.}(2015){Lyne}, {Jordan}, {Graham-Smith}, {Espinoza}, {Stappers}, \& {Weltevrede}}]{Lyne_2015}
{Lyne}, A.~G., {Jordan}, C.~A., {Graham-Smith}, F., {et~al.} 2015, \mnras, 446, 857

\bibitem[{{Lyne} {et~al.}(2000){Lyne}, {Shemar}, \& {Smith}}]{Lyne_2000}
{Lyne}, A.~G., {Shemar}, S.~L., \& {Smith}, F.~G. 2000, \mnras, 315, 534

\bibitem[{{Manchester} {et~al.}(2005){Manchester}, {Hobbs}, {Teoh}, \& {Hobbs}}]{Manchester_ATNF}
{Manchester}, R.~N., {Hobbs}, G.~B., {Teoh}, A., \& {Hobbs}, M. 2005, \aj, 129, 1993

\bibitem[{{Melatos} {et~al.}(2020){Melatos}, {Dunn}, {Suvorova}, {Moran}, \& {Evans}}]{Melatos2020_HMM}
{Melatos}, A., {Dunn}, L.~M., {Suvorova}, S., {Moran}, W., \& {Evans}, R.~J. 2020, \apj, 896, 78

\bibitem[{Morris {et~al.}(1981)Morris, Graham, \& Bartel}]{morris1981}
Morris, D., Graham, D.~A., \& Bartel, N. 1981, \mnras, 194, 7P

\bibitem[{{Naidu} {et~al.}(2015){Naidu}, {Joshi}, {Manoharan}, \& {Krishnakumar}}]{Naidu_2015}
{Naidu}, A., {Joshi}, B.~C., {Manoharan}, P.~K., \& {Krishnakumar}, M.~A. 2015, Experimental Astronomy, 39, 319

\bibitem[{{Os{\l}owski} {et~al.}(2011){Os{\l}owski}, {van Straten}, {Hobbs}, {Bailes}, \& {Demorest}}]{Osłowski_2011}
{Os{\l}owski}, S., {van Straten}, W., {Hobbs}, G.~B., {Bailes}, M., \& {Demorest}, P. 2011, \mnras, 418, 1258

\bibitem[{{Parthasarathy} {et~al.}(2019){Parthasarathy}, {Shannon}, {Johnston}, {Lentati}, {Bailes}, {Dai}, {Kerr}, {Manchester}, {Os{\l}owski}, {Sobey}, {van Straten}, \& {Weltevrede}}]{parthasarathy2019_TimingI}
{Parthasarathy}, A., {Shannon}, R.~M., {Johnston}, S., {et~al.} 2019, \mnras, 489, 3810

\bibitem[{{Parthasarathy} {et~al.}(2020){Parthasarathy}, {Johnston}, {Shannon}, {Lentati}, {Bailes}, {Dai}, {Kerr}, {Manchester}, {Os{\l}owski}, {Sobey}, {van Straten}, \& {Weltevrede}}]{parthasarathy2020_TimingII}
{Parthasarathy}, A., {Johnston}, S., {Shannon}, R.~M., {et~al.} 2020, \mnras, 494, 2012

\bibitem[{{Peng} {et~al.}(2022){Peng}, {Liu}, \& {Chou}}]{peng2018}
{Peng}, Q.-H., {Liu}, J.-J., \& {Chou}, C.-K. 2022, New Astronomy, 90, 101655

\bibitem[{{Pintore} {et~al.}(2016){Pintore}, {Bernardini}, {Mereghetti}, {Esposito}, {Turolla}, {Rea}, {Coti Zelati}, {Israel}, {Tiengo}, \& {Zane}}]{Pintore2016_anti-glitch}
{Pintore}, F., {Bernardini}, F., {Mereghetti}, S., {et~al.} 2016, \mnras, 458, 2088

\bibitem[{{Radhakrishnan} \& {Manchester}(1969)}]{Radhakrishnan_Manchester_1969}
{Radhakrishnan}, V., \& {Manchester}, R.~N. 1969, \nat, 222, 228

\bibitem[{{Ray} {et~al.}(2019){Ray}, {Guillot}, {Ho}, {Kerr}, {Enoto}, {Gendreau}, {Arzoumanian}, {Altamirano}, {Bogdanov}, {Campion}, {Chakrabarty}, {Deneva}, {Jaisawal}, {Kozon}, {Malacaria}, {Strohmayer}, \& {Wolff}}]{Ray2019_anti-glitch}
{Ray}, P.~S., {Guillot}, S., {Ho}, W. C.~G., {et~al.} 2019, \apj, 879, 130

\bibitem[{{Reichley} \& {Downs}(1969)}]{Reichley_Downs_1969}
{Reichley}, P.~E., \& {Downs}, G.~S. 1969, \nat, 222, 229

\bibitem[{{Shabanova}(1998)}]{Shabanova1998_slowglitch}
{Shabanova}, T.~V. 1998, \aap, 337, 723

\bibitem[{{Shabanova}(2005)}]{Shabanova2005}
---. 2005, \mnras, 356, 1435

\bibitem[{{Shabanova}(2010)}]{Shabanova_2010}
---. 2010, \apj, 721, 251

\bibitem[{{Shafiq Hazboun}(2020)}]{La_Forge}
{Shafiq Hazboun}, J. 2020, {La Forge}, doi:10.5281/zenodo.4152550

\bibitem[{{Shannon} \& {Cordes}(2010)}]{Shannon_2010}
{Shannon}, R.~M., \& {Cordes}, J.~M. 2010, \apj, 725, 1607

\bibitem[{{Sharma}(2017)}]{Sanjib}
{Sharma}, S. 2017, \araa, 55, 213

\bibitem[{{Shaw} {et~al.}(2021){Shaw}, {Keith}, {Lyne}, {Mickaliger}, {Stappers}, {Turner}, \& {Weltevrede}}]{Shaw2021}
{Shaw}, B., {Keith}, M.~J., {Lyne}, A.~G., {et~al.} 2021, \mnras, 505, L6

\bibitem[{{Shaw} {et~al.}(2022{\natexlab{a}}){Shaw}, {Mickaliger}, {Stappers}, {Lyne}, {Keith}, {Weltevrede}, \& {Basu}}]{ATel_shaw_J0742}
{Shaw}, B., {Mickaliger}, M.~B., {Stappers}, B.~W., {et~al.} 2022{\natexlab{a}}, The Astronomer's Telegram, 15622, 1

\bibitem[{{Shaw} {et~al.}(2022{\natexlab{b}}){Shaw}, {Stappers}, {Weltevrede}, {Brook}, {Karastergiou}, {Jordan}, {Keith}, {Kramer}, \& {Lyne}}]{Shaw_2022}
{Shaw}, B., {Stappers}, B.~W., {Weltevrede}, P., {et~al.} 2022{\natexlab{b}}, \mnras, 513, 5861

\bibitem[{{Singha} {et~al.}(2021{\natexlab{a}}){Singha}, {Basu}, {Krishnakumar}, {Joshi}, \& {Arumugam}}]{Singha_2021_agdp}
{Singha}, J., {Basu}, A., {Krishnakumar}, M.~A., {Joshi}, B.~C., \& {Arumugam}, P. 2021{\natexlab{a}}, \mnras, 505, 5488

\bibitem[{{Singha} {et~al.}(2021{\natexlab{b}}){Singha}, {Joshi}, {Arumugam}, \& {Bandyopadhyay}}]{Atel_Singha_Vela_2021}
{Singha}, J., {Joshi}, B.~C., {Arumugam}, P., \& {Bandyopadhyay}, D. 2021{\natexlab{b}}, The Astronomer's Telegram, 14812, 1

\bibitem[{{Singha} {et~al.}(2022){Singha}, {Joshi}, {Bandyopadhyay}, {Grover}, {Desai}, {Arumugam}, \& {Banik}}]{Singha_2022_SKA_Review}
{Singha}, J., {Joshi}, B.~C., {Bandyopadhyay}, D., {et~al.} 2022, Journal of Astrophysics and Astronomy, 43, 81

\bibitem[{{Sosa-Fiscella} {et~al.}(2021){Sosa-Fiscella}, {Zubieta}, {del Palacio}, {Garcia}, {Lopez-Armengol}, {Combi}, {Lousto}, {Gancio}, {Combi}, {Gutierrez}, {Bunzel}, {Hauscarriaga}, \& {PuMA Collaboration}}]{ATel_Sosa_G24_Vela}
{Sosa-Fiscella}, V., {Zubieta}, E., {del Palacio}, S., {et~al.} 2021, The Astronomer's Telegram, 14806, 1

\bibitem[{{Speagle}(2020)}]{DYNESTY2020}
{Speagle}, J.~S. 2020, \mnras, 493, 3132

\bibitem[{{Srivastava} {et~al.}(2023){Srivastava}, {Desai}, {Kolhe}, {Surnis}, {Joshi}, {Susobhanan}, {Chalumeau}, {Hisano}, {Nobleson}, {Arumugam}, {Kharbanda}, {Singha}, {Tarafdar}, {Arumugam}, {Bagchi}, {Bathula}, {Dandapat}, {Dey}, {Dwivedi}, {Girgaonkar}, {Gopakumar}, {Gupta}, {Kikunaga}, {Krishnakumar}, {Liu}, {Maan}, {Manoharan}, {Paladi}, {Rana}, {Shaifullah}, \& {Takahashi}}]{Srivastava2023}
{Srivastava}, A., {Desai}, S., {Kolhe}, N., {et~al.} 2023, \prd, 108, 023008

\bibitem[{Susobhanan {et~al.}(2021)Susobhanan, Maan, Joshi, Prabu, Desai, Nobleson, Susarla, Girgaonkar, Dey, Batra, \& et~al.}]{pinta}
Susobhanan, A., Maan, Y., Joshi, B.~C., {et~al.} 2021, \pasa, 38, e017

\bibitem[{{Swarup} {et~al.}(1991){Swarup}, {Ananthakrishnan}, {Kapahi}, {Rao}, {Subrahmanya}, \& {Kulkarni}}]{Swarup1991GMRT}
{Swarup}, G., {Ananthakrishnan}, S., {Kapahi}, V.~K., {et~al.} 1991, Current Science, 60, 95

\bibitem[{Swarup {et~al.}(1971)Swarup, Sarma, Joshi, Kapahi, Bagri, Damle, Ananthakrishnan, Balasubramanian, Bhave, \& Sinha}]{swarup1971ort}
Swarup, G., Sarma, N., Joshi, M., {et~al.} 1971, Nature Physical Science, 230, 185

\bibitem[{{Trotta}(2008)}]{Trotta2008}
{Trotta}, R. 2008, Contemporary Physics, 49, 71

\bibitem[{{Urama} \& {Okeke}(1999)}]{Urama1999}
{Urama}, J.~O., \& {Okeke}, P.~N. 1999, \mnras, 310, 313

\bibitem[{{van Haasteren} \& {Levin}(2013)}]{van_Haasteren_2013}
{van Haasteren}, R., \& {Levin}, Y. 2013, \mnras, 428, 1147

\bibitem[{{van Haasteren} {et~al.}(2009){van Haasteren}, {Levin}, {McDonald}, \& {Lu}}]{Haasteren2009}
{van Haasteren}, R., {Levin}, Y., {McDonald}, P., \& {Lu}, T. 2009, \mnras, 395, 1005

\bibitem[{{van Haasteren} \& {Vallisneri}(2014)}]{Haasteren2014}
{van Haasteren}, R., \& {Vallisneri}, M. 2014, \prd, 90, 104012

\bibitem[{{van Straten} \& {Bailes}(2011)}]{DSPSR}
{van Straten}, W., \& {Bailes}, M. 2011, \pasa, 28, 1

\bibitem[{{van Straten} {et~al.}(2012){van Straten}, {Demorest}, \& {Oslowski}}]{PSRCHIVE2012}
{van Straten}, W., {Demorest}, P., \& {Oslowski}, S. 2012, Astronomical Research and Technology, 9, 237

\bibitem[{{Younes} {et~al.}(2020){Younes}, {Ray}, {Baring}, {Kouveliotou}, {Fletcher}, {Wadiasingh}, {Harding}, \& {Goldstein}}]{Younes2020_anti-glitch}
{Younes}, G., {Ray}, P.~S., {Baring}, M.~G., {et~al.} 2020, \apjl, 896, L42

\bibitem[{{Yu} {et~al.}(2013){Yu}, {Manchester}, {Hobbs}, {Johnston}, {Kaspi}, {Keith}, {Lyne}, {Qiao}, {Ravi}, {Sarkissian}, {Shannon}, \& {Xu}}]{Yu2013}
{Yu}, M., {Manchester}, R.~N., {Hobbs}, G., {et~al.} 2013, \mnras, 429, 688

\bibitem[{{Zhou} {et~al.}(2022){Zhou}, {G{\"u}gercino{\u{g}}lu}, {Yuan}, {Ge}, \& {Yu}}]{zhou_erbil_glitch_review}
{Zhou}, S., {G{\"u}gercino{\u{g}}lu}, E., {Yuan}, J., {Ge}, M., \& {Yu}, C. 2022, Universe, 8, 641

\bibitem[{{Zhou} {et~al.}(2019){Zhou}, {Zhou}, {Zhang}, {Liu}, {Liu}, {Zhang}, {Feng}, {Zhu}, \& {Wu}}]{Zhou2019_slowglitch}
{Zhou}, S.~Q., {Zhou}, A.~A., {Zhang}, J., {et~al.} 2019, \apss, 364, 173

\bibitem[{{Zou} {et~al.}(2004){Zou}, {Wang}, {Wang}, {Manchester}, {Wu}, \& {Zhang}}]{Zou2004}
{Zou}, W.~Z., {Wang}, N., {Wang}, H.~X., {et~al.} 2004, \mnras, 354, 811

\bibitem[{{Zubieta} {et~al.}(2022{\natexlab{a}}){Zubieta}, {Furlan}, {Palacio}, {Garcia}, {Gancio}, {Lousto}, {Combi}, \& {Combi}}]{ATel_Zubieta_J1740}
{Zubieta}, E., {Furlan}, S.~B.~A., {Palacio}, S.~d., {et~al.} 2022{\natexlab{a}}, The Astronomer's Telegram, 15838, 1

\bibitem[{{Zubieta} {et~al.}(2024){Zubieta}, {Garcia}, {del Palacio}, {Araujo Furlan}, {Gancio}, {Lousto}, {Combi}, \& {Espinoza}}]{Zubieta2024}
{Zubieta}, E., {Garcia}, F., {del Palacio}, S., {et~al.} 2024, arXiv e-prints, arXiv:2406.17099

\bibitem[{{Zubieta} {et~al.}(2022{\natexlab{b}}){Zubieta}, {Del Palacio}, {Garcia}, {Gancio}, {Lousto}, {Combi}, {Combi}, {Gutierrez}, {Lopez-Armengol}, {Simaz Bunzel}, \& {Sosa-Fiscella}}]{ATel_Zubieta_J0742}
{Zubieta}, E., {Del Palacio}, S., {Garcia}, F., {et~al.} 2022{\natexlab{b}}, The Astronomer's Telegram, 15638, 1

\bibitem[{{Zubieta} {et~al.}(2023){Zubieta}, {Missel}, {Sosa Fiscella}, {Lousto}, {del Palacio}, {L{\'o}pez Armengol}, {Garc{\'\i}a}, {Combi}, {Wang}, {Combi}, {Gancio}, {Negrelli}, \& {Guti{\'e}rrez}}]{Zubieta_2023}
{Zubieta}, E., {Missel}, R., {Sosa Fiscella}, V., {et~al.} 2023, \mnras, 521, 4504

\end{thebibliography}

\clearpage
\appendix
\onecolumn

\section{Timing Noise Posteriors}
\label{appendix_TN_posterior}
The posteriors of the noise model parameters of our sample of pulsars are given below. The posterior distributions are plotted with 68\% and 95\% credible intervals for all the cases.\\

\begin{figure*}
    \centering
    \includegraphics[width=0.9\linewidth]{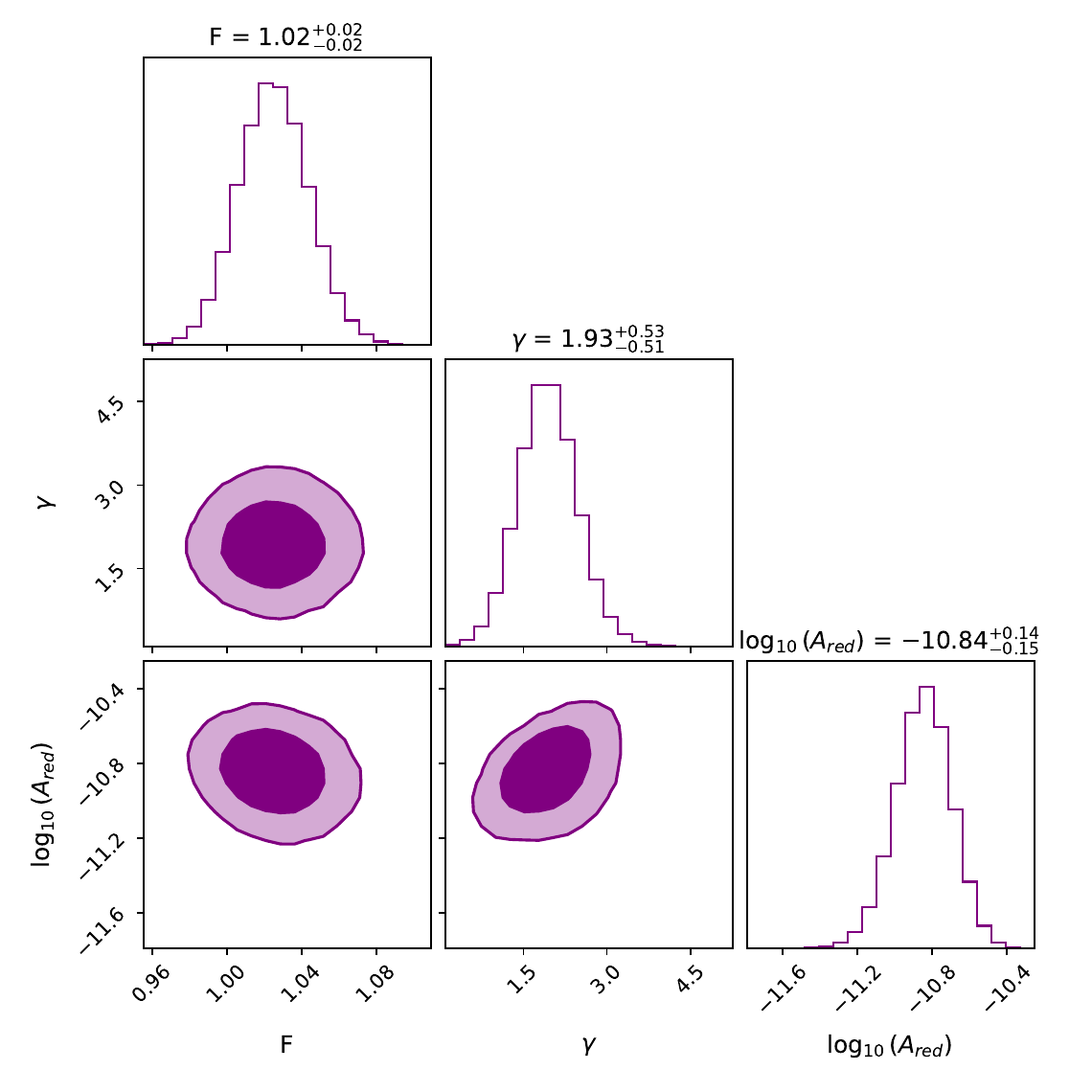}
\caption{Timing Noise posteriors of J0358+5413 with 68 and 95\% credible interval for our sample of pulsars. The symbols F, $\sigma_Q$, $A_{\rm red}$, $\gamma$ represent EFAC, EQUAD, Red noise Amplitude and Spectral index respectively.}
\label{TN_results}
\end{figure*}

\renewcommand{\thefigure}{\arabic{figure} (Cont.)}
\addtocounter{figure}{-1}
\begin{figure*}
\begin{subfigure}{0.49\textwidth}
    \centering
    \includegraphics[width=\linewidth]{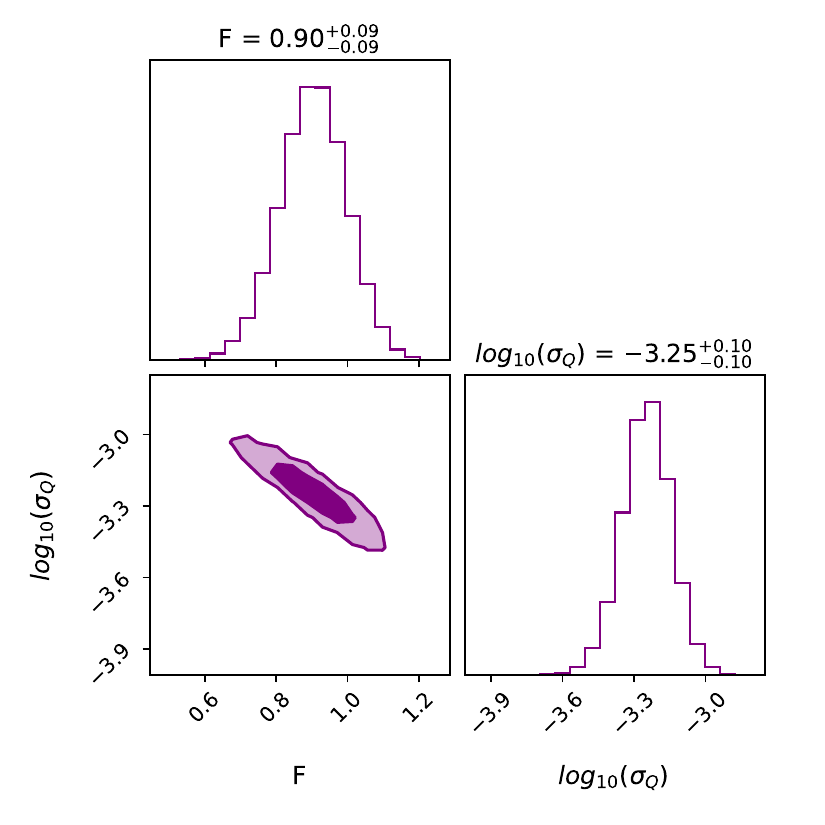}
    \caption{Timing Noise posterior of J0525+1115}
    \label{TNJ0525}
\end{subfigure}   
\begin{subfigure}{0.49\textwidth}
    \centering
    \includegraphics[width=\linewidth]{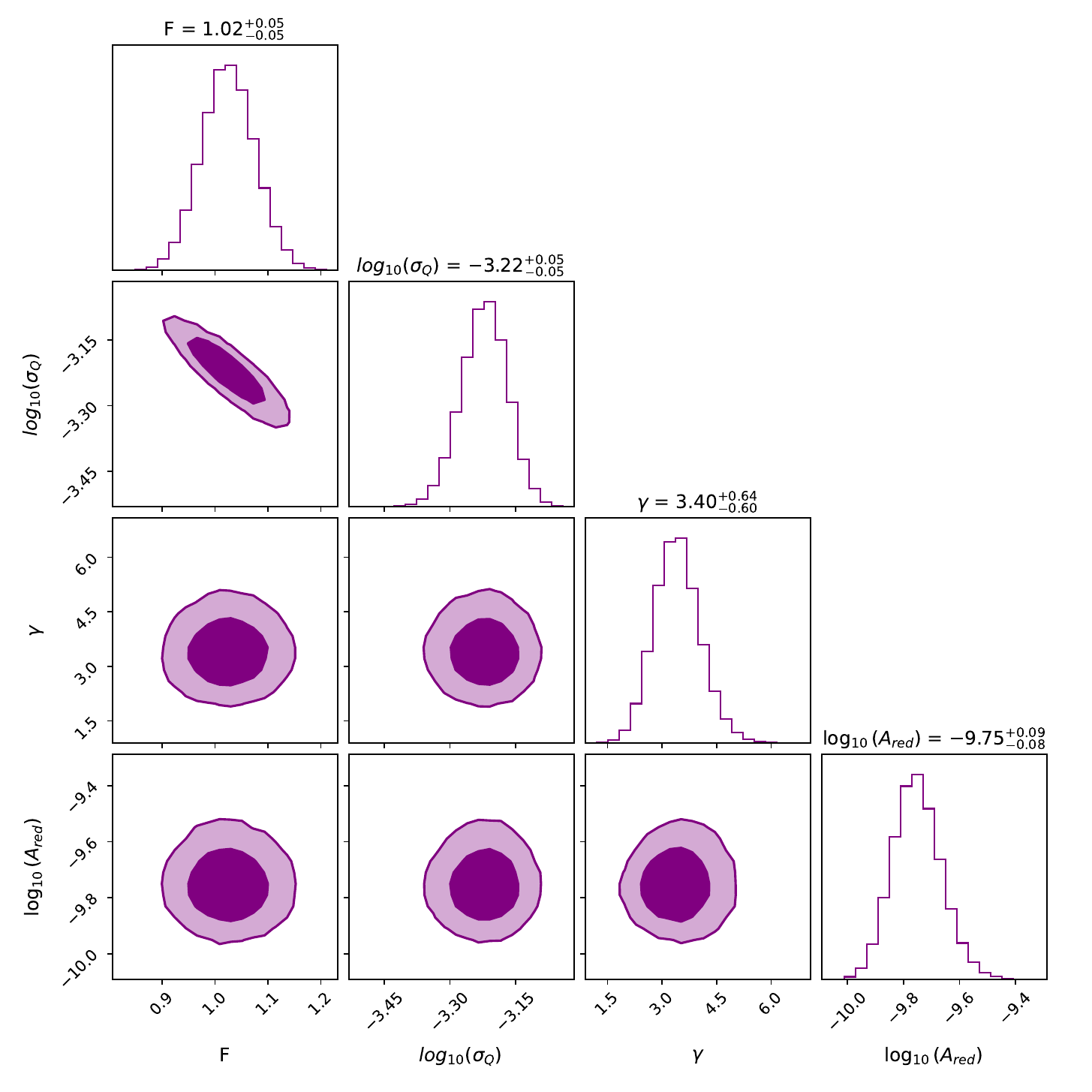}
    \caption{Timing Noise posterior of J0528+2200}
    \label{TNJ0528}
\end{subfigure}
\begin{subfigure}{0.49\textwidth}
    \centering
    \includegraphics[width=\linewidth]{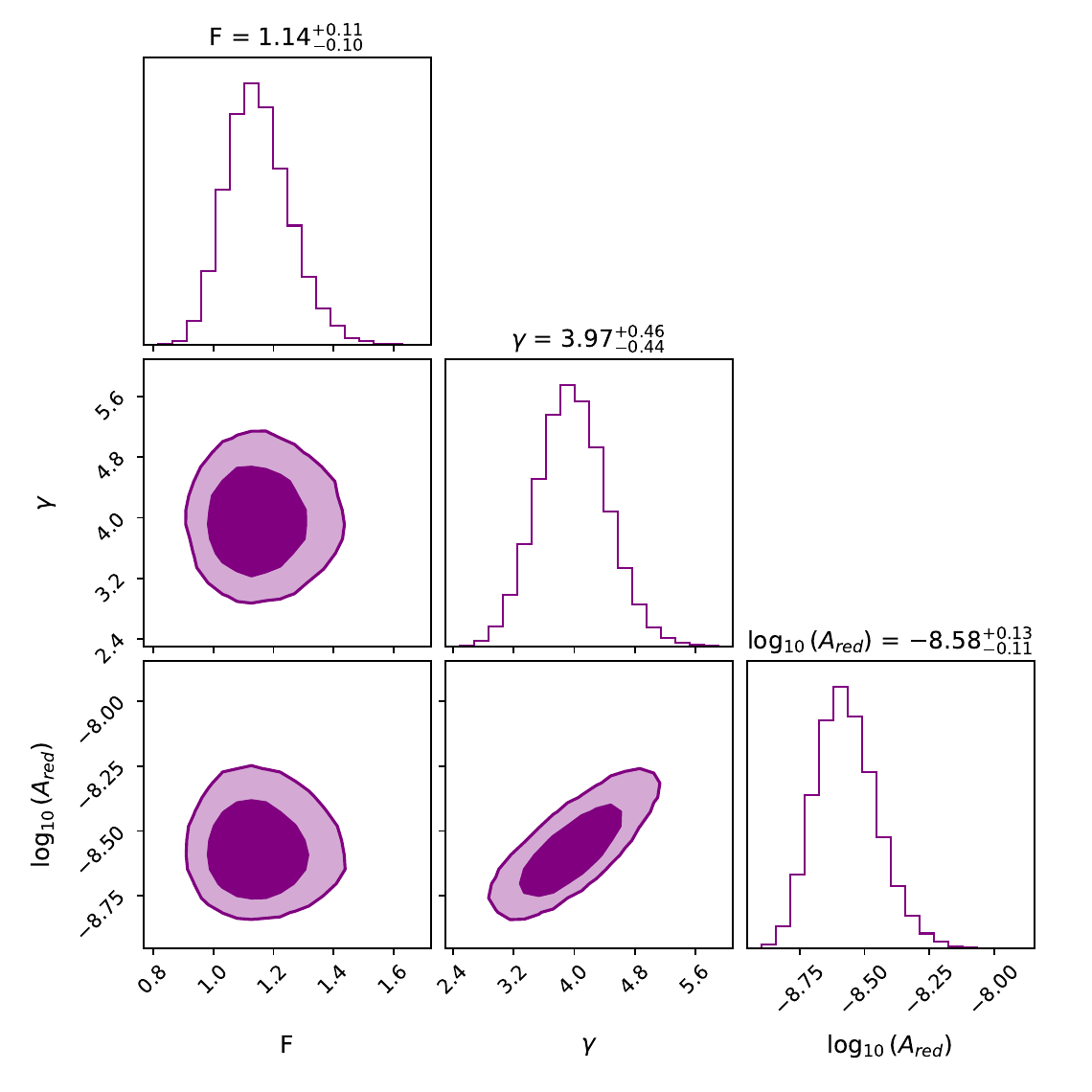}
    \caption{Timing Noise posterior of J0729--1448}
    \label{TNJ0729-1448}
\end{subfigure}  
\begin{subfigure}{0.49\textwidth}
    \centering
    \includegraphics[width=\linewidth]{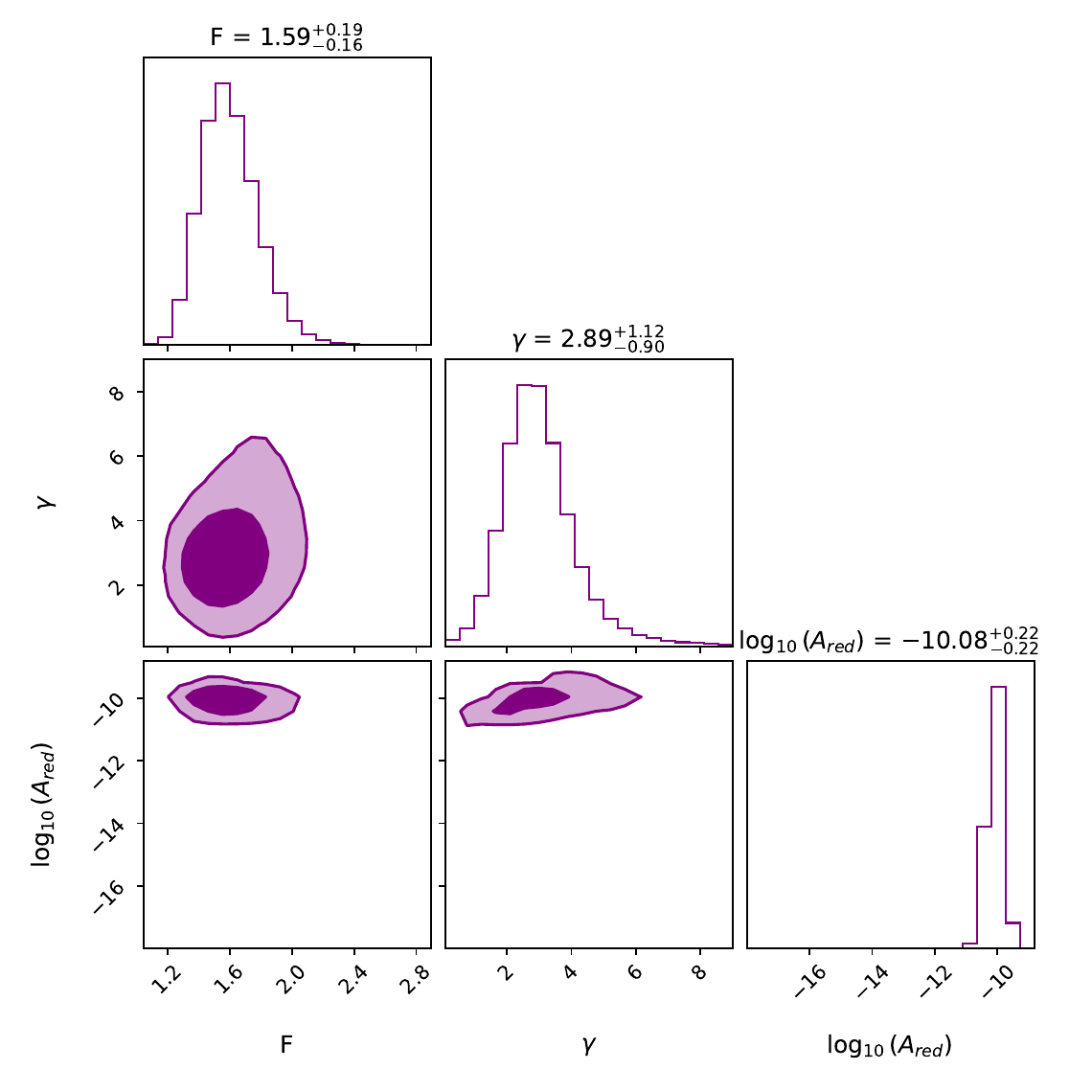}
    \caption{Timing Noise posterior of J0729--1836}
    \label{TNJ0729-1836}
\end{subfigure} 
\caption{Timing Noise posteriors with 68 and 95\% credible interval for our sample of pulsars. The symbols F, $\sigma_Q$, $A_{\rm red}$, $\gamma$ represent EFAC, EQUAD, Red noise Amplitude and Spectral index respectively.}
\end{figure*}

\renewcommand{\thefigure}{\arabic{figure} (Cont.)}
\addtocounter{figure}{-1}
\begin{figure*}
    \centering
\begin{subfigure}{0.49\textwidth}
    \centering
    \includegraphics[width=\linewidth]{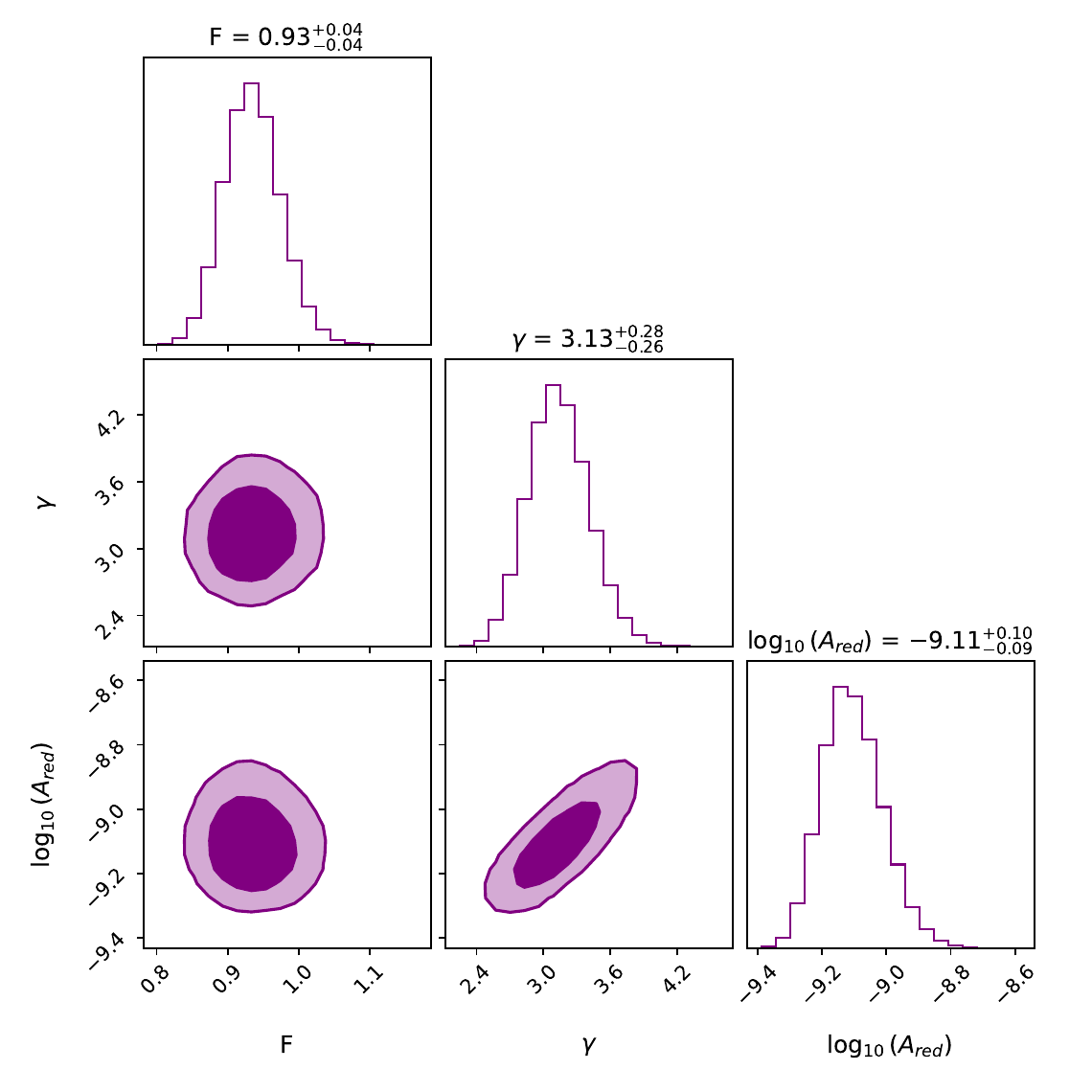}
    \caption{Timing Noise posterior of J0922+0638}
    \label{TNJ0922}
\end{subfigure}
\begin{subfigure}{0.49\textwidth}
    \centering
    \includegraphics[width=\linewidth]{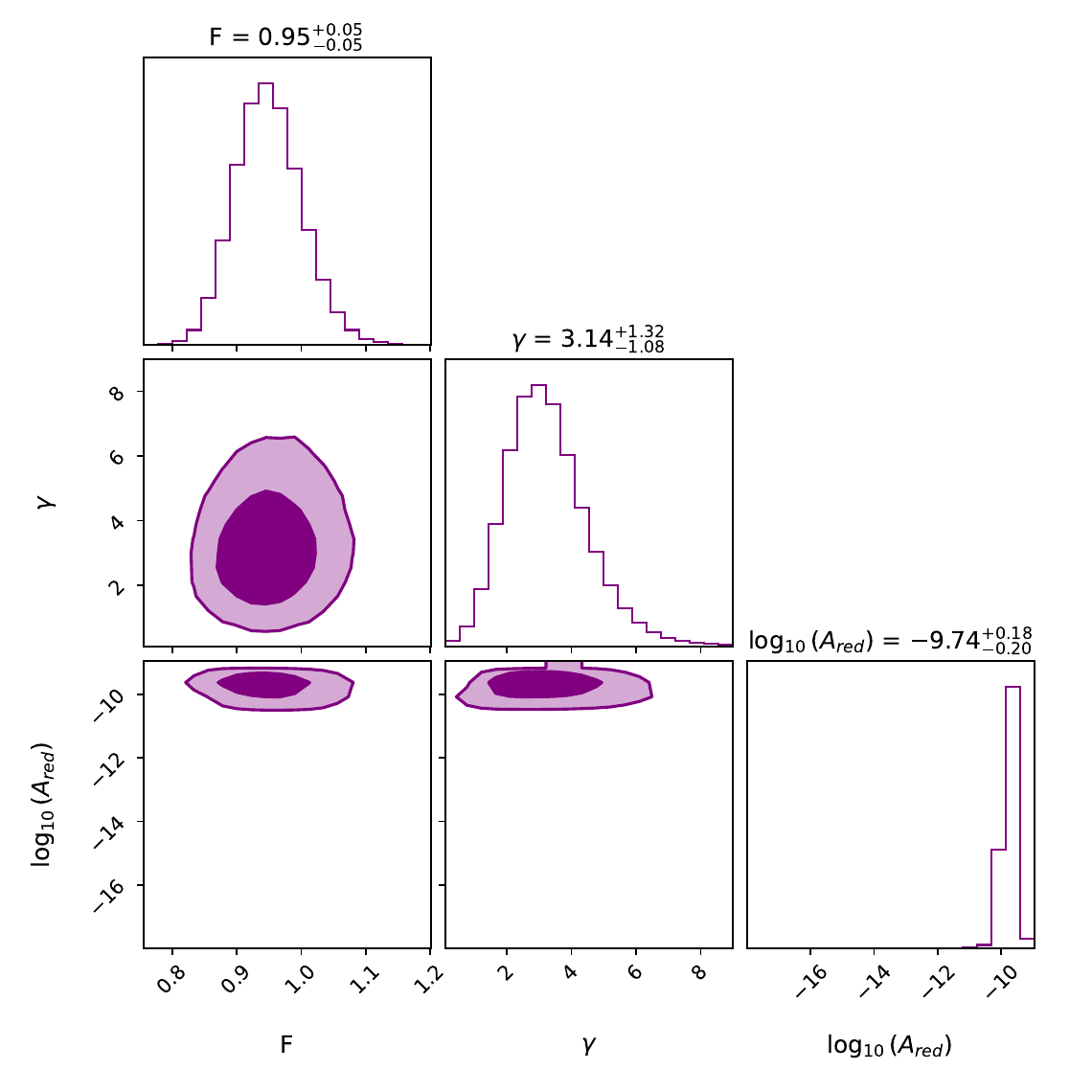}
    \caption{Timing Noise posterior of J1532+2745}
    \label{TNJ1532}
\end{subfigure}   
\begin{subfigure}{0.49\textwidth}
    \centering
    \includegraphics[width=\linewidth]{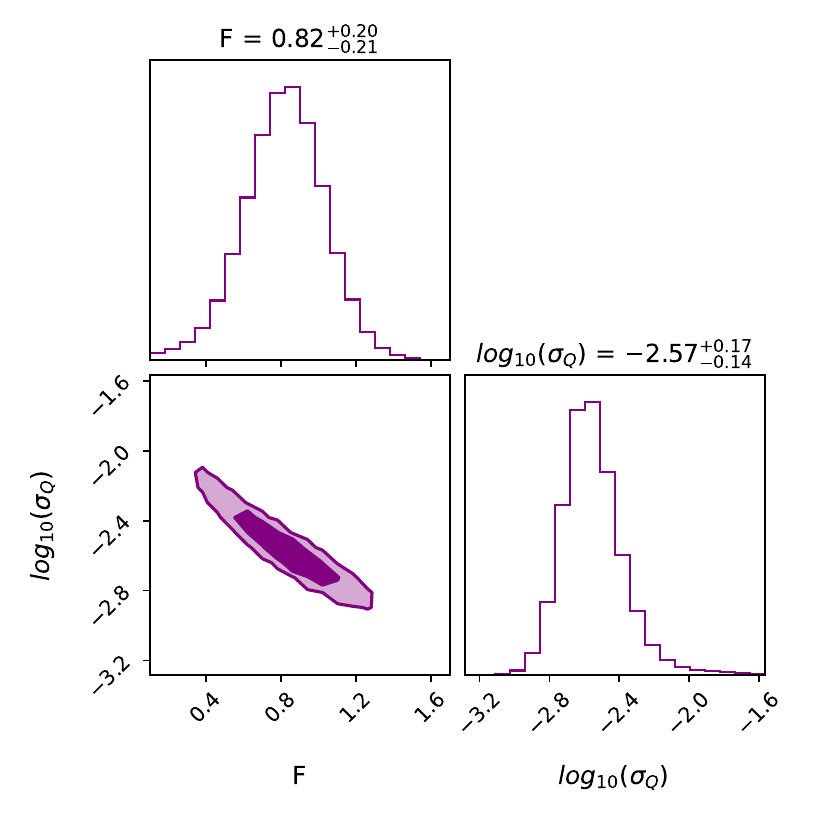}
    \caption{Timing Noise posterior of J1720--1633}
    \label{TNJ1720}
\end{subfigure}
\begin{subfigure}{0.49\textwidth}
    \centering
    \includegraphics[width=\linewidth]{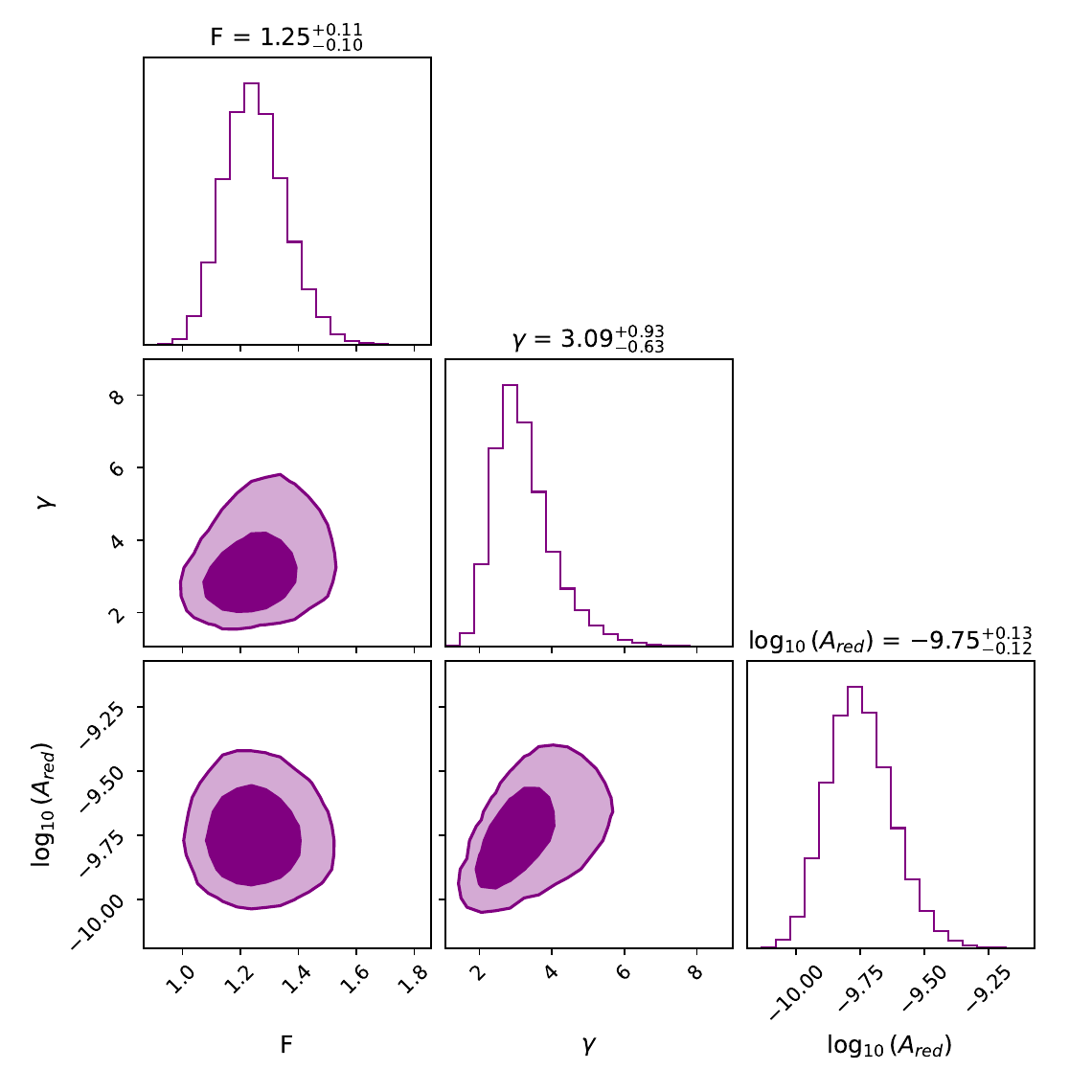}
    \caption{Timing Noise posterior of J1731--4744}
    \label{TNJ1731}
\end{subfigure}
\caption{Timing Noise posteriors with 68 and 95\% credible interval for our sample of pulsars. The symbols F, $\sigma_Q$, $A_{\rm red}$, $\gamma$ represent EFAC, EQUAD, Red noise Amplitude and Spectral index respectively.}
\end{figure*}

\addtocounter{figure}{-1}
\begin{figure*}
    \centering
\begin{subfigure}{0.49\textwidth}
    \centering
    \includegraphics[width=\linewidth]{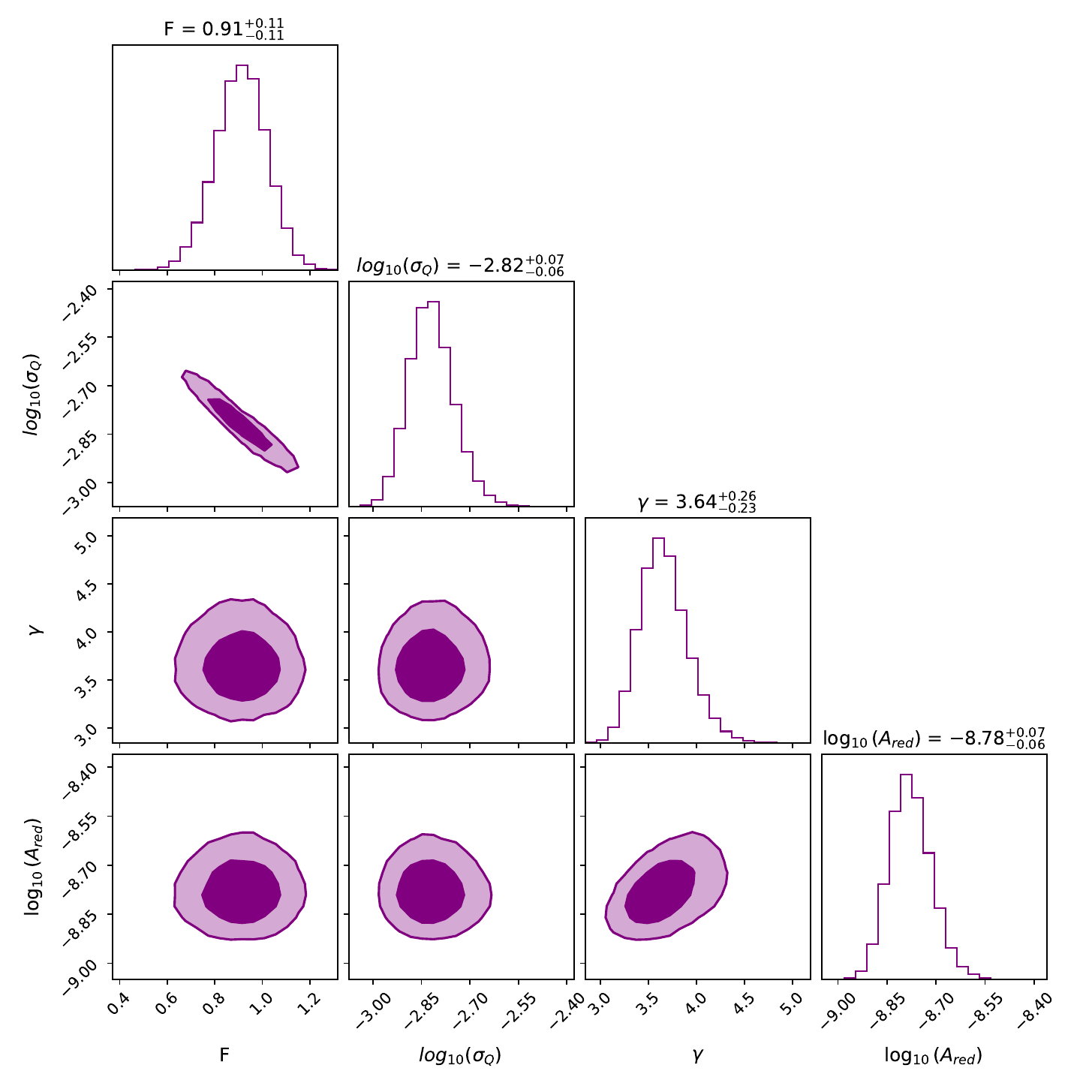}
    \caption{Timing Noise posterior of J1825--0935}
    \label{TNJ1825}
\end{subfigure} 
\begin{subfigure}{0.49\textwidth}
    \centering
    \includegraphics[width=\linewidth]{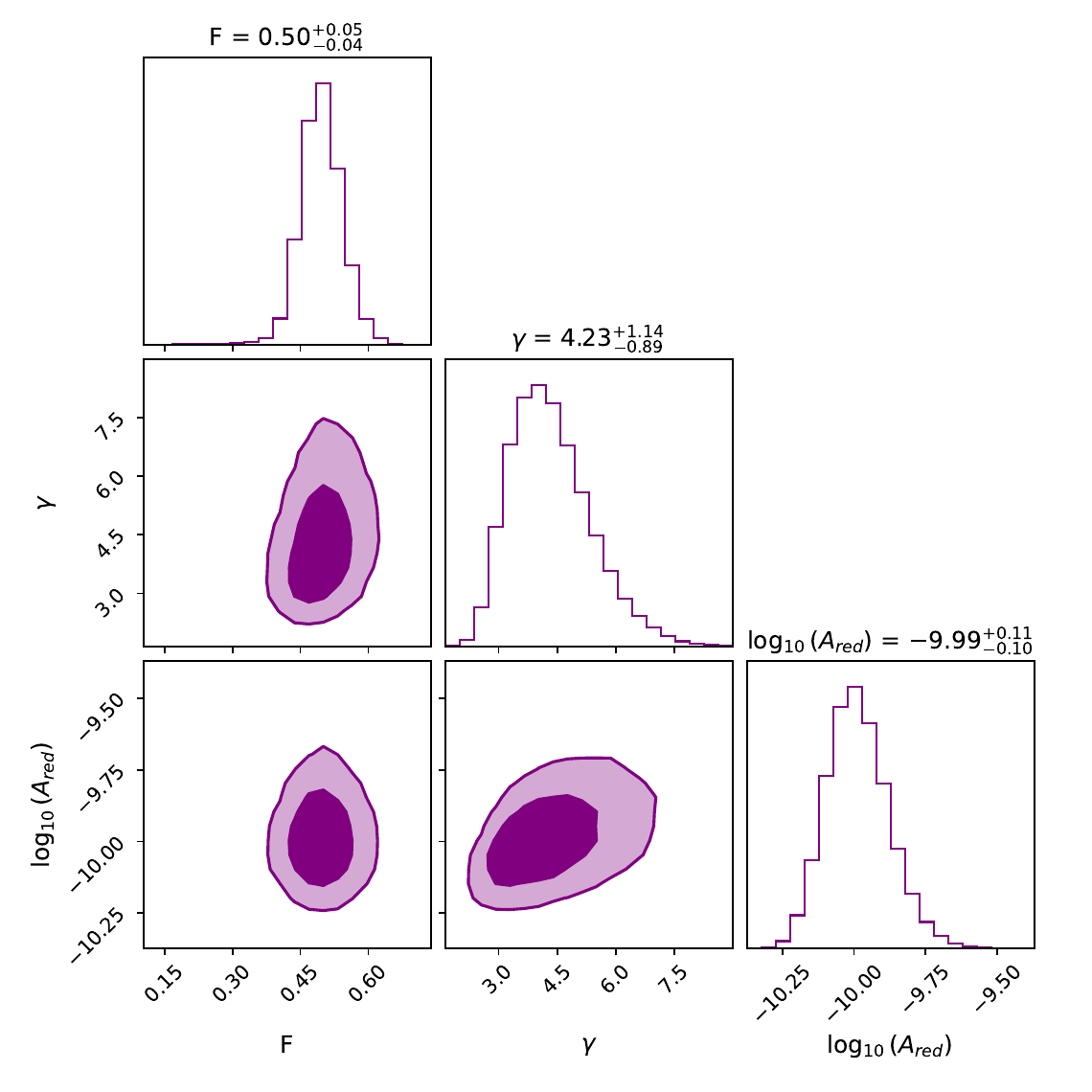}
    \caption{Timing Noise posterior of J1847--0402}
    \label{TNJ1847}
\end{subfigure}   
\begin{subfigure}{0.49\textwidth}
    \centering
    \includegraphics[width=\linewidth]{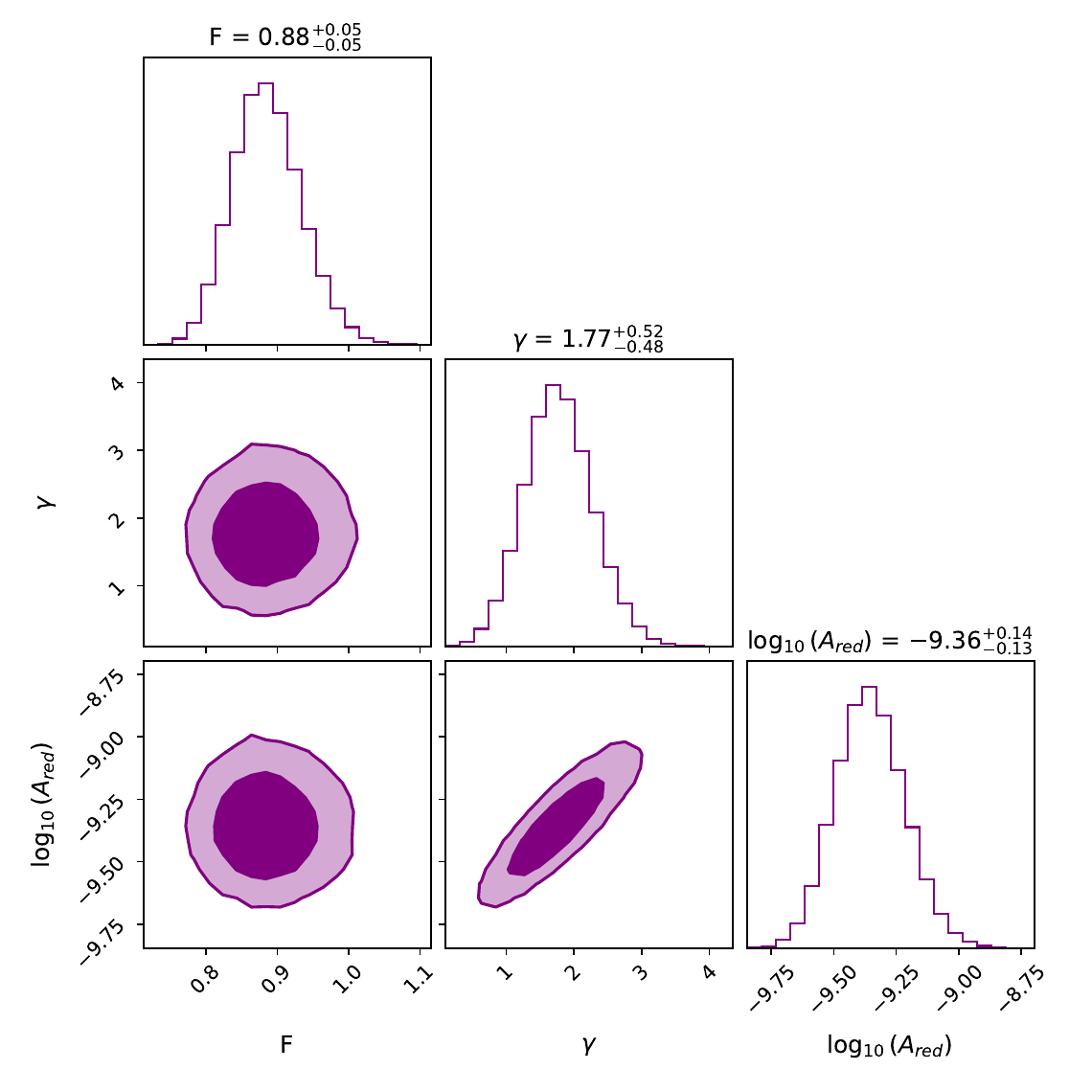}
    \caption{Timing Noise posterior of J1909+0007}
    \label{TNJ1909}
\end{subfigure}
\begin{subfigure}{0.49\textwidth}
    \centering
    \includegraphics[width=\linewidth]{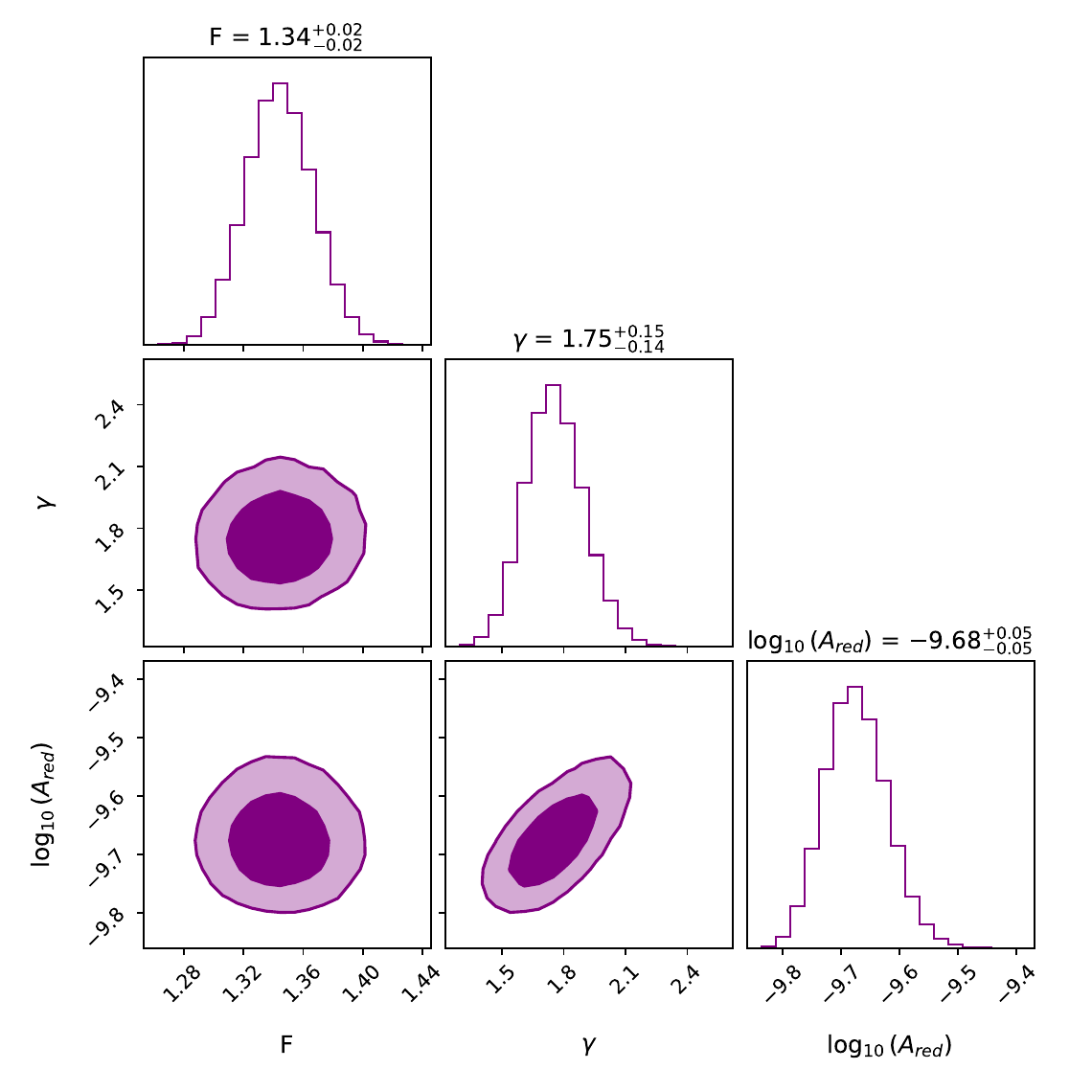}
    \caption{Timing Noise posterior of J1910--0309}
    \label{TNJ1910}
\end{subfigure}   
\caption{Timing Noise posteriors with 68 and 95\% credible interval for our sample of pulsars. The symbols F, $\sigma_Q$, $A_{\rm red}$, $\gamma$ represent EFAC, EQUAD, Red noise Amplitude and Spectral index respectively.}
\end{figure*}

\addtocounter{figure}{-1}
\begin{figure*}
    \centering
\begin{subfigure}{0.49\textwidth}
    \centering
    \includegraphics[width=\linewidth]{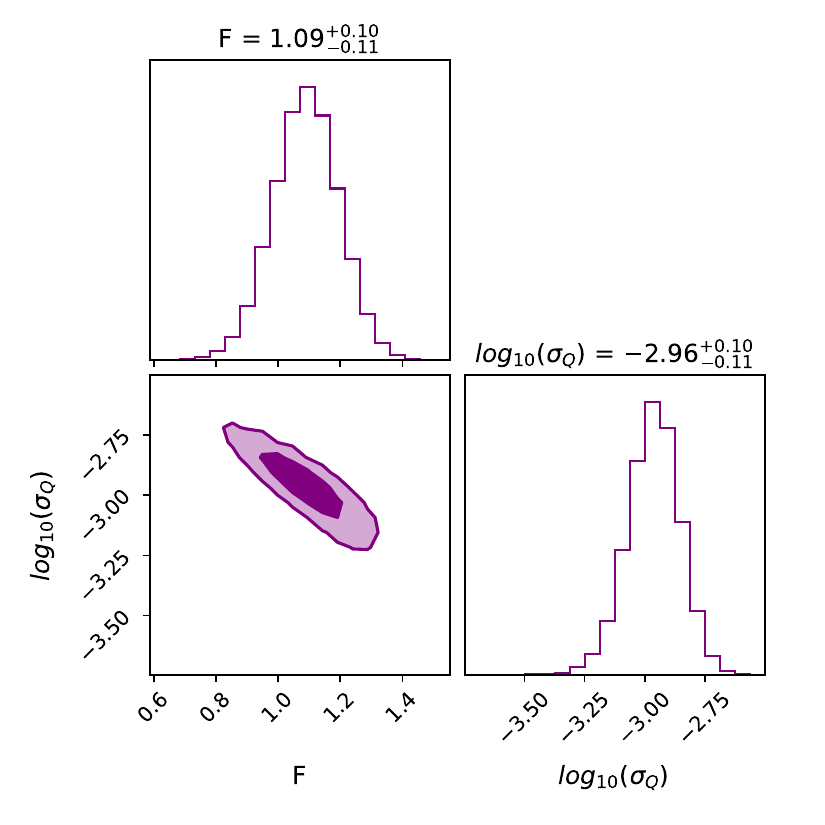}
    \caption{Timing Noise posterior of J1919+0021}
    \label{TNJ1919}
\end{subfigure}
\begin{subfigure}{0.49\textwidth}
    \centering
    \includegraphics[width=\linewidth]{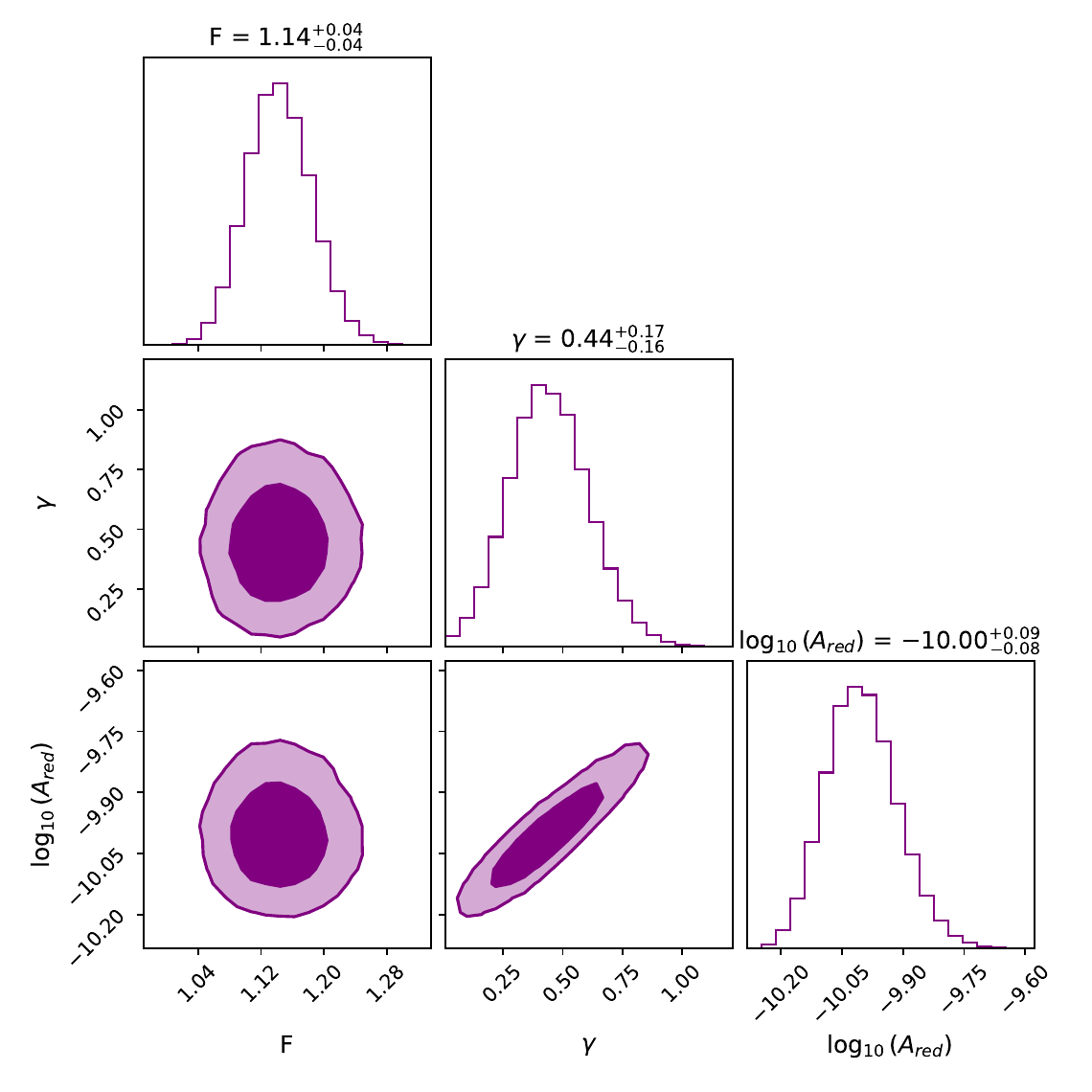}
    \caption{Timing Noise posterior of J2022+2854}
    \label{TNJ2022}
\end{subfigure}   
\begin{subfigure}{0.49\textwidth}
    \centering
    \includegraphics[width=\linewidth]{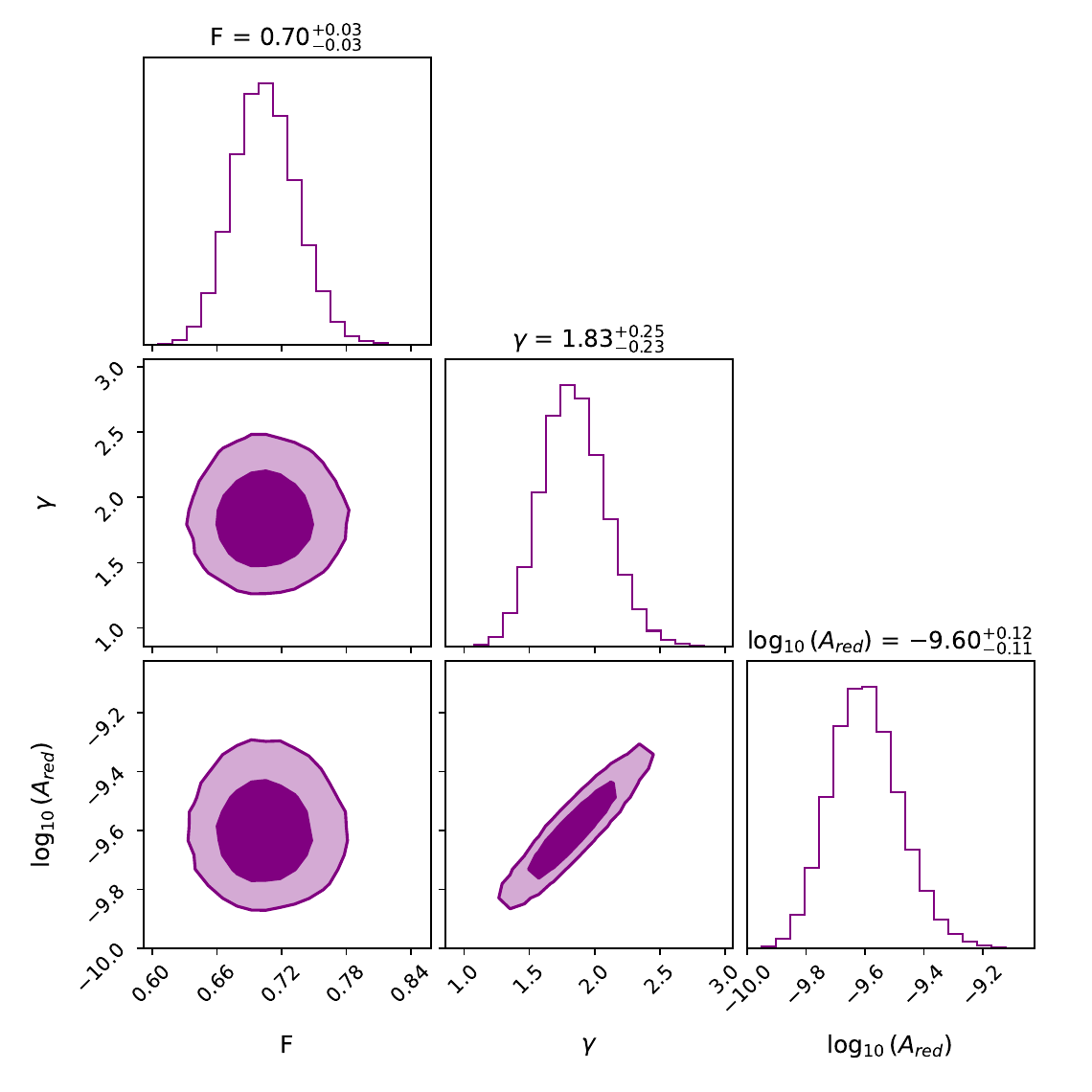}
    \caption{Timing Noise posterior of J2219+4754}
    \label{TNJ2219}
\end{subfigure}   
\begin{subfigure}{0.49\textwidth}
    \centering
    \includegraphics[width=\linewidth]{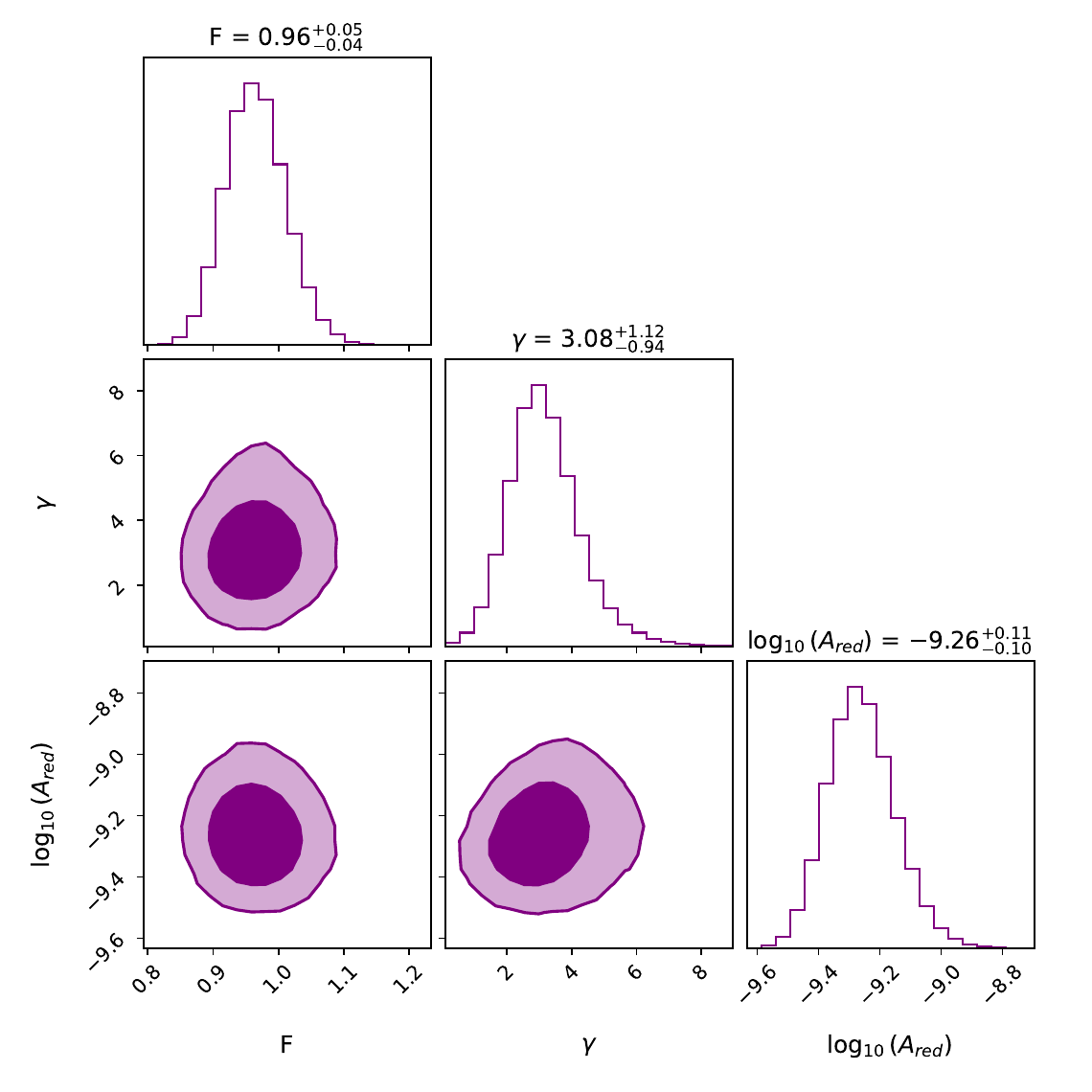}
    \caption{Timing Noise posterior of J2346--0609}
    \label{TNJ2346}
\end{subfigure}
\caption{Timing Noise posteriors with 68 and 95\% credible interval for our sample of pulsars. The symbols F, $\sigma_Q$, $A_{\rm red}$, $\gamma$ represent EFAC, EQUAD, Red noise Amplitude and Spectral index respectively.}
\end{figure*}
\renewcommand{\thefigure}{\arabic{figure}}

\begin{figure*}
    \centering
\begin{subfigure}{0.49\textwidth}
  \centering
  \includegraphics[width=\linewidth]{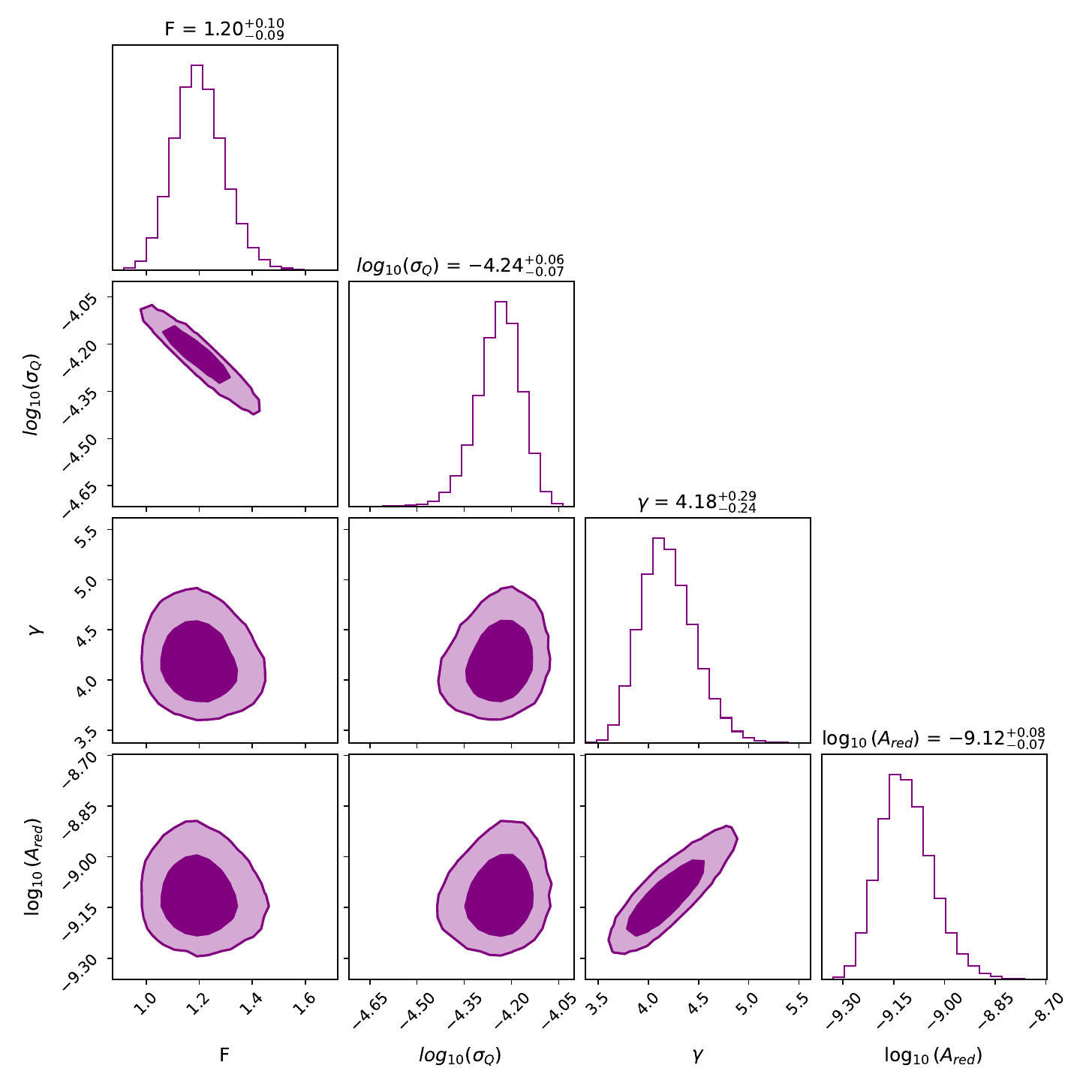}  
  \caption{Pre-glitch Timing Noise posterior of J0742--2822}
\end{subfigure}
\begin{subfigure}{0.49\textwidth}
  \centering
  \includegraphics[width=\linewidth]{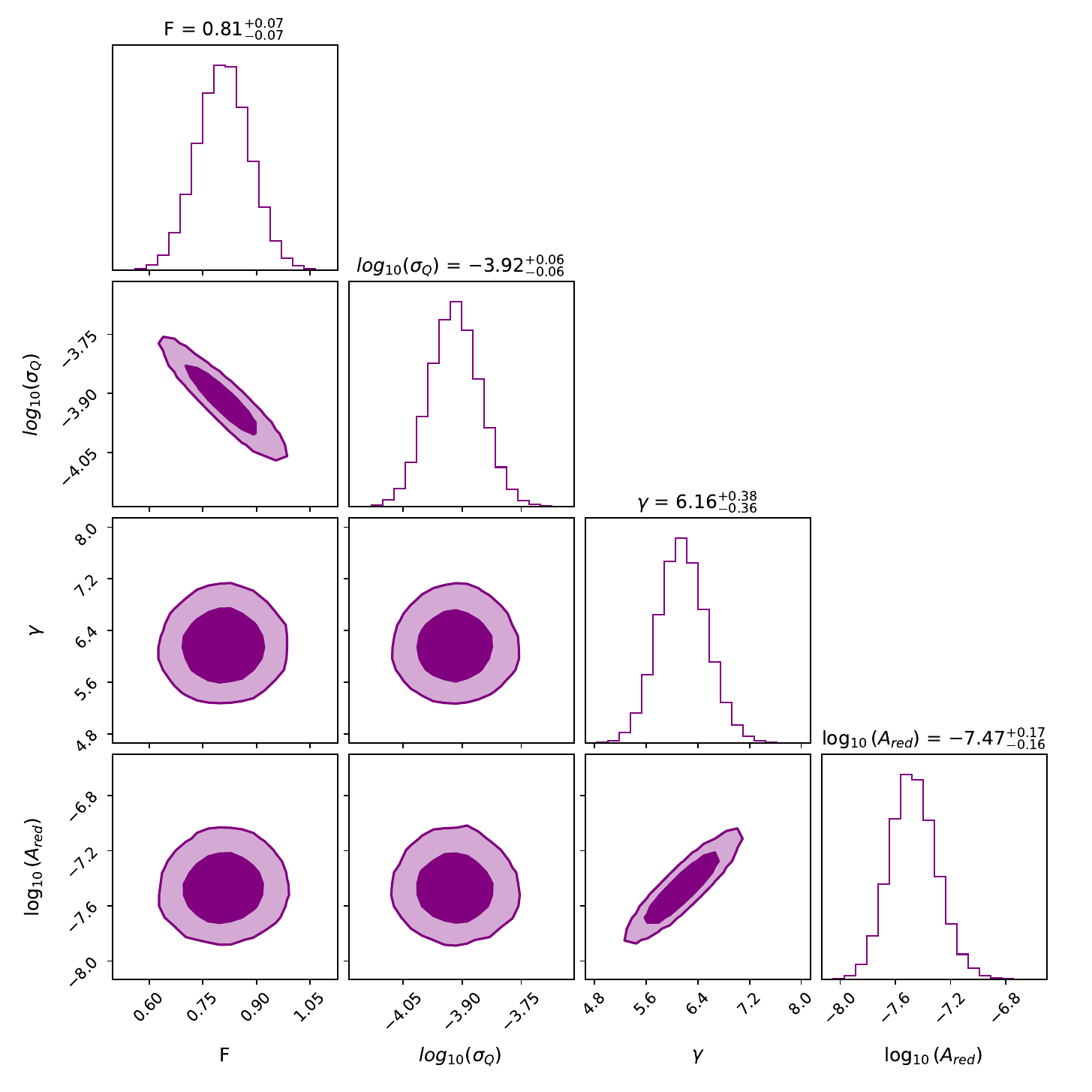} 
  \caption{Post-glitch Timing Noise posterior of J0742--2822}
\end{subfigure}   
\begin{subfigure}{0.49\textwidth}
  \centering
  \includegraphics[width=\linewidth]{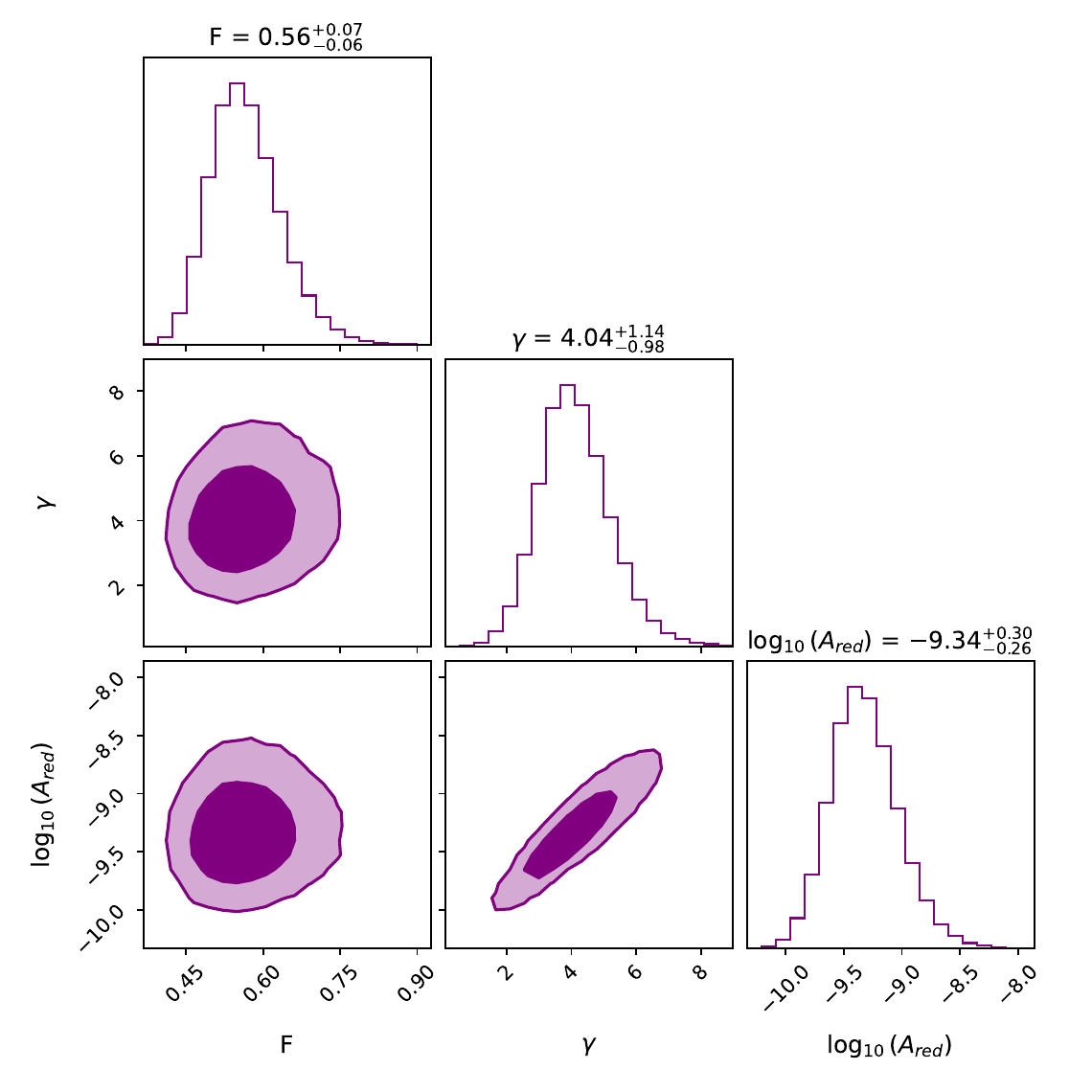}  
  \caption{Pre-glitch Timing Noise posterior of J1740--3015}
\end{subfigure}
\begin{subfigure}{0.49\textwidth}
  \centering
  \includegraphics[width=\linewidth]{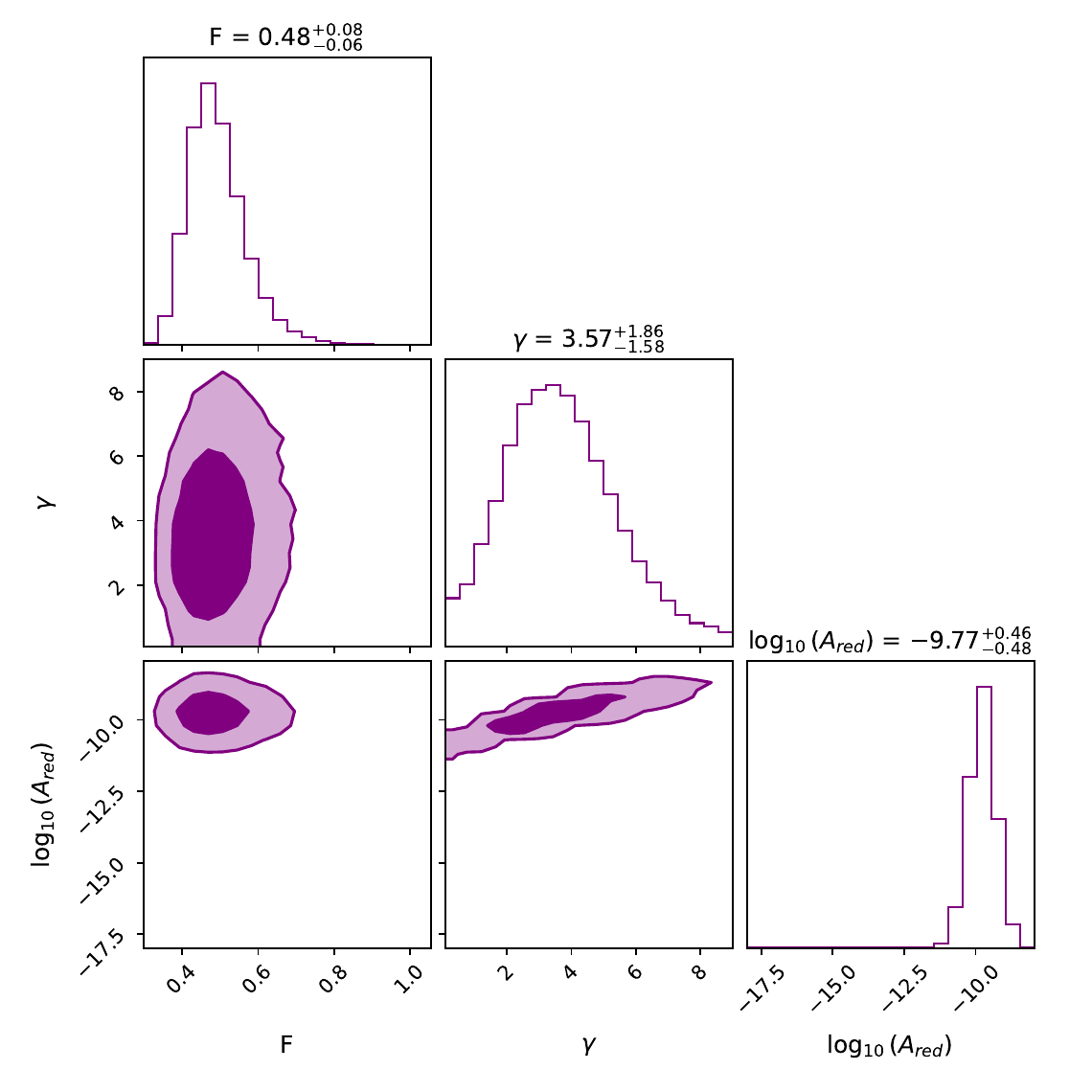} 
  \caption{Post-glitch Timing Noise posterior of J1740--3015}
\end{subfigure}   
\caption{Timing Noise posteriors with 68 and 95\% credible interval for our sample of pulsars. The symbols F, $\sigma_Q$, $A_{\rm red}$, $\gamma$ represent EFAC, EQUAD, Red noise Amplitude and Spectral index respectively.}
\label{TN_glitch}
\end{figure*}

\renewcommand{\thefigure}{\arabic{figure} (Cont.)}
\addtocounter{figure}{-1}
\begin{figure*}
    \centering
\begin{subfigure}{0.49\textwidth}
  \centering
  \includegraphics[width=\linewidth]{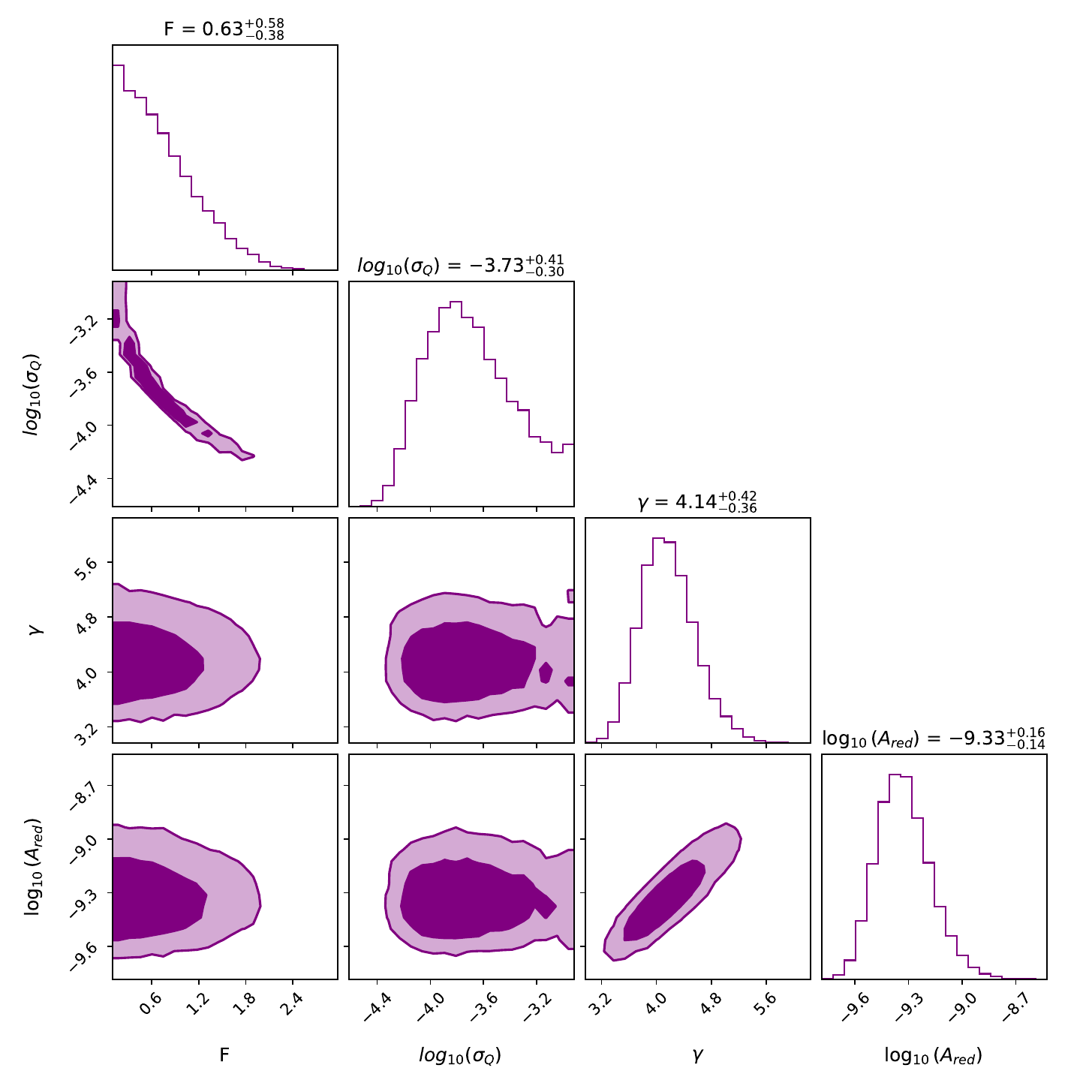}  
  \caption{Pre-glitch Timing Noise posterior for glitch at MJD 58517 in PSR J0835--4510.}
\end{subfigure}
\begin{subfigure}{0.49\textwidth}
  \centering
  \includegraphics[width=\linewidth]{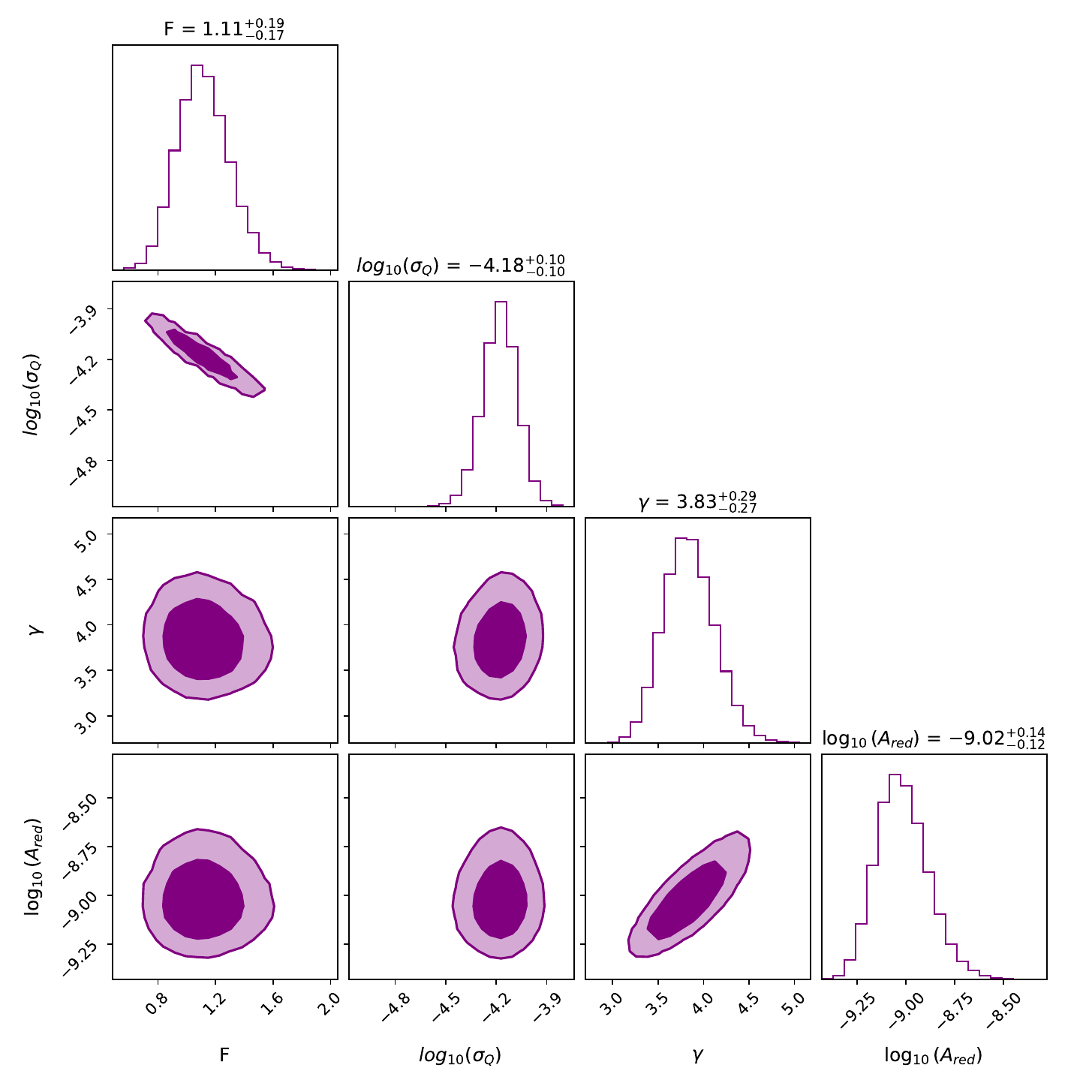} 
  \caption{Post-glitch Timing Noise posterior for glitch at MJD 58517 in PSR J0835--4510.}
\end{subfigure}   
\begin{subfigure}{0.49\textwidth}
  \centering
  \includegraphics[width=\linewidth]{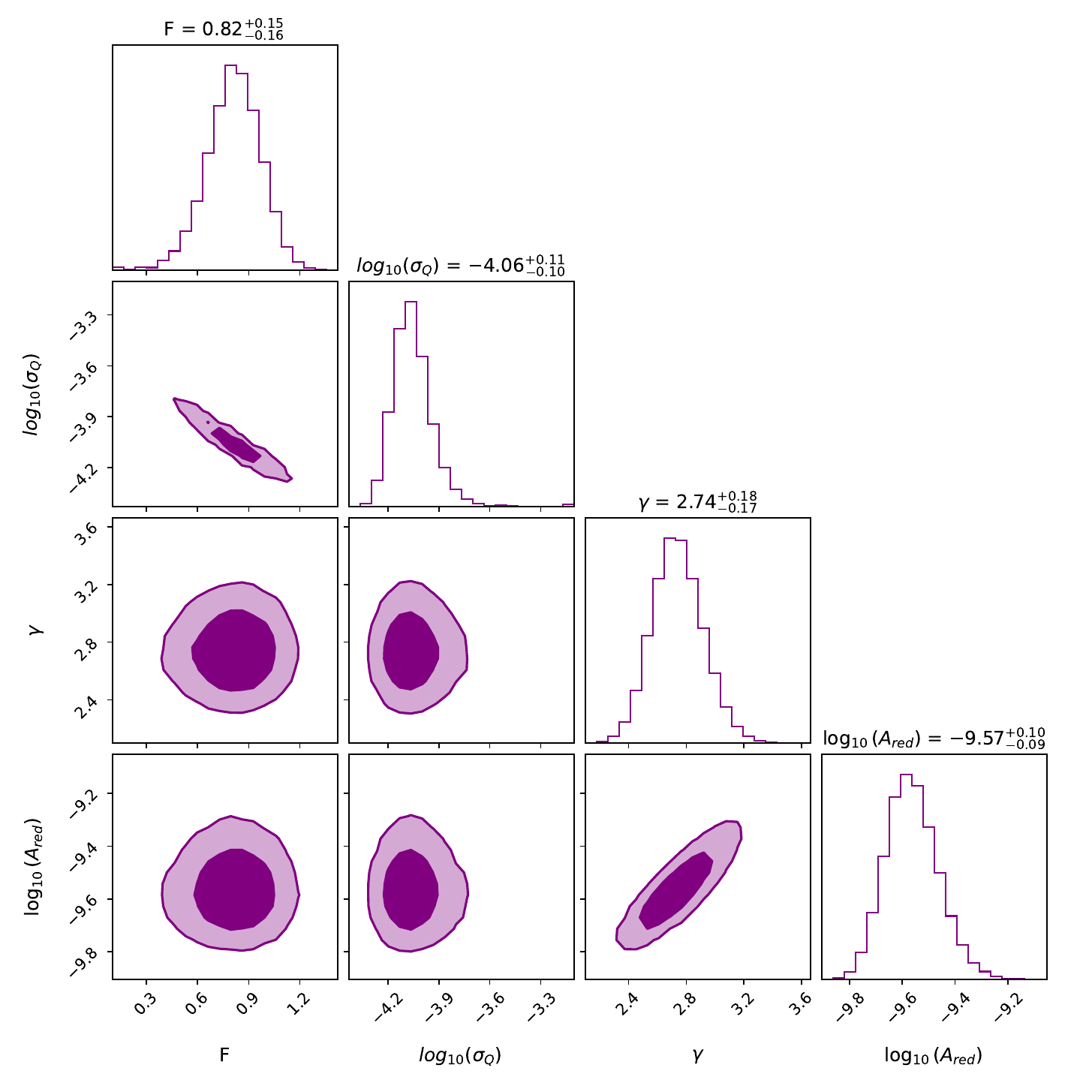}  
  \caption{Pre-glitch Timing Noise posterior for glitch at MJD 59418 in PSR J0835--4510}
\end{subfigure}
\begin{subfigure}{0.49\textwidth}
  \centering
  \includegraphics[width=\linewidth]{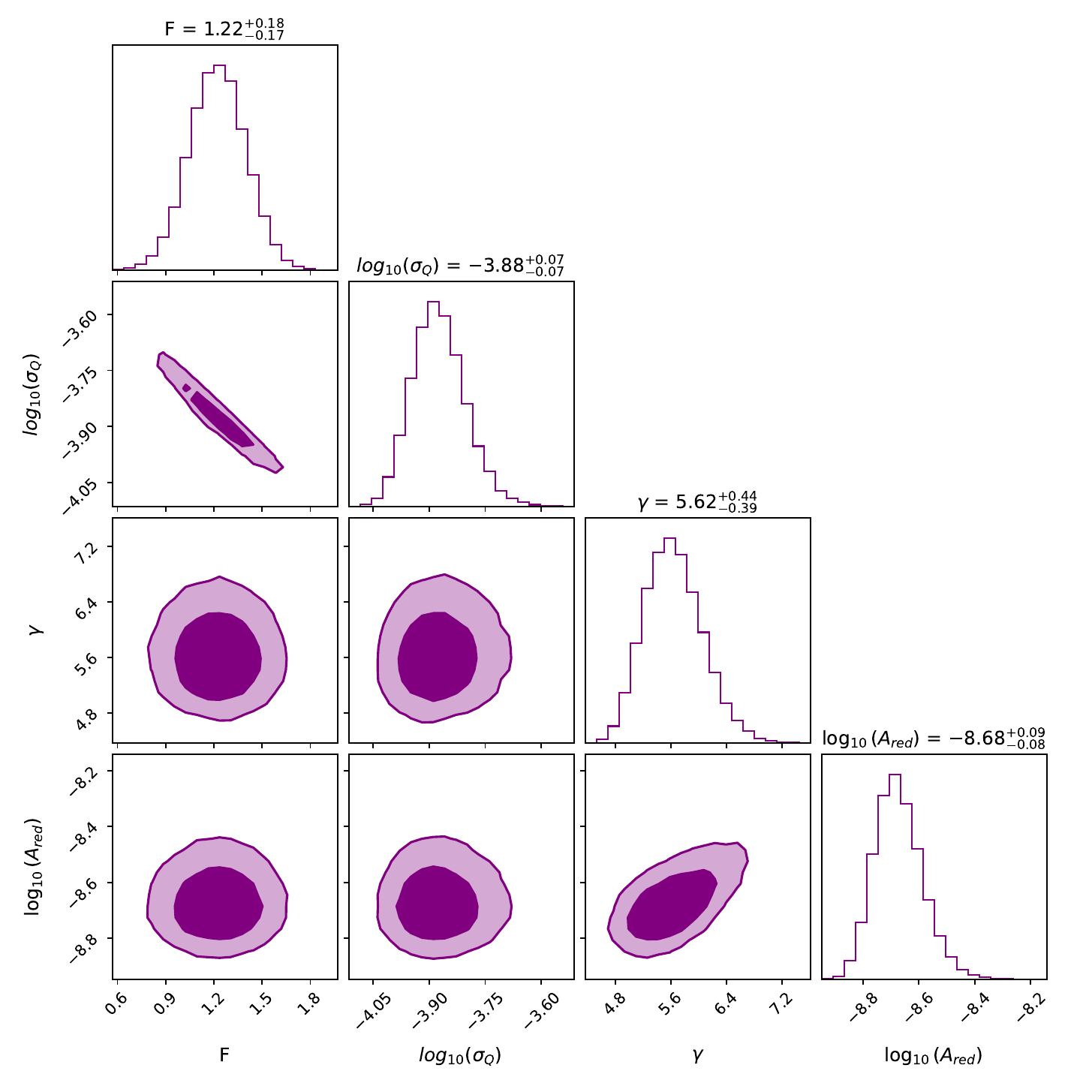} 
  \caption{Post-glitch Timing Noise posterior for glitch at MJD 59418 in PSR J0835--4510}
\end{subfigure}   
\caption{Timing Noise posteriors with 68 and 95\% credible interval for our sample of pulsars. The symbols F, $\sigma_Q$, $A_{\rm red}$, $\gamma$ represent EFAC, EQUAD, Red noise Amplitude and Spectral index respectively.}
\end{figure*}
\renewcommand{\thefigure}{\arabic{figure}}
\end{document}